\documentclass[12pt]{article} 
\pdfoutput=1

\usepackage[utf8]{inputenc}
\usepackage{jheppub}
\usepackage{amsmath}
\usepackage{amsfonts}
\usepackage{amssymb}
\usepackage{mathtools}
\usepackage{amsthm}
\usepackage{dsfont}
\usepackage{hyperref}
\usepackage[capitalise,sort]{cleveref}
\usepackage{pgfplots}
\usepackage{tabularx}
\usepackage{environ}
\usepackage[inline]{enumitem}
\usepackage{comment}
\usepackage{tikz}
\usetikzlibrary{decorations.pathmorphing,calc,fit,matrix,cd,shapes,arrows,positioning,chains,positioning,arrows.meta,hobby,patterns}
\usepackage{blkarray}
\usepackage{shadow}
\usepackage{fancybox}
\usepackage{bm}
\usepackage{abstract}
\usepackage{lscape}
\usepackage{verbatim}
\usepackage[normalem]{ulem}
\usepackage{subcaption}
\usepackage{mdframed}
\usetikzlibrary{shadows}
\pgfplotsset{
	compat=1.9,
	compat/bar nodes=1.8,
}

\makeatletter
\def\@xfootnote[#1]{%
	\protected@xdef\@thefnmark{#1}%
	\@footnotemark\@footnotetext}
\makeatother

\definecolor{prhigh}{HTML}{ff0000}
\definecolor{sechigh}{HTML}{e0fbfc}
\definecolor{prcolor}{HTML}{1d3557}
\definecolor{seccolor}{HTML}{457b9d}
\definecolor{tercolor}{HTML}{98c1d9}

\newcommand\bea{\begin{eqnarray}}
	\newcommand\eea{\end{eqnarray}}

\theoremstyle{plain}

\theoremstyle{definition}
\newtheorem{definition}{Definition}

\newtheorem{claim}{Claim}
\newtheorem{conjecture}{Conjecture}

\theoremstyle{remark}



\DeclareMathOperator{\U}{U}

\newcommand{\PP}{\mathbb{P}}
\newcommand{\RR}{\mathbb{R}}

\newcommand{\ID}{\mathds{1}}

\newcommand{\coma}{\, , \quad}
\newcommand{\fstop}{\, .}

\newcommand{\cV}{\mathcal{V}}
\newcommand{\cK}{\mathcal{K}}

\newcommand{\KK}{\text{\tiny KK}}

\newcommand{\WGC}{\text{\tiny WGC}}
\newcommand{\QG}{\text{\tiny QG}}

\newcommand{\het}{\text{\tiny het.}}
\newcommand{\IIA}{\text{\tiny IIA}}
\newcommand{\IIB}{\text{\tiny IIB}}
\newcommand{\PlD}[1]{\text{\tiny Pl,\,#1}}
\newcommand{\Plinf}{\text{\tiny Pl,\,$\infty$}}

\newcommand{\UoD}[1]{\text{\tiny U(1),\,#1}}
\newcommand{\ttiny}[1]{\text{\tiny #1}}

\newcommand{\fn}{\mathfrak{n}}

\renewcommand{\epsilon}{\varepsilon}

\makeatletter
\newsavebox{\measure@tikzpicture}
\NewEnviron{scaletikzpicturetowidth}[1]{%
	\def\tikz@width{#1}%
	\begin{lrbox}{\measure@tikzpicture}%
		\BODY
	\end{lrbox}%
	\pgfmathparse{#1/\wd\measure@tikzpicture}%
	\BODY
}
\makeatother

\makeatletter
\newcommand{\inlineitem}[1][]{%
	\ifnum\enit@type=\tw@
	{\descriptionlabel{#1}}
	\hspace{0pt}%
	\else
	\ifnum\enit@type=\z@
	\hspace{-15pt} \refstepcounter{\@listctr}\fi
	\quad\@itemlabel\hspace{0pt}%
	\fi}
\makeatother
\DeclareMathAlphabet{\mathdutchcal}{U}{dutchcal}{m}{n}

\def\fnote#1#2{\begingroup\def\thefootnote{#1}\footnote{#2}
	\addtocounter{footnote}{-1}\endgroup}

\tikzset{
	partial ellipse/.style args={#1:#2:#3}{
		insert path={+ (#1:#3) arc (#1:#2:#3)}
	}
}

\tikzset{cross/.style={cross out, draw=black, fill=none, minimum size=2*(#1-\pgflinewidth), inner sep=0pt, outer sep=0pt}, cross/.default={2pt}}

\tikzset{
	pics/torus/.style n args={3}{
		code = {
			\providecolor{pgffillcolor}{rgb}{1,1,1}
			\begin{scope}[
				yscale=cos(#3),
				outer torus/.style = {draw,line width/.expanded={\the\dimexpr2\pgflinewidth+#2*2},line join=round},
				inner torus/.style = {draw=pgffillcolor,line width={#2*2}}
				]
				\draw[outer torus] circle(#1);\draw[inner torus] circle(#1);
				\draw[outer torus] (180:#1) arc (180:360:#1);\draw[inner torus,line cap=round] (180:#1) arc (180:360:#1);
			\end{scope}
		}
	}
}

\tikzset{
	pics/hole/.style n args={2}{
		code = {
			\draw[fill=white] (0,0) arc(120:60:#1 and #2)  arc(-60:-120:#1 and #2);
			\draw (0,0) arc(-120:-130:#1 and #2) (#1,0) arc(-60:-50:#1 and #2);
		}
	}
}

\makeatletter
\newcommand*{\itemequation}[3][]{%
	\item
	\begingroup
	\refstepcounter{equation}%
	\ifx\\#1\\%
	\else  
	\label{#1}%
	\fi
	\sbox0{#2}%
	\sbox2{$\displaystyle#3\m@th$}%
	\sbox4{\@eqnnum}%
	\dimen@=.5\dimexpr\linewidth-\wd2\relax
	\ifcase
	\ifdim\wd0>\dimen@
	\z@
	\else
	\ifdim\wd4>\dimen@
	\z@
	\else 
	\@ne
	\fi 
	\fi
	\@latex@warning{Equation is too large}%
	\fi
	\noindent   
	\rlap{\copy0}%
	\rlap{\hbox to \linewidth{\hfill\copy2\hfill}}%
	\hbox to \linewidth{\hfill\copy4}%
	\hspace{0pt}
	\endgroup
	\ignorespaces 
}
\makeatother  

\hypersetup{
	pdftitle={The Minimal Weak Gravity Conjecture},    
	pdfauthor={\textcopyright\ Cesar Fierro Cota, Alessandro Mininno, Timo Weigand, Max Wiesner},     
	pdfsubject={hep-th},   
	pdfcreator={pdfLaTex},   
	pdfproducer={LaTex}, 
	pdfkeywords={},
	colorlinks=true
	,urlcolor=blue
	,anchorcolor=blue
	,citecolor=blue
	,filecolor=blue
	,linkcolor=blue
	,menucolor=blue
	,linktocpage=true
}

\allowdisplaybreaks[1] 

\crefname{figure}{Figure}{Figures}
\crefname{table}{Table}{Tables}
\crefname{definition}{Definition}{Definitions}
\crefname{proposition}{Proposition}{Propositions}
\crefname{claim}{Claim}{Claims}
\crefname{conjecture}{Conjecture}{Conjectures}

\renewenvironment{abstract}
{\small
	\begin{center}
		\bfseries \abstractname\vspace{-.5em}\vspace{0pt}
	\end{center}
	\list{}{%
		\setlength{\leftmargin}{4mm}
		\setlength{\rightmargin}{\leftmargin}%
	}%
	\item\relax}
{\endlist}

\begin{document}
	\pagestyle{plain}

	\setlength{\sboxrule}{0.5em}
	\setlength{\sboxsep}{1.5em} 
	\setlength{\sdim}{7pt}

	\makeatletter
	\@addtoreset{equation}{section}
	\makeatother
	\renewcommand{\theequation}{\thesection.\arabic{equation}}
	\pagestyle{empty}
	
	\rightline{ZMP-HH/23-21}
	\vspace{1.0cm}
	
	\begin{center}
		{\large \bf
			The Minimal Weak Gravity Conjecture
		} 
		
		\vskip 9 mm
		
		Cesar Fierro Cota,${}^1$ Alessandro Mininno,${}^1$ Timo Weigand,${}^{1,2}$ Max Wiesner${}^{3}$
		
		\vskip 9 mm

		\small ${}^{1}$\textit{II. Institut f\"ur Theoretische Physik, Universit\"at Hamburg, Luruper Chaussee 149,\\ 22607 Hamburg, Germany} 
		
		\vspace{2mm}
		
		\small ${}^{2}$\textit{Zentrum f\"ur Mathematische Physik, Universit\"at Hamburg, Bundesstrasse 55, \\ 20146 Hamburg, Germany  }   \\[3 mm]
		
		\small ${}^{3}$\textit{Jefferson Physical Laboratory, Harvard University, 17 Oxford Street, \\ Cambridge, MA 02138, USA} \\ [3mm]
		
		\fnote{}{\hspace{-0.75cm} cesar.fierro.cota at desy.de, \\ alessandro.mininno at desy.de, \\  timo.weigand at desy.de,  \\ mwiesner at fas.harvard.edu}
		
	\end{center}

	\begin{abstract}
		
		We examine the minimal constraints imposed by the Weak Gravity Conjecture (WGC) on the particle spectrum of a quantum gravity theory.
		Towers of super-extremal states have previously been argued to be required for
		consistency of the WGC under circle reduction. 
		At the same time, there exist classes of theories where no tower of super-extremal particle states below the black hole threshold has been established with current techniques. 
		We resolve this tension by arguing for the existence of a minimal radius for circle reductions of generic quantum gravity theories.
		Below this threshold, the notion of a circle compactification breaks down, bypassing the need for a tower of super-extremal states to satisfy the WGC after circle reduction.
		Based on this we propose that if a theory satisfies the WGC at the particle level below the black hole threshold, these states are sufficient for consistency under dimensional reduction, even in absence of a tower of super-extremal particles.
		Apart from general arguments, we provide independent evidence for this main result in F-, M- and string theory compactifications.
		According to the Emergent String Conjecture the only exception to the appearance of a minimal radius arises in asymptotically weak-coupling limits for heterotic strings, which aligns with the appearance of a weakly coupled super-extremal tower of particle states. This observation motivates a Minimal Weak Gravity Conjecture which states that towers of super-extremal particles occur if and only if they are required by consistency of the WGC under dimensional reduction. 
		
	\end{abstract} 
	
	\newpage
	\setcounter{page}{1}
	\pagestyle{plain}
	\renewcommand{\thefootnote}{\arabic{footnote}}
	\setcounter{footnote}{0}
	
	\tableofcontents

	\section{Introduction and Summary}
	\label{sec:intro}
	In quantum gravity effective field theories (EFTs) are widely believed to be subject to 
	non-trivial consistency conditions beyond the ones imposed by quantum field theory alone. The goal of the Swampland program~\cite{Vafa:2005ui} is to identify these conditions and to extract their implications for the low-energy physics (see~\cite{Brennan:2017rbf,Palti:2019pca,vanBeest:2021lhn,Grana:2021zvf,Agmon:2022thq} for reviews). 
	As a general property, quantum gravity (QG) theories require massive states beyond those described by the EFT for their consistency. Constraining the spectrum of massive states is, therefore, a key component of the Swampland program. Of particular interest for this work is the Weak Gravity Conjecture (WGC)~\cite{Arkani-Hamed:2006emk}, which gives a concrete constraint on the spectrum of massive states (see~\cite{Palti:2020mwc,Harlow:2022ich} for reviews of the WGC). In its mildest form, the WGC asserts that, in a consistent theory of quantum gravity coupled to a $\U(1)$ gauge theory, there must exist a charged state whose charge-to-mass ratio exceeds the charge-to-mass ratio of an extremal black hole (BH) charged under the same $\U(1)$ gauge theory. In other words, there must exist a state with mass $m$ and charge $q$ such that 
	\begin{equation}\label{WGCmild}
		\frac{g_{\ttiny{U(1)}}^2q^2}{m^2} \geq \left.\frac{g_{\ttiny{U(1)}}^2Q^2}{M^2}\right|_{\ttiny{BH}}\,,
	\end{equation}
	where $g_{\ttiny{U(1)}}$ is the coupling of the $\U(1)$ gauge theory. The argument typically invoked to motivate this conjecture is that extremal black holes should be able to decay. In theories with multiple $\U(1)$ factors, the requirement that any extremal charged black hole must be able to decay then amounts to the condition that the \textit{convex hull} of the charge-to-mass ratio of the super-extremal states contains the black hole region~\cite{Cheung:2014vva}. This generalization is referred to as the Convex Hull Condition (CHC) in the following. In principle, there are two ways to satisfy the WGC in an EFT: the super-extremal states satisfying \eqref{WGCmild} are either particle-like states or charged black holes that become super-extremal due to higher-derivative corrections to the Einstein-Maxwell action. Still, it is an interesting question if, and under which conditions, the WGC is, in fact, satisfied by particle-like excitations without having to resort to super-extremal black holes. 
	
	Even though, at first sight, a single super-extremal particle state seems to be enough to satisfy the WGC, there can possibly arise problems for compactifications of the theory. More precisely, after dimensional reduction on a circle there is an additional Kaluza--Klein (KK) $\U(1)$ gauge theory, and the CHC including this additional $\U(1)$ requires super-extremal states for which the convex hull of charge-to-mass ratios contains the black hole region. As shown in \cite{Heidenreich:2015nta}, this condition is violated for arbitrarily small circle radii if  the original higher-dimensional theory contains only a single (or, in fact, finite number of) super-extremal state(s). To avoid an inconsistency, \cite{Heidenreich:2015nta} proposed the tower Weak Gravity Conjecture (tWGC), which postulates a tower of super-extremal states with arbitrarily large charge along any ray in the charge lattice of the original theory (see also~\cite{Heidenreich:2016aqi,Montero:2016tif,Andriolo:2018lvp}). 
	
	If an infinite tower of super-extremal states with increasing charge and mass must exist in order for the WGC to be consistent under dimensional reduction, this would immediately rule out the possibility that the WGC can be satisfied strictly at the particle level. This is because the states in this tower would inevitably cross into the black hole region for large enough mass (provided the black hole threshold is not at infinite energies). However, there exists a caveat to this conclusion that was already discussed in the original works on the tWGC~\cite{Heidenreich:2015nta,Andriolo:2018lvp}, namely that the existence of an infinite tower can be circumvented in case there exists a minimal radius for a circle compactification of a theory with a finite number of super-extremal states. 
	
	Establishing the existence of super-extremal states in consistent theories of gravity in full generality is a rather difficult task, not to mention an entire tower of such states. Even in controlled settings such as supersymmetric string theory/M-theory and compactifications thereof, the exact spectrum of massive, charged states is not known outside certain weakly coupled or BPS protected sectors. In these sectors, the tower Weak Gravity Conjecture has been tested in a variety of settings. Such tests have been performed, for example, in M-theory compactifications on Calabi--Yau threefolds, where a tower of BPS states can be identified in certain regions of the charge lattice \cite{Alim:2021vhs,Gendler:2022ztv}. In case the charge-to-mass ratio is not protected by supersymmetry, towers of super-extremal states can be established in certain weak coupling limits, as has been systematically analyzed in F-/M-theory compactifications to six, five, and four dimensions with minimal supersymmetry in~\cite{Lee:2018urn,Lee:2018spm,Lee:2019tst,Klaewer:2020lfg,Cota:2022yjw,Cota:2022maf}. Nevertheless, there are notable instances where a tower of super-extremal states cannot straightforwardly be established in all directions of the charge lattice (at least not at the particle level). This is the case if there is no suitable weak coupling limit or the charge-to-mass ratio is not protected by a BPS bound~\cite{Alim:2021vhs,Cota:2022maf,Cota:2022yjw}. Even in these cases, there typically exist finitely many charged particle states whose existence is established. Thus, one can ask whether, in controlled string theory settings, the states whose existence can be established reliably with current techniques suffice to satisfy the WGC also after dimensional reduction. In case there is no obvious tower of super-extremal states, this can only be possible if the notion of a circle compactification of the theory breaks down below a minimal radius such that the analysis of \cite{Heidenreich:2015nta} is no longer applicable. The goal of this work is to investigate whether there indeed exists a minimal radius for circle compactifications in the presence of gauge factors for which no tower of super-extremal states can be identified, and hence to understand under which conditions the WGC can be satisfied already at the particle level. In addition to developing general arguments,
	we will test them as quantitatively as possible
	in supersymmetric compactifications of F-/M-theory.

	\subsubsection*{Summary of the Results}
	
	A key aspect of our analysis is the concept of a minimal radius for circle compactifications. If we consider a $D$-dimensional theory compactified on a circle, there exists a minimal circle radius $r_\ttiny{min.}$ below which we cannot view the resulting $(D-1)$-dimensional theory as a circle reduction of a higher-dimensional gauge theory. This happens whenever the KK scale associated with the circle compactification reaches the black hole threshold of the theory, which may differ from the Planck scale of the $(D-1)$-dimensional theory. We illustrate how this leads to a sensible definition of the minimal radius in circle compactifications of F- and M-theory on Calabi--Yau threefolds. In both cases, we generically identify a minimal radius for the circle and we confirm that the presence of such a minimal radius is sufficient to claim that the CHC is satisfied in the compactified theory, even without a tower of super-extremal states. 
	An exception occurs for theories that undergo emergent heterotic string limits \cite{Lee:2019wij}: Although a minimal radius can be defined for the KK $\U(1)_\KK$, a tower of states is necessary to satisfy the CHC with respect to the winding $\U(1)_w$. This is a consequence of the existence of T-duality for perturbative strings. Equivalently, in such theories, the minimal radius depends on the emergent heterotic string coupling, which goes to zero in the emergent string limits. By contrast, as we will see, perturbative open string theories do not require a tower of super-extremal states in their weak coupling limit.
	
	In general, one may wonder under which conditions the WGC is already satisfied at the particle level without having to resort to small (super-)extremal black holes to account for the super-extremal states. As we will argue in this paper, the WGC is satisfied at the particle level (though not necessarily by a tower of states) whenever we are dealing with genuine 0-form gauge theories\footnote{\label{footnote:pformsymmetries}We stick to the standard convention according to which the field strength of a $p$-form gauge symmetry is a $(p+2)$-form.} coupled to gravity for which it cannot be decided within the regime of validity of the EFT that the gauge sector may arise from a higher-form or a defect gauge theory. This is summarized in \cref{claim:WGC}.
	Given such a genuine 0-form gauge theory, the next interesting question concerns the conditions under which there can in fact exist towers of super-extremal particle states with arbitrarily high charge. We argue that such a tower of particles can only arise if a given gauge group either has a certain weak coupling limit or allows for a strong coupling limit. The precise necessary condition is formulated in \cref{claim:existencetower}.
	
	The main claim of the paper is then reached by combining the conditions for the existence of the (tower of) super-extremal states with the existence of a minimal radius for circle compactifications: Upon circle reduction the CHC is always satisfied even in the absence of a tower of super-extremal states for any value of the circle radius which allows for an interpretation of a $D$-dimensional gauge theory on a circle. The latter is in turn a necessary condition for the CHC for the KK $\U(1)_\KK$ to impose a sensible constraint.
	This claim is formulated in full generality as our Conjecture \ref{Conjecture1}. To support it, we illustrate the consistency of the WGC in the absence of towers of super-extremal particle states in explicit F-/M-theory examples. Moreover, whenever a tower of super-extremal particle states is needed for consistency of the WGC under dimensional reduction, we can indeed identify the necessary states with our current techniques.

	Finally, with Conjecture~\ref{Conjecture2} we propose a stronger version of Conjecture \ref{Conjecture1} which claims that towers of (super-)extremal particle states below the black hole threshold exist \textit{if and only if} they are required by consistency of the WGC under dimensional reduction. This happens only in emergent string limits, KK reductions with KK gauge bosons, and strongly coupled limits with exactly extremal states. 
	The conjectural part is that these are the only instances with super-extremal particle towers that occur.

	To summarize, string theory provides us with either a tower of states whose charge-to-mass ratio can be computed given our current techniques or with a minimal radius for the circle compactification precisely in such a way that the WGC is consistent under dimensional reduction.
	
	\subsubsection*{Structure of the Paper}
	
	The paper is structured as follows. In Section \ref{sec:review}, we review the basics of the WGC and its consistency under circle compactifications that led to the tower WGC~\cite{Heidenreich:2015nta}. In Section \ref{sec:necessitytower}, we introduce the notion of minimal radius for circle compactifications, with particular focus on F-/M-theory compactifications on Calabi--Yau threefolds. In Section \ref{sec:particlesandWGC}, we argue for which gauge theories a tower of super-extremal particles can exist, and under which conditions the WGC is satisfied already at the particle level. Using the example of Calabi--Yau threefold compactifications of F-/M-theory, we then show in Section~\ref{sec:absencetower} that even in the absence of a tower of super-extremal states, the CHC is satisfied under dimensional reduction due to the presence of a minimal radius for circle reductions. Finally, in Section \ref{sec:Conclusions}, we discuss our results and speculate about extending our Conjecture~\ref{Conjecture1} to Conjecture~\ref{Conjecture2}. Appendix \ref{app:Conventions} summarizes our conventions together with some aspects of the string dualities used in this paper.
	
	\section{Review: Weak Gravity Conjecture and Dimensional Reduction}
	\label{sec:review}
	
	In this section, we review the general ideas that led to the formulation of the different versions of the WGC \cite{Arkani-Hamed:2006emk}.
	For a more detailed overview of the literature, we also refer to the original works on the tower WGC \cite{Heidenreich:2015nta,Heidenreich:2016aqi,Montero:2016tif,Andriolo:2018lvp} and the reviews \cite{Palti:2019pca,vanBeest:2021lhn,Grana:2021zvf,Harlow:2022ich,Reece:2023czb}.
	
	\subsubsection*{The Weak Gravity Conjecture for Charged Particles}
	
	Let us consider an effective theory of gravity given by the Einstein-Maxwell action in Einstein frame\footnote{\label{footnote:FstarFconv}In our convention $F_{p+2}\wedge \star F_{p+2}=\frac{1}{(p+2)!}F_{\mu_1\ldots \mu_{p+2}}F^{\mu_1\ldots \mu_{p+2}}$.} 
	\begin{equation}\label{eq:Schargedparticles}
		S \supset \frac{M_{\PlD{D}}^{D-2}}{2}\int_{\mathcal{M}_D}  R\star\ID -\frac{1}{2g_{\UoD{D}}^2}\int_{\mathcal{M}_D} F_2\wedge \star F_2\coma
	\end{equation}
	where $M_{\PlD{D}}$ is the $D$-dimensional Planck mass and $g_{\UoD{D}}$ is the $\U(1)$ gauge coupling with mass dimension $\frac{4-D}{2}$.
	If such an action is obtained by some compactification of a higher-dimensional quantum gravity theory, then $g_{\UoD{D}}$ will be a function of the moduli defining the volumes of the compactification space. 
	
	The charge of a particle of mass $m$ under such a $\U(1)$ gauge symmetry is
	\begin{equation}
		q = \frac{1}{g_{\UoD{D}}^2}\int_{\mathcal{N}_{D-2}} \star F_2 \fstop
	\end{equation}
	The WGC \cite{Arkani-Hamed:2006emk} postulates that for any $\U(1)$ gauge field coupled to gravity, there must exist an object whose charge-to-mass ratio is larger than the charge-to-mass ratio of an extremal black hole charged under the same gauge fields, i.e.
	\begin{equation}
		\frac{
			q^2}{m^2} \geq \left.\frac{
			Q^2}{M^{2}}\right|_{\text{ext.}}\fstop
	\end{equation}
	One can relate the charge-to-mass ratio of an extremal black hole with the Planck mass of the $D$-dimensional theory, such that
	\begin{equation}\label{extremalgamma}
		\left.\frac{g_{\UoD{D}}^2
			Q^2}{M^{2}}\right|_{\text{ext.}} \equiv \gamma \frac{1}{{M_{\PlD{D}}^{D-2}}}\fstop
	\end{equation}
	In case the gauge coupling does not depend on the moduli the extremality factor is simply given by the charge-to-mass ratio of extremal Reissner--Nordstr\"om black holes, $\gamma =\frac{D-3}{D-2}$~\cite{Heidenreich:2015nta,Harlow:2022ich}. Using \eqref{extremalgamma}, the WGC condition reads
	\begin{equation}
		\frac{\left(g_{\UoD{D}}^2M_{\PlD{D}}^{D-4}\right)
			q^2}{m^2} \geq \gamma \frac{1}{{M_{\PlD{D}}^{2}}}\fstop
	\end{equation}

	\subsubsection*{The Magnetic Weak Gravity Conjecture}
	
	Together with the usual WGC, one can also introduce the so-called magnetic WGC that quantifies the scale up to which the EFT is expected to be valid in the weak coupling limit of the gauge theory. This scale is set by the mass of magnetic monopoles in Maxwell theory, which is proportional to the inverse of the gauge coupling $g_{\UoD{D}}$. It is then natural to define the magnetic WGC bound to be
	\begin{equation}
		\Lambda_\ttiny{WGC}^2 \lesssim \left(g_{\UoD{D}}^2M_{\PlD{D}}^{D-4}\right)M_{\PlD{D}}^2\fstop
		\label{eq:LambdaWGCD}
	\end{equation}

	\subsubsection*{The Convex Hull Condition}
	
	For multiple $\U(1)$s there is a generalization of the WGC that is called the Convex Hull Condition \cite{Cheung:2014vva}. For a particle $i$ charged under multiple  $\U(1)$s we can define the vector of charge-to-mass ratios as
	\begin{equation}
		{\bf{z}}_i = \frac{M_{\PlD{D}}^{\frac{D-2}{2}}}{m_i}\gamma^{-1/2}\left(g_{\UoD{D,\,1}}q_1,\ldots,g_{\UoD{D,\,N}}q_N \right) \equiv \frac{M_{\PlD{D}}^{\frac{D-2}{2}}}{m_i} {\bf{q}}_i\,.
	\end{equation}
	Then the CHC implies that the convex hull formed by the vectors of charge-to-mass ratios of all the multiparticle states must include the unit ball.
	
	\subsubsection*{The Tower Weak Gravity Conjecture}
	
	Based on the CHC, \cite{Heidenreich:2015nta,Heidenreich:2016aqi,Andriolo:2018lvp} have argued for the need for an infinite tower of particle states that satisfy the WGC. This stronger version of the WGC is commonly referred to as the tower WGC \cite{Heidenreich:2015nta,Heidenreich:2016aqi, Montero:2016tif,Andriolo:2018lvp}. 
	
	Let us briefly review the argument of \cite{Heidenreich:2015nta}. Consider the dimensional reduction of \eqref{eq:Schargedparticles} on a circle. The gauge sector of the lower-dimensional theory now contains an additional vector, the KK photon. In the conventions of \cite{Heidenreich:2015nta,Harlow:2022ich}, the KK photon gauge coupling in the $(D-1)$-dimensional theory is given by
	\begin{equation}
		\frac{1}{g_{\ttiny{KK,\,D--1}}^2} = \frac{1}{2}r_{S^1}^2M_{\PlD{D--1}}^{D-3} \text{ , or, analogously, } g_{\ttiny{KK,\,D--1}}^2M_{\PlD{D--1}}^{D-5} = \frac{2}{r_{S^1}^2M_{\PlD{D--1}}^{2}}
		\coma
	\end{equation}
	where we have defined $r_{S^1}$ to be the radius of the compactification circle. The WGC bound for a particle with charge $q_\KK$ and mass $m_{D-1}$ in the dimensional reduced theory is \cite{Heidenreich:2015nta}
	\begin{equation}
		m^2_{D-1} \leq \frac{1}{2}g_{\ttiny{KK,\,D--1}}^2q_\KK^2M_\PlD{D--1}^{D-3}\coma    
	\end{equation}
	which implies
	\begin{equation}
		m^2_{D-1} \leq \frac{q_\KK^2}{r_{S^1}^2}\fstop
	\end{equation}
	
	This bound can be compared with the spectrum of KK modes for a particle with mass $m_D$ in the original theory, i.e. 
	\begin{equation}\label{massformulaKK}
		m^2_{D-1} = m_D^2 + \frac{1}{r_{S^1}^2}\left(q_\KK-\frac{q\theta'}{2\pi}\right)^2\coma
	\end{equation}
	where $\theta'$ is a background axion VEV coming from the compactification of the 1-form connection of the $D$-dimensional theory, taking values $0\leq \theta'<2\pi$.\footnote{\label{footnote:thetadefinition}In the following sections, we will define $\theta'=2\pi \theta$, and work with $\theta\in [0,1[$.} 
	The KK tower from a massless particle
	in the $D$-dimensional theory hence always satisfies the WGC bound for the KK $\U(1)_\KK$, and saturates it for $\theta'=0$. 
	Since every gravitational theory contains the graviton as a massless particle, there always exists an infinite tower of super-extremal KK modes.
	
	A potential problem arises when we consider the other $\U(1)$ present in the theory and their mixing with the KK $\U(1)_\KK$. Let us define 
	\begin{equation}
		z_D = g_{\UoD{D}}M_{\PlD{D}}^{\frac{D-2}{2}}\gamma^{-1/2}\frac{|q|}{m_D}\coma
		\label{eq:zDvector}
	\end{equation}
	the charge-to-mass ratio for a particle of charge $q$ and mass $m_D$ in the original $D$-dimensional EFT. In the $(D-1)$-dimensional EFT, the charge-to-mass ratio for a particle charged under both KK $\U(1)_\KK$ and the original $\U(1)$ gauge theory is given by
	\begin{equation}
		{\bf{z}} = \frac{1}{m_{D-1}} \left(g_{\UoD{D--1}}\gamma^{-1/2}M_{\PlD{D--1}}^{\frac{D-3}{2}}q, \frac{1}{r_{S^1}}\left(q_\KK-\fn_D\right) \right)\coma
	\end{equation}
	where $\fn_D = \frac{q \theta'}{2\pi}=q\theta$ (see Footnote \ref{footnote:thetadefinition}). This vector, in fact, refers to a full tower of KK modes because there exists a state for all possible integer charges $q_\KK\in \mathbb{Z}$. We can also express this vector in terms of the $D$-dimensional EFT data as
	\begin{equation}\label{boldz}
		{\bf{z}} = \frac{1}{\left(m_{D}^2r_{S^1}^2+\left(q_\KK-\fn_D\right)^2\right)^{1/2}} \left(m_{D}r_{S^1}z_D, q_\KK - \fn_D \right)\fstop
	\end{equation}
	
	As analyzed in \cite{Heidenreich:2015nta}, the CHC for $\U(1)_\KK$ and the original $\U(1)$  requires a state in the original theory whose charge-to-mass ratio $z_D$ satisfies the inequality  
	\begin{equation}
		(m_Dr_{S^1})^2 \geq \frac{1}{4z_D^2(z_D^2-1)} + \frac{\fn_D(1-\fn_D)}{z_D^2}\fstop
		\label{eq:mDrvszD}
	\end{equation}
	Given any state with a fixed value of  $z_D$, this inequality is violated once $r_{S^1}$ drops below a certain value. Therefore, there exists a minimal value of the radius below which the CHC is not satisfied for any value of $z_D$.
	Already in \cite{Heidenreich:2015nta}, the authors noticed that there are two ways to solve this problem:
	\begin{enumerate}
		\item Either there exists a minimal radius $r_{\ttiny{min.}}$ below which the EFT breaks down since, e.g. the cutoff $\Lambda_D$ of the higher-dimensional theory is reached, $r_{\ttiny{min.}}\sim \Lambda_D^{-1}$. 
		\item Or there exists an infinite tower of super-extremal, i.e. $z_D\geq 1$, charged particles in the original $D$-dimensional theory with arbitrarily large mass $m_D$ and charge. In this case, the CHC of the lower-dimensional theory would be satisfied for all values of $r_{S^1}$. 
	\end{enumerate}
	The first solution may be considered unsatisfactory, as it would imply that the regime $r_{S^1}<r_{\ttiny{min.}}$ can never be considered within the validity of the EFT. However, from experience in string theory, we know that the small radius limit can also usually be described within the EFT by taking into account additional degrees of freedom. Hence, if the CHC in the lower-dimensional theory is to be valid for any value of $r_{S^1}$, we are led to requiring an infinite tower of super-extremal states in the higher-dimensional theory.  
	This argument gave rise to the formulation of the tower WGC, which indeed demands the existence of an infinite number of super-extremal states in the $D$-dimensional EFT to guarantee that the CHC is satisfied.

	\section{Minimal Radius in Quantum Gravity}
	\label{sec:necessitytower}
	
	The arguments of \cite{Heidenreich:2015nta,Andriolo:2018lvp} reviewed at the end of the previous section
	imply that when no tower of super-extremal states exists, the CHC may in principle be violated after circle reduction. However, as also reviewed, it was already stressed in the original work \cite{Heidenreich:2015nta} that in the full quantum gravity theory the radius might be bounded by a minimal value set by the scale at which the EFT breaks down. This would mean that, in the small radius limit, the charge and mass of the particle states could receive corrections compared to the analysis underlying \eqref{eq:mDrvszD}, which may prevent a violation of the CHC even for small radii.

	From a general quantum gravity point of view it is very natural to expect that the radius of a circle compactification should not become arbitrarily small --- or more precisely that below a certain minimal radius the notion of a compactification circle loses its meaning. To see this, compactify a $D$-dimensional theory of gravity on a circle of radius $r_{S^1}$ and consider the regime where $2 \pi r_{S^1} M_{\PlD{D--1}}\leq 1$. To deserve the name of a circle reduction, the effective theory must exhibit a Kaluza--Klein tower of particles, which by definition must lie below the
	black hole threshold defined as the mass of the smallest black hole, $M_\ttiny{BH, min.}$. 
	In a $D$-dimensional theory, the minimal mass of a $D$-dimensional black hole is determined by the quantum gravity cutoff $\Lambda_\QG$ via the relation (see, e.g. \cite{Dvali:2007hz,Agmon:2022thq}) 
	\begin{equation}\label{BHthresholdemstring}
		\frac{M_{\ttiny{BH, min.}}}{M_\PlD{D}}= \left(\frac{M_\PlD{D}}{\Lambda_\QG} \right)^{D-3} \,.
	\end{equation}
	Here, the quantum gravity cutoff is defined as the inverse radius of the smallest black hole, $r^{-1}_{\ttiny{H, min.}}\sim \Lambda_{\QG}$~\cite{Dvali:2007hz}.\footnote{The species scale as the quantum gravity cutoff and its relation to minimal black holes in quantum gravity have recently been analyzed in great detail in \cite{Long:2021jlv,Marchesano:2022axe,Bedroya:2022twb,Castellano:2022bvr,vandeHeisteeg:2022btw,Cribiori:2022nke,vandeHeisteeg:2023ubh,vandeHeisteeg:2023uxj,Cribiori:2023ffn,Cribiori:2023sch,Calderon-Infante:2023ler,Calderon-Infante:2023uhz,vandeHeisteeg:2023dlw,Castellano:2023aum,Castellano:2023jjt,Castellano:2023stg}.}
	Unless the limit of small KK radius gives rise to a tower of weakly coupled light states, the quantum gravitational cutoff of the $(D-1)$-dimensional theory is $\Lambda_{\QG} = M_{\PlD{D--1}}$.\footnote{For large radii the quantum gravity cutoff is instead set by the cutoff of the higher-dimensional theory, which can be taken to be $M_\PlD{D}$.} In such a situation, the bound 
	\begin{equation} \label{eq:minrad1}
		\frac{1}{2\pi r_{S^1}} \sim M_{\ttiny{KK}} \stackrel{!}{\leq} M_{\ttiny{BH, min.}} \sim M_{\PlD{D--1}}\,
	\end{equation}
	clashes with the assumption of having a small circle at $2 \pi r_{S^1} M_{\PlD{D--1}}\leq 1$. To treat the $(D-1)$-dimensional theory as a KK reduction on a small circle, it should be possible to satisfy the above bound parametrically; otherwise, we regard $2\pi r_{S^1} = M^{-1}_{\PlD{D--1}}$ as a ``minimal radius". 
	Indeed, we claim this to be the correct picture whenever the small radius limit itself is not associated with a weakly coupled tower. This assumption is satisfied, for example, for generic 6d/5d  circle compactifications of F-theory.  We will give independent evidence for the existence of a minimal radius in this framework in Section \ref{sec:minimalradiusFM}.  
	
	Notice that the above reasoning assumes that the black hole threshold $M_\ttiny{BH, min.}$ does not differ significantly from $M_{\PlD{D--1}}$. This is not necessarily the case if the quantum gravity cutoff $\Lambda_{\QG}$ drops parametrically below the Planck scale. To parametrically decrease the cutoff, we need to consider a limit in the quantum gravity field space where a large number of states become massless. Such limits correspond to infinite distance boundaries of the field space which, according to the Emergent String Conjecture~\cite{Lee:2019wij} are either decompactification limits or emergent string limits. In a pure decompactification limit, the black hole threshold is set by the mass of the smallest \emph{higher-dimensional} black holes given by the higher-dimensional Planck mass, which we identify with $\Lambda_\QG$ in decompactification limits. On the other hand, in emergent string limits the mass of the smallest black hole is set by the scale of the string-black hole transition such that (in the $(D-1)$-dimensional theory)
	\begin{equation}
		\frac{M_\ttiny{BH, min.}}{M_\PlD{D--1}}=\left(\frac{M_\PlD{D--1}}{M_s} \right)^{D-4}\,,
	\end{equation}
	where we identified the quantum gravity cutoff with the string scale, i.e. $\Lambda_\QG=M_s$ \cite{Dvali:2009ks}. In the emergent string limit, $M_s\ll M_\PlD{D--1}$ so that the minimal radius compatible with \eqref{eq:minrad1} can be very small and, in fact, go to zero in the limit $M_s/M_\PlD{D--1}\rightarrow 0$. 
	
	There are two scenarios in which we can thus consider very small radii. The first occurs when $r_{S^1} \to 0$ itself corresponds to an emergent string limit in which a weakly coupled string description arises. This is the case for circle compactifications of M-theory to Type IIA string theory, where in the small radius regime the new QG cutoff is the emergent Type IIA string scale. In this case, there is no notion of a minimal radius. Nevertheless, for M-theory compactifications on Calabi--Yau threefolds we will argue in Section \ref{sec:minradius5to4} that upon further circle compactification there exists a minimal radius below which the theory loses its nature as a KK theory, albeit for a different reason from the one that the KK tower exceeds the black hole threshold. 
	
	The second scenario is that of a circle compactification of perturbative string theory ($g_s\ll 1$). Here, the notion of the minimal radius is a bit more subtle. Since in the perturbative regime, $g_s\ll 1$, the string scale is parametrically below the Planck scale by \eqref{eq:minrad1}, we can, in principle, also consider radii smaller than $M_\PlD{D--1}$. In perturbative string theory there are winding states that become light in the small radius limit as a consequence of T-duality. It turns out that the analysis differs for the heterotic string, where the winding tower is charged, and for Type II strings with D-branes.
	
	Let us begin with the heterotic string. At first sight, T-duality effectively induces a minimal radius $r_{S^1}\sim \sqrt{\alpha'}$ with $\ell_s = 2 \pi \sqrt{\alpha'}$ the string length, since below this radius the effective theory can be treated as a reduction on a dual circle with radius $r_{\tilde{S}^1}={\alpha'}/r_{S^1}$. However, such a duality only exists in the presence of a 2-form in the higher-dimensional theory; upon circle reduction, the latter gives rise to a winding $\U(1)_w$ in the lower-dimensional theory. Due to T-duality, this winding $\U(1)_w$ should be treated on the same footing as the KK $\U(1)_\KK$ as long as the winding modes are below the black hole threshold. In perturbative string limits $g_s\to 0$, the black hole threshold \eqref{BHthresholdemstring} lies above the Planck scale and, in particular, scales with $g_s$, as $M_{\ttiny{BH, min.}} \sim {1}/({\sqrt{\alpha'} g_s^2)}$.  Therefore, the winding states have masses well below $M_{\ttiny{BH, min.}}$ for any value of the radius, as long as $g_s\to 0$. Viewing the winding $\U(1)_w$ effectively as a KK $\U(1)_\KK$ for small radii tells us that the minimal radius imposed by T-duality is \textit{not} the minimal radius to which we refer in this work. 
	Equivalently, rather than restricting to the regime $r_{S^1} \geq \sqrt{\alpha'}$ and taking into account also $\U(1)_w$, one can consider the entire range of $r_{S^1}$ subject to the constraint $M_\KK \leq M_{\ttiny{BH, min.}}$ for $\U(1)_\KK$. The resulting minimal radius scales as $r_{S^1} \sim g_s^2 \sqrt{\alpha'}$ and hence goes to zero in the perturbative limit. We will analyze this in more detail in Section \ref{sec:6dto5dheterotic}.
	For the D-brane gauge theories in perturbative Type II string theory, on the other hand, the theory can only be interpreted as being coupled to gravity above the self-dual radius in a sense that we will discuss in Section \ref{sec:pertDbranes}. Here, the self-dual radius hence does play the role of a minimal radius.
	
	In the remainder of this section, we analyze circle compactifications of F-theory and M-theory as two representative classes of theories and argue in more detail for the appearance of a minimal radius.

	\subsection{Minimal Radius via F-/M-theory Duality}
	\label{sec:minimalradiusFM}
	
	We start by arguing for a minimal radius in circle compactifications of six-dimensional $\mathcal{N}=(1,0)$ theories arising from F-theory on elliptically fibered Calabi--Yau threefolds $\pi:X_3\rightarrow B_2$.
	The argument we will present is complementary to the general considerations around \eqref{eq:minrad1} and is based on the details of the effective action.

	The effective action of F-theory compactifications is most conveniently described as a decompactification limit of M-theory in one dimension lower. 
	The standard duality between F-theory on $B_2\times S^1$ and M-theory compactified on $X_3$ identifies the radius, $r_{S^1}$, of the $S^1$ with the volume, $\text{vol}(T^2)$, of the generic elliptic fiber of $X_3$ as\footnote{\label{footnote:conventionvolumes}In the following, it is important to keep track of the units in which volumes, masses, etc. are measured. Therefore, we denote by $\text{vol}(\dots)$ \textit{dimensional} volumes. On the other hand, we use $\mathcal{V}_{\dots}$ for \textit{dimensionless} volumes measured in M-theory units and $\mathcal{V}_{\dots,\,s}$ for volumes measured in string units.} 
	\begin{equation}\label{circleradius}
		r_{S^1} M_\ttiny{11d} = \frac{1}{2\pi\text{vol}(T^2) M_\ttiny{11d}^2} 
		\,.
	\end{equation}
	Here $M_\ttiny{11d}$ is the fundamental M-theory scale.\footnote{We explain our conventions in Appendix \ref{app:Conventions}.} In terms of the string scale, $M_s$, which F-theory inherits from Type IIB string theory, the radius is instead given by 
	\begin{equation}
		r_{S^1}^2 M_s^2= \frac{g_\IIB}{(2\pi)^2\text{vol}(T^2)M_s^2}\,,
		\label{eq:radiustorusrelIIB}
	\end{equation}
	where $g_\IIB$ is the Type IIB string coupling. 
	From the above expression, it is clear that the six-dimensional F-theory limit of M-theory corresponds to the limit of small elliptic fiber volume. More generally, the generic fiber can split into multiple components over certain loci in the base, associated with additional fibral curves $C^a$. If we denote their volume by $t^a= \text{vol}(C^a) M_\ttiny{11d}^2$ and the volume of the base curves, $C^\alpha_b$, by $t^\alpha=\text{vol}(C^\alpha_b) M_\ttiny{11d}^2$, the F-theory limit corresponds to rescaling~\cite{Grimm:2010ks}
	\begin{equation}\label{Ftheorylimit}
		t^a \rightarrow \lambda^{-1}\, t^a_0\,, \qquad t^\alpha \rightarrow \lambda^{1/2}\, t^\alpha_0 \,,\qquad \lambda \rightarrow \infty\,. 
	\end{equation}
	Equipped with this dictionary, we now argue that there is a minimal radius below which the notion of a circle compactification breaks down in six-dimensional F-theory on $S^1$. To this end, note that from the point of view of the five-dimensional $\mathcal{N}=1$ supergravity, the radius $r_{S^1}$ of the KK circle is part of a vector multiplet. 
	In view of the identification in \eqref{circleradius}, the decrease of the radius $r_{S^1}$ corresponds to a motion in the K\"ahler moduli space of $X_3$ which increases the volume of the elliptic fiber ${\mathbb E}$ in M-theory units, $\tau = \mathcal{V}_\mathbb{E}$, while keeping the total volume of $X_3$ in M-theory units, $\mathcal{V}_{X_3}$, fixed. This is because the latter sits in a hypermultiplet, rather than a vector multiplet, and changing $\mathcal{V}_{X_3}$ can hence not be interpreted as changing the radius of the KK circle. 
	
	In the F-theory limit \eqref{Ftheorylimit}, the volume of $X_3$ factorizes as
	\begin{equation}\label{factorizedVB3}
		\mathcal{V}_{X_3} \rightarrow \tau\, \mathcal{V}_{B_2} \,,
	\end{equation}   
	where $\mathcal{V}_{B_2}$ is the volume of the base $B_2$ in M-theory units. If \eqref{factorizedVB3} was exact, we could in fact take the limit opposite to \eqref{Ftheorylimit}, i.e. $\tau\rightarrow \infty$, which by \eqref{circleradius} we could identify as the limit $r_{S^1}\rightarrow 0$. Eq. \eqref{factorizedVB3} is indeed exact in the case of a trivial fibration and hence in the absence of 7-branes and gauge theories in the six-dimensional theory. However, in the presence of 7-branes,  \eqref{factorizedVB3} receives corrections away from the strict F-theory limit and the volume of $X_3$ takes the general form (in M-theory units)
	\begin{equation} \label{VX3parametrization}
		\mathcal{V}_{X_3} = \alpha \tau^3 + \beta \tau^2 + \tau \mathcal{V}_{B_2} \,,
	\end{equation}
	where $\alpha$ and $\beta$ depend on topological invariants of $X_3$ and on (ratios of) the remaining $h^{1,1}(X_3) -1$ K\"ahler moduli of $X_3$. 
	We will give explicit expressions for $\alpha$ and $\beta$ in \eqref{eq:genVolintheta}.
	For our purposes, we can restrict the discussion to the slice in K\"ahler moduli space where $\mathcal{V}_{X_3}=1$ because in all relevant expressions the explicit value of $\mathcal{V}_{X_3}$ will drop out. 
	
	Taking the limit $r_{S^1} \to 0$, if possible at all, would correspond to a limit $\tau \to \infty$ at $\mathcal{V}_{X_3}=1$ and an infinite distance limit in the classical K\"ahler moduli space of $X_3$ at fixed volume. For general Calabi--Yau threefolds, such limits have been classified in \cite{Lee:2019wij} and fall into one of the following two categories:
	\begin{enumerate}
		\item  $X_3$ 
		admits a surface fibration whose 
		generic fiber $\mathcal{F}$ is a K3 or abelian surface such that 
		\begin{equation}
			\mathcal{V}_{\mathcal F} \sim \frac{1}{\lambda} \to 0 \qquad  {\rm while}  \qquad \mathcal{V}_{C} \sim \lambda \to 0 \,,
		\end{equation}
		where $C$ denotes the base of the surface fibration;
		\item 
		$X_3$ admits a genus-one fibration
		with base ${\tilde B_2}$ and generic fiber $\mathcal{E}$ such that 
		\begin{equation}
			\mathcal{V}_{{\tilde B}_2} \sim \mu \to \infty \qquad  {\rm while} \qquad \mathcal{V}_\mathcal{E} \sim \frac{1}{\mu} \to 0\,;
		\end{equation}
		furthermore $\mathcal{V}_{\mathcal{E}}$ is parametrically smaller than the square-root of the volume of any K3 or abelian surface fiber, which may also shrink in the limit.\footnote{The latter condition is meant to distinguish the second case from the first: If there is both a surface fibration and a genus-one fibration one subsumes  the limit into the two different cases depending on which of the fibers shrinks at the faster rate.}
	\end{enumerate}
	Applied to our genus-one fibered Calabi--Yau threefold $X_3$ with fiber $\mathbb E$, 
	this leaves us with the following possibilities for a limit $\mathcal{V}_\mathbb{E} =: \tau \to \infty$ at $\mathcal{V}_{X_3}=1$:
	\begin{enumerate} [label={\alph*.},ref={\alph*}]
		\item \label{fibrationcase1} $X_3$ admits, in addition to the genus-one fibration with fiber ${\mathbb E}$ and $\tau = \mathcal{V}_\mathbb{E}$, a K3 or abelian surface fibration incompatible with the
		genus-one fibration of $X_3$. To delineate this case from the one below, we also assume that the K3 or abelian surface fiber does not admit a genus-one fibration compatible with the surface fibration.
		Then $\mathbb E$ must be a multiple cover of the rational curve $C$ acting as the base of the surface fibration. 
		\item \label{fibrationcase2}
		$X_3$ admits, in addition to the original genus-one fibration with fiber ${\mathbb E}$, another incompatible genus-one fibration with fiber $\mathcal{E}$ and base ${\tilde B}_2$.
		The original fiber $\mathbb E$ is contained in ${\tilde B}_2$ or a multi-cover thereof or in a multi-cover of its base if $\tilde B_2$ itself admits a fibration structure.

	\end{enumerate}
	In Case \ref{fibrationcase1}, the limit $\tau \to \infty$ describes a five-dimensional emergent string limit in which a solitonic critical heterotic or Type II string (obtained by wrapping an M5-brane on the K3 or abelian surface fiber, respectively) becomes asymptotically weakly coupled. 
	The tension of this string is determined by the volume of the shrinking surface fiber $\mathcal{F}$,
	\begin{equation}
		M^2_\ttiny{het./II} = 16 \pi^2 \mathcal{V}_{\mathcal F} M^2_\ttiny{11d} \,.
	\end{equation}
	Geometrically, such limits are only possible if both $\alpha,\beta \to 0$ in the parametrization \eqref{VX3parametrization} (or if they vanish already identically). 
	What is important for us is that the geometric modulus $\tau \to \infty$ describes, in the new duality frame set by the emergent string, a weak coupling limit of a genuinely five-dimensional heterotic or Type II theory and not the limit $r_{S^1} \to 0$ of the original 6d/5d KK reduction of F-theory. To see this, we notice that the KK scale of the original circle compactification from six to five dimensions is related to the black hole threshold \eqref{BHthresholdemstring} in the five-dimensional emergent string limit via
	\begin{equation}
		M_\KK = (2\pi \tau) M_\ttiny{11d} \sim \left(\frac{M_{\ttiny{11d}}}{\Lambda_\QG}\right)^2 M_\ttiny{11d} = M_\ttiny{BH, min.}\,.
	\end{equation}
	Here we used $\Lambda_\QG = M_\ttiny{het./II} = \tau^{-1/2}M_\ttiny{11d}$ in the heterotic string limit. Therefore, the KK scale does not lie parametrically below the black hole threshold, as we would expect in the case where the circle compactification is a sensible description of the geometry. 
	
	In Case \ref{fibrationcase2}, there are two possibilities:
	The first option is that the limit $\tau \to \infty$ describes a decompactification limit of the five-dimensional M-theory along a different circle of radius $\tilde r M_\ttiny{11d} = \frac{1}{\mathcal{V}_\mathbb{E}}$. This is the case provided that $\mathcal{V}_\mathbb{E}$ is parametrically smaller than the square root of any surface fiber volume that may shrink. This limit corresponds to a \textit{different} six-dimensional F-theory limit for the five-dimensional M-theory on $X_3$.
	Clearly, the KK scale associated with this different circle compactification parametrically drops
	below the original KK scale:
	\begin{equation}
		M^2_\ttiny{KK, 1} = (2 \pi \tau)^2 M^2_\ttiny{11d} \gg  (2 \pi \mathcal{V}_\mathbb{E})^2 M^2_\ttiny{11d}  =M^2_\ttiny{KK, 2} \,.
	\end{equation}
	The second possibility is that we again encounter a genuinely five-dimensional emergent string limit. 
	For this to occur, ${\tilde B_2}$ must itself admit a fibration structure such that $X_3$ also exhibits a K3 or abelian surface fibration whose generic fiber volume shrinks faster than $\mathcal{V}^2_\mathbb{E}$.
	In any event, a limit of Case \ref{fibrationcase2} requires that at least $\alpha \to 0$ (or $\alpha =0$ in the first place).

	These considerations show that even if a geometric limit $\tau \to \infty$ at $\mathcal{V}_{X_3}=1$ can be taken, it cannot be interpreted as a limit $r_{S^1} \to 0$ for the original 6d/5d KK reduction of the six-dimensional F-theory. We interpret this as pointing to the existence of a lower bound on the radius $r_{S^1}$ of a circle reduction from six to five dimensions for the specific quantum gravity obtained by F-theory on $X_3$.

	The problem of finding the minimal possible radius $r_\ttiny{min.}$ is then equivalent to determining the maximal possible fiber volume $\tau_\ttiny{max.}$ 
	compatible with the condition $\mathcal{V}_{X_3}=1$, and for which $\tau$ still enjoys the interpretation of the inverse of the original KK radius. 
	If $\alpha$ and $\beta$ in \eqref{VX3parametrization} are both bounded from below by a $\mathcal{O}(1)$ number throughout the moduli space, an upper bound on $\tau$ is simply obtained by solving for
	\begin{equation} \label{maxtauprescr1}
		\max_{\tau\geq 0}\left(\,\tau \left|\,\alpha \tau^3 + \beta \tau^2 + \tau \mathcal{V}_{B_2} =1\right.\right) \,.
	\end{equation}
	If the limit falls under Case \ref{fibrationcase1}, where $\alpha \to 0$ and $\beta \to 0$, this prescription is not  sufficient. Rather, we need to impose the condition that the KK scale for the 6d/5d circle reduction is below the string scale associated with the genuinely five-dimensional emergent string.
	This amounts to demanding
	\begin{equation}
		M_\KK^2 \leq M^2_\ttiny{het./II}    \Longrightarrow   (2 \pi \tau_\ttiny{max.})^2 =  16 \pi^2 \mathcal{V}_{\mathcal F} \,.
	\end{equation}
	Since in this case the fiber $\mathcal{F}$ is an $m$-fold cover of $B_2$ this means that 
	\begin{equation}
		\tau^2_\ttiny{max.} = 4 m \,  \mathcal{V}_{B_2} \,.
	\end{equation}
	Together with \eqref{maxtauprescr1} this implies that 
	\begin{equation} \label{taumax-emstr}
		\tau_\ttiny{max.} \simeq (4m)^{\frac{1}{3}} \,.
	\end{equation}
	A limit of Case \ref{fibrationcase2} can either be an emergent string limit or a decompactification limit.
	Let us first analyze the latter case. A constraint on $\tau$ now arises by demanding that the original KK scale, $M_\ttiny{KK, 1}$, to be below the KK scale set by the volume of $\mathcal{E}$, i.e.
	\begin{equation}
		M_\ttiny{KK, 1} \leq M_\ttiny{KK, 2}   \Longrightarrow  \tau_\ttiny{max.} \leq \mathcal{V}_\mathcal{E} \,.
	\end{equation}
	At the same time, the second fibration structure imposes that in the limit of small $\mathcal{E}$,
	\begin{equation}
		\mathcal{V}_{X_3} \simeq \mathcal{V}_\mathcal{E} \mathcal{V}_{\tilde{B}_2} = \mathcal{V}_\mathcal{E}   (x \tau^2 + y \tau)  \,,
	\end{equation}
	where $x = \mathcal{O}(1)$ and $y$ depends on other curve moduli on $\tilde B_2$. 
	Together, this gives the constraint
	\begin{equation}
		x \tau^3_\ttiny{max.} + y \tau^2_\ttiny{max.} = 1 \,.
	\end{equation}
	For $x \neq 0$, this implies $\tau_\ttiny{max.} = \mathcal{O}(1)$. If $x=0$, on the other hand, this means that
	${\tilde B}_2$ is fibered over a rational base curve $C$ and
	${\mathbb E}$ is a multi-cover of this curve.
	Let us denote the fiber of ${\tilde B_2}$ by $\Sigma$. For the limit to correspond to a decompactification limit, rather than to an emergent string limit, we must be in the regime 
	\begin{equation}
		\mathcal{V}_\mathbb{E}^2 \leq \mathcal{V}_\mathbb{E}\mathcal{V}_{\Sigma}\coma
	\end{equation}
	as otherwise the KK scale associated with 
	$\mathcal{V}_\mathbb{E}$ sits above the 
	emergent string scale associated with the surface fiber obtained by fibering $\mathbb{E}$ over $\Sigma$. On the other hand, imposing that
	$\tau_\ttiny{max.} = \mathcal{V}_\mathcal{E}$ together with $\mathcal{V}_{X_3} \simeq \mathcal{V}_\mathcal{E} \mathcal{V}_\mathbb{E}\mathcal{V}_\Sigma = 1$ gives $\tau^3_\ttiny{max.} = 1$. 
	The emergent string limit, on the other hand, is analyzed as before.

	\subsubsection*{Illustration: Elliptic Fibration with One Extra Section} 
	
	We now illustrate these considerations
	in an F-theory model
	with abelian gauge symmetry, which lends itself to quantitative tests of the WGC.
	
	Abelian gauge factors in F-theory arise if $X_3$ admits additional rational sections $S_A: B_2 \rightarrow X_3$ besides the zero section $S_0$ \cite{Morrison:1996na} (see the reviews \cite{Weigand:2018rez,Cvetic:2018bni} for details), where $A=1,\dots, r$ with $r$ the rank of the abelian gauge sector. For simplicity of notation, we henceforth ignore any non-abelian gauge factor that might be present in addition to the abelian gauge groups.\footnote{This does not affect the generality of our findings because we could equally well extend our results to the Cartan subgroups of the abelian gauge groups. We refrain from spelling this out here, merely to keep the notation simple.}  The abelian gauge theories in F-theory arise by lifting the M-theory $\U(1)$ factors obtained by reducing the M-theory 3-form $C_3$ over curves in $X_3$ to six dimensions. In general, for a basis of 2-forms dual to divisors $\{D_a\}$, $C_3$ gives rise to $\U(1)$ gauge fields as 
	\begin{equation}\label{U1sMtheory}
		C_3 = (2\pi)^{-1}M_{\ttiny{11d}} A^a_1 \wedge D_a \,. 
	\end{equation}
	The gauge field $A^0_1$ obtained from the zero section $S_0$, or more precisely from the combination 
	\begin{equation}\label{hatS0}
		\widehat{S}_0 = S_0 + \pi^\ast( \bar K_{B_2}),
	\end{equation}
	is identified with the KK $\U(1)_\KK$ associated to the circle compactification. 
	All other abelian (non-Cartan) $\U(1)_A$ potentials in F-theory arise from the expansion 
	\begin{equation}
		C_3 = (2\pi)^{-1}M_{\ttiny{11d}} A^A_1 \wedge \sigma(S_A) +\dots \,, 
	\end{equation}
	where $S_A$ is a rational section and the Shioda map 
	\begin{equation} \label{Shiodamap}
		\sigma(S_A) = S_A - S_0 - \pi^{-1} (\pi_*((S_A-S_0)\cdot S_0))
	\end{equation}
	ensures that the gauge potential has no admixture with this KK $\U(1)_\KK$ and also do not have any component along the base $B_2$.
	
	In the F-theory limit \eqref{Ftheorylimit}, the gauge theory associated to $\sigma(S_A)$ now localizes on certain divisors of the base $B_2$. In fact, it can be shown that in F-theory the dynamics of the abelian gauge factors is encoded in the so-called height-pairing divisors defined as
	\begin{equation}\label{heightpairing}
		b_{AB} = - \pi_*(\sigma(S_A)\cdot \sigma(S_B))\,.
	\end{equation}
	In the following we denote by $\U(1)_6$ an abelian gauge factor that is already present in the six-dimensional parent theory. 
	
	Let us be a bit more concrete and consider, for simplicity, an elliptically fibered Calabi--Yau threefold $\pi : X_3 \to B_2$ that allows for a single additional rational section $S_Q$ apart from the zero section $S_0$. 
	Its K\"ahler form can be expanded in the Shioda--Tate--Wazir basis as
	\begin{equation}\label{ShiodaTW}
		J = \tau \widehat{S}_0  + z \sigma( S_Q) + \pi^\ast j\,,
	\end{equation} 
	where $j$ is the K\"ahler form of $B_2$, which, in turn, has the expansion
	\begin{equation}
		j = \sum_{\alpha =1}^{h^{1,1}(B_2)} v^\alpha j_\alpha^{\mathrm{b}}\,.
	\end{equation}
	For simplicity, let us abbreviate the Shioda homomorphism for the section $S_Q$ as
	\begin{equation}
		\sigma(S_Q) = S_Q - S_0 - \pi^*D_\ttiny{perp.}^{\mathrm{b}}\,,
	\end{equation}
	where $D_\ttiny{perp.}^{\mathrm{b}} \subset B_2$ is some base divisor.

	The volume of $X_3$ in M-theory units then follows as  
	\begin{align}
		\begin{split}
			\mathcal{V}_J(X_3) = & \,\frac{1}{6} c_1(B_2)^2 \tau^3  + \frac{1}{2}  \mathcal{V}_j\left(\bar{K}\right)\tau^2 +  \mathcal{V}_j(B_2)\tau + \varepsilon  z^3 + \delta \tau  z^2 -\frac{1}{2} \mathcal{V}_j(b) z^2   + \\
			& + \mathcal{V}_j (\tilde{C}) \tau z  + \frac{1}{2} \left(\tilde{C} \cdot_{B_2} \bar{K}\right) \tau^2 z \fstop
		\end{split}
		\label{eq:volCY}
	\end{align}
	Here $b = - \pi_*\left(S_Q \cdot S_Q\right)$ is the height-pairing curve, $\bar{K}$ is the anti-canonical class of $B_2$, and $\tilde{C} \coloneq  \bar{K} - D_\ttiny{perp.}^{\mathrm{b}} + \pi_\ast \left(S_Q \cdot S_0\right) $.
	Moreover, the numerical coefficients for this K\"ahler parameter polynomial can be obtained by using the intersection theory relations of elliptic fibrations~\cite{Weigand:2018rez}. In terms of the data of the base $B_2$, these numbers read
	\begin{align}
		\begin{split}
			\varepsilon &= D_\ttiny{perp.}^{\mathrm{b}} \cdot_{B_2} \left(\bar{K} + \pi_\ast \left(S_0 \cdot S_Q\right) \right) \coma\\
			\delta &= \left(D_\ttiny{perp.}^{\mathrm{b}}\right)^2- \bar{K}^2 -2 \bar{K} \cdot_{B_2} D_\ttiny{perp.}^{\mathrm{b}} - \pi_\ast \left(S_0 \cdot S_Q\right) \cdot_{B_2} \left(2 D_\ttiny{perp.}^{\mathrm{b}} + \bar{K}\right) \fstop
		\end{split} 
	\end{align}
	Furthermore, we can identify the Wilson line parameter $\theta'$ appearing in \eqref{massformulaKK} as
	\begin{equation}
		\frac{z}{\tau} = \frac{\theta'}{2\pi}=\theta
	\end{equation} 
	because $z$ is the Coulomb branch parameter for the gauge theory. The total volume then reads
	\begin{equation}\label{eq:genVolintheta}
		\text{\footnotesize $\displaystyle  \mathcal{V}_J(X_3)= \tau^3\left(\epsilon \theta^3+ \delta\, \theta^2  +\frac{\theta}{2} (\tilde{C} \cdot_{B_2} \bar{K})+\frac{1}{6} c_1(B_2)^2 \right) -
			\frac{1}{2} \tau^2 \left(\theta^2\mathcal{V}_j(b)-2\theta \mathcal{V}_j(\tilde{C})-\mathcal{V}_j (\bar{K})\right)
			+\tau \mathcal{V}_j(B_2)$ .}
	\end{equation}
	This gives an explicit expression for $\alpha$ and $\beta$ that appeared in \eqref{VX3parametrization} and which, if non-vanishing, enter the explicit estimate 
	\eqref{maxtauprescr1}
	for $\tau_{\ttiny{max.}}$.
	
	However, we note that for special values of $\theta$, the terms cubic and quadratic in $\tau $ may vanish. 
	To see this, we start by writing the generic elliptic fiber, ${\mathbb E}$, as a linear combination of fibral Mori cone generators $\mathcal{C}^a$ 
	\begin{equation}\label{Eexpansion}
		{\mathbb  E} = \sum_a \alpha_a \mathcal{C}^a\,,\quad \alpha_a\geq 0\,.
	\end{equation}
	On the other hand, in the presence of a $\U(1)$ gauge symmetry in six dimensions, there generically exists another fibral curve $C_z$ which admits a similar expansion
	\begin{equation}\label{Czexpansion}
		C_z = \sum_a \beta_a \mathcal{C}^a \,,
	\end{equation}
	where $\alpha_a \neq \beta_a$ for at least one $a$. To the curves $\mathcal{C}^a$, we can associate the K\"ahler moduli $v^a$ as in \eqref{Ftheorylimit} and, as before, from the perspective of F-theory on $B_2\times S^1$, we interpret the ratio of the volume of $C_z$ and the volume of $\mathbb{E}$ as the Coulomb branch parameter $\theta$.

	Let us assume that there exists a fibral curve $\mathcal{C}^1$ whose volume, $v^1$, only appears linearly in  $\cV_{X_3}$. Then we can take the limit $v^1\rightarrow \infty$ by co-scaling to zero all other volumes. In this way, we can achieve $\tau\rightarrow \infty$ without changing $\cV_{X_3}$. However, we cannot interpret this limit as the five-dimensional limit of a six-dimensional theory compactified on a circle. 
	Since $v^1$ appears only linearly, its dual divisor is the fiber of a surface fibration. For concreteness, let us assume that this divisor does not admit a genus-one fibration, so that we are in a situation as in Case \ref{fibrationcase1}.
	We obtain that the limit $v^1\rightarrow \infty$ with suitable co-scaling is an emergent string limit \cite{Lee:2019wij}. If the emergent string is a heterotic string, this dual heterotic string is compactified on $K3\times S^1$ but the radius of the $S^1$ does not change in the limit $v^1\rightarrow \infty$. Therefore, the limit $v^1\rightarrow \infty$ cannot be interpreted as the small radius limit for a circle compactification of a six-dimensional theory. Similarly, the mass of the M2-branes charged under the $\U(1)$ which in the F-theory limit takes the role of  $\U(1)_\KK$ is given by
	\begin{equation}
		\frac{M_\ttiny{$M2|_{\mathbb E}$}}{M_\PlD{5}} = n v^1 \,,
	\end{equation}
	which lies at (or above) the black hole threshold \eqref{BHthresholdemstring} in the dual heterotic theory,
	\begin{equation}
		\frac{M_{\ttiny{BH, min.}}}{M_\PlD{5}} \sim \left(\frac{M_\PlD{5}}{\Lambda_{\QG}}\right)^2 \sim v^1\,.
	\end{equation}
	Here, we used the fact that in the dual heterotic limit the QG cutoff is set by the tension of the heterotic string 
	\begin{equation}
		\frac{\Lambda_{\QG}^2}{M_\PlD{5}^2} = \frac{T_\ttiny{het.}}{M_\PlD{5}^2} \sim \frac{1}{v^1}\,. 
	\end{equation}
	This means that all states carrying charge under this $\U(1)$ have masses in the black hole region, unlike what would be expected for KK states. Finally, due to the co-scaling, which is necessary to reach $v^1\rightarrow \infty$ without changing $\cV_{X_3}$, the parameter $\theta$ is fixed to 
	\begin{equation}
		\theta \stackrel{v^1\rightarrow \infty}{\longrightarrow} \frac{\alpha_1}{\beta_1}\,,
	\end{equation}
	where $\alpha_1$ and $\beta_1$ are the expansion coefficients appearing in \eqref{Eexpansion} and \eqref{Czexpansion}. While, in the large radius limit, $\theta$ is the Coulomb branch parameter of the $\U(1)$ in five dimensions, which can be chosen freely, this is not possible in the limit $v^1\rightarrow \infty$, further illustrating that in this limit the interpretation of $\tau$ as the radius of an $S^1$ is not valid. 
	
	To summarize, even if $\tau>\tau_\ttiny{max.}$ (as estimated in \eqref{taumax-emstr}) is possible, it does not correspond to a small radius limit for a circle compactification of a six-dimensional theory to five dimensions. The $\U(1)$ associated with the generic fiber ${\mathbb E}$ has the interpretation of a KK $\U(1)_\KK$ only for $\tau<\tau_\ttiny{max.}$. 
	
	\subsection{Minimal Radius via M-theory/Type IIA Duality}
	\label{sec:minradius5to4}
	
	Let us now consider a circle compactification of the five-dimensional theory
	obtained from M-theory on a Calabi--Yau threefold $X_3$. The resulting four-dimensional $\mathcal{N}=2$ supersymmetric theory is dual to Type IIA string theory on the same Calabi--Yau. We aim to understand the small radius limit of the KK circle, $r_{S^1} M_\PlD{5} \to 0$,
	in the regime where the Planck scale of the original theory is unchanged as $r_{S^1}$ decreases.
	In view of the relation $M^3_\PlD{5} = 4\pi M^3_{\ttiny{11d}} \mathcal{V}_{X_3}$,
	with $\mathcal{V}_{X_3}$ the volume measured in units of $M_{\ttiny{11d}}$,
	this translates into taking
	\begin{equation} \label{eq:limit_rS1}
		r_{S^1} M_{\ttiny{11d}} \to 0 \qquad \text{at} \qquad \mathcal{V}_{X_3} = \text{const.} \,.
	\end{equation}
	This limit corresponds to the regime of asymptotically vanishing ten-dimensional Type IIA string coupling $g_{\ttiny{IIA}}$ because
	\begin{equation} \label{eq:gIIA23}
		g_{\ttiny{IIA}}^{2/3} = 2 \pi M_{\ttiny{11d}} r_{S^1}  \,. 
	\end{equation}
	Furthermore, via
	\begin{equation}
		\frac{M_{{s}}}{M_{\ttiny{11d}}} = g_{\ttiny{IIA}}^{1/3} \,,
	\end{equation}
	the Calabi--Yau volumes measured in units of $M_{s}$ and $M_{\ttiny{11d}}$ are related as 
	\begin{equation} \label{eq:4ddilaton}
		\frac{ \mathcal{V}_{X_3,\,s}}{g^2_{\ttiny{IIA}}} = \mathcal{V}_{X_3} \,,
	\end{equation}
	where we defined $ \mathcal{V}_{X_3,\,s} =\text{vol}(X_3)M_s^6$ and $ \mathcal{V}_{X_3} =\text{vol}(X_3)M_\ttiny{11d}^6$ (cf. Footnote \ref{footnote:conventionvolumes}).  
	The combination on the left denotes the four-dimensional dilaton from the perspective of the four-dimensional KK reduction, which, according to
	\eqref{eq:limit_rS1}, remains constant along the limit of interest.
	This, however, means that as $r_{S^1} M_{\ttiny{11d}}$ decreases, the volume in string units must be co-scaled as
	\begin{equation} \label{eq:VX3rs1}
		\mathcal{V}_{X_3,\,s}  \sim (2\pi)^3(r_{S^1} M_{\ttiny{11d}})^3   \quad \to \quad  0 \,.
	\end{equation}
	This motion in the moduli space is unproblematic as long as $r_{S^1} M_{\ttiny{11d}} \geq 1$. 
	However, the total Calabi--Yau volume cannot decrease below the string scale due to Type IIA $\alpha'$-corrections. This is a manifestation of the fact that the stringy K\"ahler moduli space differs considerably from its classical counterpart; in particular, the regime $\mathcal{V}_{X_3,\,s} \ll 1$ is not part of the actual moduli space. To see this, we notice that at the quantum level, in general, we identify 
	\begin{equation}
		\cV_{X_3,\,s} = e^{-\cK_\ttiny{K}(X_3)}\,,
	\end{equation}
	where $\cK_\ttiny{K}(X_3)$ is the component of the K\"ahler potential depending on the K\"ahler moduli of $X_3$. In the large volume limit, this simply reduces to 
	\begin{equation}
		e^{-\cK_\ttiny{K}(X_3)} \rightarrow \frac{1}{6} \int_{X_3} J\wedge J \wedge J -\frac{\chi(X_3) \zeta(3)}{4\pi^3}=\cV_{X_3,\,s}\,.
	\end{equation}
	Via mirror symmetry, the K\"ahler potential for the K\"ahler sector of $X_3$ is identified with the K\"ahler potential for the complex structure sector of the mirror $Y_3$ of $X_3$,
	\begin{equation}
		\label{eq:eKY3}
		e^{-\cK_{\rm c.s}(Y_3)} = i \int_{Y_3} \Omega \wedge \bar{\Omega}\,,
	\end{equation}
	where $\Omega$ is the holomorphic $(3,0)$ form on $Y_3$. 
	For $\cV_{X_3,\,s}$ to approach zero, there must exist a regime in the complex structure moduli space of $Y_3$ where $|X^0|^{-2} \int_{Y_3} \Omega \wedge \bar{\Omega} \to 0$ for $X^0$ the fundamental period of $\Omega$. This limit leads to a singularity in the complex structure moduli space of $Y_3$. These singularities are classified and correspond to infinite distance singularities, conifold-like singularities, or orbifold points. In the latter case, the K\"ahler potential must remain finite, since the orbifold singularity can simply be removed by going to the covering space of the complex structure moduli space. In an infinite distance limit, $e^{-\cK_{\rm c.s.}(Y_3)}$ diverges, whereas for a conifold point in a compact Calabi--Yau threefold, it takes the form 
	\begin{equation}
		\frac{1}{|X^0|^2}e^{-\cK_{\rm c.s.}(Y_3)} = -|\mu|^2 \log |\mu|^2 + \text{const.}\,,
	\end{equation}
	where the conifold locus is located at $\mu=0$ and the constant is of $\mathcal{O}(1)$; its exact value depends on the precise geometry, e.g. on the dimension of the moduli space. Therefore, the quantum volume of $X_3$ at the mirror of the conifold point is also finite, although it does not need to be minimized at the conifold point. In general, the minimum of $\cV_{X_3,\,s}$ will be somewhere in the bulk of the moduli space. Since $\cV_{X_3,\,s}$ cannot vanish at this minimum, we expect
	\begin{equation} \label{eq:BoundonVX3}
		\cV_{X_3,\,s}\geq \alpha \,, \qquad \alpha \sim \mathcal{O}(1) \,,
	\end{equation}
	everywhere in moduli space. In Section~\ref{sec:absencetower}, we compute the lower bound on $\cV_{X_3,\,s}$ in explicit examples.

	Assuming henceforth a lower bound of the form \eqref{eq:BoundonVX3}, we see from \eqref{eq:4ddilaton} that decreasing $g_{\ttiny{IIA}}$ below $\left(\mathcal{V}_{X_3}\right)^{\frac{1}{2}}$ necessarily takes us out of the vector multiplet moduli space. By the same argument as for the reduction from six to five dimensions studied in Section \ref{sec:minimalradiusFM},  $g_\ttiny{IIA}$ hence loses its interpretation as a KK radius, which would have to lie in a vector multiplet.
	This indicates that one cannot extrapolate $r_{S^1}$ beyond a minimal value, $r_{S^1}^\ttiny{min.}$, which can be inferred from
	\begin{equation} \label{eq:rmin5d}
		g_\IIA^{2/3} \sim (2\pi) r_{S^1}^\ttiny{min.} M_{\ttiny{11d}} = \frac{\alpha^{\frac{1}{3}}}{ \left(\mathcal{V}_{X_3}\right)^{\frac{1}{3}}} \quad \Longrightarrow\quad  r_{S^1}^\ttiny{min.} M_{\ttiny{11d}} = \frac{\alpha^{\frac{1}{3}}}{(2\pi) \left(\mathcal{V}_{X_3}\right)^{\frac{1}{3}}}\,.
	\end{equation}
	Here, we used \eqref{eq:gIIA23} and \eqref{eq:4ddilaton}. This does not mean, of course, that the coupling $g_{\ttiny{IIA}}$ cannot be made arbitrarily small. Rather, the notion of a KK circle ceases to be appropriate in the following sense. The tower of BPS states from D0-branes has a tower mass that scales as
	\begin{equation}
		M_{\ttiny{D0}}=\frac{M_s}{g_\IIA}\quad  \Longrightarrow \quad \frac{M_{\ttiny{D0}}}{M_\PlD{4}} \sim \frac{1}{\left(\mathcal{V}_{X_3,\,s}\right)^{1/2}} \coma 
	\end{equation}
	where $M_\PlD{4}^2=4\pi M_s^2\mathcal{V}_{X_3,\,s} g_\IIA^{-2}$. 
	For large $r_{S^1}$, that is, $\mathcal{V}_{X_3,\,s}\geq 1$, we can interpret this formula, via \eqref{eq:VX3rs1}, in the sense that 
	\begin{equation}
		\frac{M_{\ttiny{D0}}}{M_\PlD{4}} \sim  \frac{1}{r_{S^1}^{3/2} M_\ttiny{11d}^{3/2}}\coma
	\end{equation}
	and decreasing the radius makes this tower heavier at
	the correct rate for it to be interpreted as a KK tower.
	But due to the bound \eqref{eq:BoundonVX3}, 
	the tower cannot become arbitrarily heavy in units of the four-dimensional Planck scale. In this sense, it is no longer justified to interpret the D0-brane tower as a KK tower associated with a freely scalable circle. We can also interpret this fact geometrically: for large radii, there is a hierarchy between the size of the curves inside $X_3$ and the circle. However, as we decrease the radius, eventually we reach a point where all cycles (including the radius) are of string/M-theory size. As is well known, in this stringy regime classical geometry is no longer a good approximation to the actual theory, and in particular the differentiation between circles and curves in $X_3$ is no longer sensible. 
	
	We conclude that there is effectively a minimal radius, beyond which we cannot interpret the geometry as a circle compactification. Consequently, all the equations for the four-dimensional quantities that treat them via KK reduction from five dimensions are not applicable for smaller values of $g_\IIA$. 
	
	\subsubsection*{Comments on M-theory/Type IIA Duality}
	
	To avoid confusion, let us contrast the behavior just described for circle reductions of M-theory on Calabi--Yau threefolds to the behavior of circle reductions of eleven-dimensional M-theory. In particular, we aim to show that in the latter setup, the Type II string coupling can be interpreted as a radius of a circle compactification for any value. Hence, the effect discussed previously is inherent to compactifications to lower dimensions. To see this, notice that, in ten dimensions, the tower of BPS D0-branes scales as
	\begin{equation}
		\frac{M_{\ttiny{D0}}}{M_\PlD{10}} \sim \frac{1}{g_\IIA^{3/4}}\coma 
	\end{equation}
	since $M_\PlD{10}^8=4\pi M_s^8 g_\IIA^{-2}$. This should be contrasted with the lower-dimensional situation, where the mass of the D0-brane tower in four-dimensional Planck units was given in terms of the volume. In the ten-dimensional case, on the other hand, there is nothing bounding the mass of D0-branes, as we can make $g_\IIA$ arbitrarily small. In particular, even in the weak coupling limit the mass scale of the D0-branes remains below the black hole threshold,
	\begin{equation}
		\frac{M_{\ttiny{D0}}}{M_\PlD{10}} \ll \frac{1}{g_{\IIA}^{7/4}} = \frac{M_{\PlD{10}}^7}{M_s^7} = \frac{M_\ttiny{BH, min.}}{M_\PlD{10}}\,,
		\label{eq:MD0compared}
	\end{equation}
	for $g_\IIA \leq 1$.\footnote{Even if it is not needed for our discussion, for completeness, we notice that \eqref{eq:MD0compared} is not satisfied for $g_\IIA> 1$. However, for those kinds of limit, the species scale is the eleven-dimensional Planck scale, $M_\ttiny{11d}$, so we need to compute the new BH threshold, namely
		\begin{equation}
			\frac{M_\ttiny{BH, min.}}{M_\PlD{10}}=\frac{M_{\PlD{10}}^7}{M_\ttiny{11d}^7} = g_\IIA^{7/12}\coma
		\end{equation}
		where we used \eqref{eq:mapMtoIIA-2}. This BH threshold is above the mass scale of the D0-branes precisely for $g_\IIA>1$.}
	This means that in ten dimensions, we can always interpret the D0-brane tower as a KK tower for any value of the radius.

	\section{Particle States and the Weak Gravity Conjecture}
	\label{sec:particlesandWGC}
	
	In Section \ref{sec:necessitytower}, we have argued that for an $S^1$ reduction of a quantum gravity theory, there typically exists a minimal radius below which the notion of a circle compactification breaks down. 
	To understand the implications of this observation for the tower WGC, we first elaborate on the different versions of the WGC that one might consider.
	The WGC in its mildest form does not necessarily require a new \textit{particle} in the spectrum of a quantum gravity theory; rather, it merely postulates some super-extremal state in the theory. Such a state can very well be an extremal black hole whose charge-to-mass ratio receives higher-derivative corrections to make it slightly super-extremal. In general, the WGC can thus be satisfied in two ways in a quantum gravity theory: 
	\begin{enumerate}[label={{\itshape\roman*}.},ref={{\itshape\roman*}}]
		\item\label{case:particlesvsBH}There exists an additional super-extremal particle state with mass below the black hole threshold \eqref{BHthresholdemstring}. 
		\item\label{case:higherderivative}The higher-derivative corrections to the two-derivative Einstein-Maxwell action render extremal black holes of smaller mass super-extremal compared to larger ones; these are then the states required by the mild version of the WGC.
	\end{enumerate}
	In Case \ref{case:particlesvsBH}, the mass of the particle state may even lie above the (quantum gravitational) cutoff $\Lambda_\QG$ of the EFT as long as it is below the black hole threshold.\footnote{This occurs for the perturbative heterotic string, where super-extremal excitations have masses at or above the heterotic string scale, which sets the species scale and hence the quantum gravity cutoff, even though they lie below the black hole threshold, at least for small charges.}
	The term \textit{particle} will always include this option and will be used in contradistinction from states above the black hole threshold.
	This twofold way to satisfy a Swampland condition is, in fact, common also for other Swampland constraints. For example, the No Global Symmetry condition \cite{Banks:1988yz,Banks:2010zn,Harlow:2018tng,Harlow:2018jwu} can be satisfied either already at the EFT level by a particle state explicitly breaking a would-be global symmetry or only be satisfied in the full theory of quantum gravity. In the latter case, the symmetry breaking manifests itself in the higher-derivative corrections to Einstein's theory. 
	
	One may now ask whether, in a given theory, the WGC is actually satisfied already at the particle level in the above sense. Furthermore, can the WGC be satisfied by a finite number of super-extremal particle states in a way that is consistent under dimensional reduction, or can the WGC be only consistently satisfied at the particle level in the presence of a \textit{tower} of super-extremal particles? By a tower of particles, we refer to a collection of particles with increasing charge and mass below the black hole threshold, interpolating between the particle and the black hole regime. For large masses, the particles of the tower hence transition into the black hole region, where they become black hole microstates. In particular, for the particles in a tower, a parametric separation of their masses is not possible, so that we cannot include just a subset of these particles in the EFT. This is summarized in
	
	\begin{mdframed}[backgroundcolor=white,shadow=true,shadowsize=4pt,shadowcolor=seccolor,roundcorner=6pt]
		\begin{definition} \label{Def-tower}
			A gauge theory coupled to a quantum gravity theory exhibits a tower of super-extremal particle states if, for every charge $q$ in the charge lattice, there exists a region in the scalar field space where there is a super-extremal particle state below the black hole threshold of charge $n \, q$, for any $n \in \mathcal{I}_q$, with $\mathcal{I}_q$ an infinite index set.
		\end{definition}
	\end{mdframed}
	
	\vspace{2mm}

	\noindent We are, in this work, ultimately interested in the question if the tower WGC applied to particle towers in the above sense holds for any quantum gravity or whether it can be consistent if this strong version of the WGC is violated.
	
	\subsection{Existence of Super-extremal Particle States}
	\label{sec:ExistenceSuperExremalParticles}
	
	Before we address whether there always exists a tower of super-extremal particle states in the sense of Definition~\ref{Def-tower}, we first aim to understand under which conditions the WGC is satisfied at the particle level at all. The original WGC is formulated for 0-form gauge theories (cf. Footnote \ref{footnote:pformsymmetries}) in quantum gravity. Let us be inspired by string theory to see under what conditions such 0-form gauge theories are expected to be constrained by the WGC.
	In effective theories arising from string compactifications most 0-form gauge symmetries in fact arise either from defect theories realized on branes or from higher-form gauge symmetries, e.g. from the RR-forms of ten-dimensional string theory, reduced on the compact manifold. For defect theories, the WGC is not expected to give strong constraints because such theories are effectively decoupled from gravity.\footnote{Indeed, defect gauge theories are exempt from many quantum gravity conjectures, for example there is no bound on the rank of a lower-dimensional brane stack in a string compactification.} For higher-form symmetries, on the other hand, it is the higher-form version of the WGC that should hold by virtue of (super-)extremal extended objects, rather than the WGC for 0-form symmetries. 
	
	Although many gauge theories in string theory originate from defect or higher-form symmetries, it is not always possible to resolve this origin from the perspective of the EFT. To see this recall that in an EFT coupled to gravity there is a minimal length scale set by the quantum gravity cutoff of the theory,
	\begin{equation}
		\ell_\ttiny{min.} = \frac{1}{\Lambda_\QG}\,. 
	\end{equation}
	On the other hand, upon compactification, we can also associate a length scale to defect and higher-form symmetries. Suppose that we have a $d$-dimensional defect theory in a $D$-dimensional theory of gravity with $(D-d)$-compact dimensions. Then a natural length scale associated with the defect theory is the diameter of the $(D-d)$-dimensional space, $\mathcal{S}_\ttiny{perp.}$, transverse to the defect. For instance, as we will consider below, in a $D$-dimensional theory of gravity with $d$-dimensional spacetime filling D-branes, $\mathcal{S}_\ttiny{perp.}$ is the $(D-d)$-dimensional compact space in which the branes are point-like. More generally, if a gauge theory is realized on a divisor $\Sigma$ in a compact manifold, then the space $\mathcal{S}_\ttiny{perp.}$ is represented by curves with positive intersection with $\Sigma$. We then define the length scale associated to the defect theory as 
	\begin{equation}\label{ellperphigherform}
		\ell_\ttiny{perp.} = \text{diam}(\mathcal{S}_\ttiny{perp.})\,. 
	\end{equation}
	If this transverse space is large compared to $\ell_\ttiny{min.}$, from the EFT perspective the theory is a defect theory. On the contrary, once $\ell_\ttiny{perp.} < \ell_\ttiny{min.}$ there is no way to resolve the directions transverse to the defect within the EFT, and one encounters a genuinely $d$-dimensional gauge theory coupled to gravity. Similarly, suppose that a $p$-form symmetry ($p>0$) is reduced over a compact $p$-cycle $\mathcal{C}$ to an effective 0-form symmetry. Then in the EFT, the diameter $\ell_\ttiny{perp.}$ of $\mathcal{C}$ sets the scale at which additional polarizations of the $p$-form symmetry can be detected. If $\ell_\ttiny{perp.}< \ell_\ttiny{min.}$ there is no way to detect this within the validity of the EFT, and the reduced $p$-form symmetry should be viewed as a genuine 0-form symmetry in the effective theory of gravity.
	
	In addition, the WGC is expected to only constrain genuine gauge theories that are not decoupled from gravity. This is the case as long as the WGC scale,
	\begin{equation} \label{WGC-scale}
		\Lambda^2_\ttiny{WGC}\coloneq g_\UoD{D}^2 M_\PlD{D}^{D-2} \,,
	\end{equation}
	is not parametrically above the quantum gravity cutoff. If, instead, $\Lambda_\WGC\gg \Lambda_\QG$, the gauge theory effectively decouples from gravity. In general, the gravity decoupling limit corresponds to a regime where the gauge interaction is parametrically larger than the gravitational interaction, i.e., the gauge coupling is parametrically larger than the gravitational coupling $G_\ttiny{N,D} = \frac{1}{8 \pi} M^{2-D}_{\ttiny{Pl,D}}$. However, since these two couplings have different dimensions, their ratio $ \frac{g_\UoD{D}^2}{G_\ttiny{N,D}}\sim \Lambda_\WGC^2 $ is a dimensionful quantity. We hence need to define a reference scale in order to determine whether this ratio is large or small. In theories of gravity, a natural choice for such scale is given by the quantum gravity cutoff $\Lambda_\QG$. Accordingly, we consider a gauge theory decoupled from gravity if $\frac{g_\UoD{D}^2}{G_\ttiny{N,D}}$ is large in units of $\Lambda_\QG$ which amounts to $\Lambda_\WGC\gg \Lambda_\QG$.

	We claim that it is for genuine 0-form gauge theories coupled to gravity that the WGC must be satisfied at the particle level:
	
	\begin{mdframed}[backgroundcolor=white,shadow=true,shadowsize=4pt,shadowcolor=seccolor,shadowsize=4pt,roundcorner=6pt]
		\begin{claim}\label{claim:WGC}
			For a 0-form gauge theory that is coupled to gravity ($\Lambda_\WGC\lesssim \Lambda_\QG$), the Weak Gravity Conjecture must be satisfied by particle-like states (massless or massive) if it is a {\it genuine} 0-form gauge theory, i.e. if within the validity of the EFT, the 0-form symmetry cannot be resolved to be a defect or higher-form symmetry ($\ell_\ttiny{perp.} < \ell_\ttiny{min.}$). 
		\end{claim}
	\end{mdframed}
	
	\vspace{2mm}
	
	If the theory is effectively a defect theory (and hence decoupled from gravity), the WGC may still be satisfied at the particle level, though it does not a priori have to. To illustrate this, consider two D-branes in string theory that fill the $d$-dimensional spacetime and are point-like in a $(D-d)$-dimensional compact space $\mathcal{S}$. Let $\ell_\ttiny{perp.}$ be the diameter of this compact space. Within $\mathcal{S}$, the two D-branes can be separated by a distance $\ell_\ttiny{sep.}$ that satisfies 
	\begin{equation}\label{ellsepellperp}
		\ell_\ttiny{sep.} \leq \ell_\ttiny{perp.}\,,
	\end{equation}
	up to $\mathcal{O}(1)$ factors that depend on the precise definition of the diameter. Charged states now arise from strings stretched between the two D-branes with mass 
	\begin{equation}
		m \sim M_s^2 \ell_\ttiny{sep.}\,. 
	\end{equation}
	For $\ell_\ttiny{perp.}<\ell_\ttiny{min.}$ the mass of these charged states always lies below the black hole threshold, since 
	\begin{equation}
		m = M_s^2 \ell_\ttiny{sep.} \leq M_s^2 \ell_\ttiny{perp.} \leq M_\PlD{D}^2 \ell_\ttiny{perp.} \leq M_\PlD{D}^2 \ell_\ttiny{min.} \leq  M_\PlD{D} \left( \ell_\ttiny{min.} M_\PlD{D} \right)^{D-3} = M_\ttiny{BH, min.}\,. 
	\end{equation}
	The first inequality follows from \eqref{ellsepellperp}, and for the second inequality we used the fact that the string scale has to be lower than any higher $D$-dimensional Planck scale. Finally, we used $\ell_\ttiny{min.} M_\PlD{D}\geq 1$ and, in the last step, identified the minimal $D$-dimensional black hole mass \eqref{BHthresholdemstring} for $D>3$. Thus, for a defect theory for which we cannot resolve the transverse space within the EFT, the massive charged string states indeed are particle-like. On the other hand, for $\ell_\ttiny{perp.}\geq \ell_\ttiny{min.}$ the above conclusion does not hold, and the string states can have a mass above the black hole threshold for sufficiently large $\ell_\ttiny{sep.}$. Hence, for a defect theory, there may not exist charged particle-like states that can satisfy the WGC.\footnote{For non-abelian theories, massless super-extremal states are provided by the gauge bosons, but for abelian theories this trivial way of satisfying the WGC at the particle level is not available.} 
	
	Similarly, reductions of higher-form symmetries for which we can detect the higher-form nature of the gauge theory within the EFT do not need to satisfy the 0-form WGC at the particle level. To illustrate this, consider M-theory compactifications on a Calabi--Yau threefold $X_3$. As reviewed in Section \ref{sec:minimalradiusFM}, in such theories, $\U(1)$ gauge theories arise from reducing the M-theory $C_3$ form on curves in $X_3$. Given a basis $\{C^a\}$ of curves in $X_3$, we can associate to any curve class $[C]\in H_2(B_2)$ a linear combination of $\U(1)$ gauge factors, 
	\begin{equation}\label{U1C}
		\U(1)_C = c_a \U(1)^a \,,\qquad \text{with} \qquad C=c_a C^a\,. 
	\end{equation}
	For simplicity, assume that $\{C^a \}$ is a basis of Mori cone generators and consider $\U(1)^a$. The states charged under $\U(1)^a$  correspond to M2-branes wrapping curves in multiples of $[C^a]$, with mass  
	\begin{equation}
		m_\ttiny{M2} = M_\ttiny{11d}^3 \text{vol}(C^a)\,. 
	\end{equation}
	Let us focus on the vector multiplet sector where we can set $\cV_{X_3}=1$ so that $M_\ttiny{11d}=M_\PlD{5}$. From \eqref{ellperphigherform} we infer that for $\U(1)^a$ the length scale $\ell_\ttiny{perp.}$ is set by the diameter of $C^a$. For simplicity, let us assume that 
	\begin{equation}
		\ell_\ttiny{perp.} = \sqrt{\text{vol}(C^a)}\,. 
	\end{equation}
	For $\ell_\ttiny{perp.}\geq \ell_\ttiny{min.}$, we then find 
	\begin{equation}
		m_\ttiny{M2} = M_\PlD{5}^3 \ell_\ttiny{perp.}^2 \geq M_\PlD{5} (M_\PlD{5} \ell_\ttiny{min.})^2 = M_\ttiny{BH, min.}^{5D} \geq  M_\ttiny{BH, min.}\,. 
	\end{equation}
	In the second step, we identified the mass of the minimal five-dimensional black hole, which is always an upper bound for the black hole threshold $M_\ttiny{BH, min.}$. We hence see that, in the situations where we can resolve the higher-form nature of the gauge theory within the EFT ($\ell_\ttiny{perp.}\geq \ell_\ttiny{min.}$), the charged states indeed have masses at or above the black hole threshold.
	
	It is interesting to note that \cref{claim:WGC} can be naturally extended to any $p$-form gauge theory coupled to gravity, and it is easy to generalize the motivations used to justify it to a general gauge symmetry. However, in the following sections, we will mainly focus on the existence of a tower of particle-like states satisfying the WGC.
	
	\subsection{Existence of the Tower of Super-extremal Particle States}
	\label{sec:existencetower}
	
	We now turn to whether, for all genuine 0-form gauge theories, there necessarily exists a tower of super-extremal particles with arbitrarily high charge. To this end, recall that at a generic point in field space, the super-extremal particles furnishing the tower of states have to transition into extremal black holes for sufficiently large charges. However, it is more instructive to reverse this statement: For there to exist a tower of particle-like super-extremal states at any point in field space, extremal black holes must be able to transition into particles. In other words, there must exist a limit in field space where the mass of the tower of extremal black holes (charged under a fixed $\U(1)$) drops parametrically below the black hole threshold, defined as the mass of the smallest black hole in the theory. 
	
	The mass $M_\ttiny{BH, ext.}$ and charge $Q$ of an extremal black hole are related as
	\begin{equation}\label{extbound}
		g_\UoD{D}^2 M_\PlD{D}^{D-2} Q^2 = \gamma\, M_\ttiny{BH, ext.}^2 \quad \Longrightarrow \quad Q^2 \frac{g_\UoD{D}^2 M_\PlD{D}^{D-2}}{\gamma} = M_\ttiny{BH, ext.}^2\,,
	\end{equation}
	up to curvature corrections arising for small charges. The mass of a black hole with an arbitrarily large charge can drop below the black hole threshold \eqref{BHthresholdemstring} of the theory in some regions of scalar field space if either the gauge coupling becomes arbitrarily small or the extremality coefficient $\gamma$ becomes arbitrarily large. 
	
	As for the first possibility, it has been shown~\cite{Lee:2018spm,Lee:2018urn,Lee:2019wij,Klaewer:2020lfg,Cota:2022yjw} in the context of various classes of string or M-theory theory compactifications that every weakly coupled gauge theory indeed exhibits a tower of super-extremal particle states in the asymptotic region. 
	To give a more general, bottom-up criterion when this can happen, note first that in a weakly coupled theory, the black hole extremality coefficient $\gamma$ should be of order one. Therefore, by \eqref{extbound}, the mass scale of extremal black holes is set by $\Lambda^2_\ttiny{WGC}$ as defined in (\ref{WGC-scale}), and so is the scale of the tower of would-be particles in the weak coupling limits. 
	For this tower to be composed of particles that are below the black hole threshold, it is necessary that $\Lambda^2_\ttiny{WGC}\ll M_\PlD{D}^2$.
	But this is not yet enough. The reason is that, in general, weak coupling limits correspond to infinite distance limits in field space, and in these limits the black hole threshold coincides with the asymptotic Planck scale $M_\Plinf$.
	In fact, according to the Emergent String Conjecture~\cite{Lee:2019wij}, in these infinite distance limits, the theory reduces to a dual theory which is either a weakly coupled string theory in the same number of dimensions, corresponding to an emergent string limit, or a higher-dimensional theory of gravity, the endpoint of a decompactification limit.
	In the first case,  $M_\Plinf$  is identified with $M_\PlD{D}$, while in the second case, $M_\Plinf$ coincides with the higher-dimensional Planck scale.
	Since the weak coupling limit should be described in terms of a dual theory, the states in the potential particle-like tower must also become particle-like in the dual theory. This requires, as stressed above, that the masses of the states lie below the Planck scale $M_\Plinf$ of the asymptotic theory.  If this is not the case, the gauge theory does not actually become weakly coupled in the dual frame, and hence there cannot be a tower of super-extremal particle states in this region of the field space. This leads to a criterion for when a weakly coupled super-extremal particle tower can arise, namely that $\Lambda^2_\ttiny{WGC}$ drops parametrically below $M^2_\Plinf$.
	
	Let us now discuss the other possibility to obtain a tower of light super-extremal states, which requires $\gamma$ to diverge. As already noted, we expect that $\gamma\rightarrow \infty$ cannot be achieved in a weak coupling limit. Instead, $\gamma\rightarrow \infty$ requires strong coupling dynamics, i.e. some limit for which $g_\UoD{D}^2 M_\PlD{D}^{D-4}\rightarrow \infty$. However, although it is a necessary requirement, strong coupling does not automatically imply $\gamma\rightarrow \infty$. To see this, note first that, for certain gauge groups, strong coupling can also be achieved by taking an infinite distance limit. An example is a non-perturbative gauge sector in heterotic string theory, which becomes strongly coupled in the limit of vanishing heterotic coupling.\footnote{The appearance of collateral strongly coupled subsectors in infinite distance limits has been investigated systematically in \cite{Marchesano:2023thx}.}  As we will explain now, for strong coupling limits at infinite distance, $\gamma$ does in fact not automatically diverge.
	
	To make the argument, note first that for
	a strongly coupled gauge theory we need to have $\gamma\gg \mathcal{O}(1)$ in the case where 
	\begin{equation}\label{reqlargegamma}
		\Lambda_\WGC \gg M_\ttiny{BH, min.}\,.
	\end{equation}
	To see this, we can consider a black hole with mass $M>M_\ttiny{BH, min.}$, with a single unit of charge $q=1$. For this black hole, the extremality bound reads 
	\begin{equation}
		\Lambda^2_\WGC \leq \gamma M^2\,,
	\end{equation}
	which cannot be satisfied in the limit $\Lambda_\WGC \rightarrow \infty$ for $\gamma$ finite. Thus, in case the hierarchy \eqref{reqlargegamma} is realized, charged black holes exist only if $\gamma\gg 1$. 
	On the other hand, we will now argue that \eqref{reqlargegamma} does not hold in infinite distance strong coupling limits, which explains why in such limits $\gamma$ does not generically diverge.
	To this end, suppose that a gauge theory becomes strongly coupled in an infinite distance limit. We invoke once again the Emergent String Conjecture, identifying every infinite distance limit as either a decompactification limit or an emergent string limit. In the former case, one obtains a higher-dimensional theory, and we can evaluate 
	\begin{equation} \label{LambdaWGCdecomp}
		\Lambda_\WGC^2 = g_\UoD{d}^2 M_\PlD{d}^{d-2} = g_\UoD{d}^2 r_{S^1}^{D-d} M_\PlD{D}^{D-2}=g_\UoD{D}^2 M_\PlD{D}^{D-2} \,. 
	\end{equation}
	Here, $g_\UoD{d}$ and $g_\UoD{D}$ are the $d$- and $D$-dimensional gauge couplings, respectively. In a decompactification limit, the black hole threshold is set by the higher-dimensional Planck scale, and in view of \eqref{LambdaWGCdecomp} it is clear that \eqref{reqlargegamma} only holds if an additional strong coupling limit is taken in the higher-dimensional theory. Hence, even though from the lower-dimensional perspective a gauge theory may become strongly coupled, this does not necessarily lead to $\gamma\gg \mathcal{O}(1)$ unless the decompactification limit is superimposed with an additional limit. If this limit is again a decompactification limit, we can repeat the above argument to still obtain $\gamma\sim \mathcal{O}(1)$. Alternatively, the limit may be an emergent string limit. For the gauge theory to become strongly coupled as a consequence of an emergent string limit, it has to be a non-perturbative gauge sector in the $D$-dimensional string theory. In the heterotic theory, the gauge coupling of a non-perturbative gauge sector scales as 
	\begin{equation}\label{gnonpert}
		g_\ttiny{$\U(1)_{\rm n.p.}$}^2 M_\het^{D-4} \sim \frac{1}{g_\het^2}\,,
	\end{equation}
	where $M_\het$ and $g_\het$ are, respectively, the string scale and the $D$-dimensional string coupling of the emergent heterotic string. In emergent string limits the black hole threshold is set by 
	\begin{equation}\label{Mminhet}
		\frac{M_\ttiny{BH, min.}^2}{M_\PlD{D}^2} \sim \left(\frac{M_\PlD{D}}{M_\het}\right)^{2D-6} =  \left(\frac{1}{g_\het^2}\right)^{\frac{2D-6}{D-2}}\,. 
	\end{equation}
	On the other hand, from \eqref{gnonpert} we find 
	\begin{equation}
		\frac{\Lambda_\WGC^2}{M_\PlD{D}^2} =  g_\ttiny{$\U(1)_{\rm n.p.}$}^2 M_\het^{D-4} \left(\frac{M_\PlD{D}}{M_\het}\right)^{D-4}\sim \left(\frac{1}{g_\het^2}\right)^{\frac{2D-6}{D-2}} \sim \frac{M_\ttiny{BH,min}^2}{M_\PlD{D}^2}\,. 
	\end{equation}
	We have obtained that non-perturbative heterotic gauge sectors in the heterotic string also do not satisfy \eqref{reqlargegamma} in the emergent string limit. 
	To conclude, a strong coupling limit with $\gamma\rightarrow \infty$ in the $D$-dimensional theory requires taking a finite distance strong coupling limit. 
	We can combine these considerations in the following way:
	
	
	\begin{mdframed}[backgroundcolor=white,shadow=true,shadowsize=4pt,shadowcolor=seccolor,roundcorner=6pt]
		\begin{claim}\label{claim:existencetower}
			Consider a $D$-dimensional $\U(1)$ gauge theory in an effective theory of gravity and let $\mathcal{M}$ be the field space of the EFT with $B$ its boundary. Then there can exist an infinite tower of super-extremal states with arbitrarily high charge that become particle-like excitations in certain regions of the field space if either of the following is true:
			\begin{enumerate}[label={C2.\arabic*.},ref={C2.\arabic*}]
				\item\label{case:claim1-i} For every $\epsilon>0$, there exists a $p_\epsilon\in \mathcal{M}\backslash B$ such that 
				\begin{equation} \label{Claim1.1}
					\left.\left(g^2_\UoD{D} M_\PlD{D}^{D-4}\right)\right|_{p_\epsilon}<\epsilon \qquad {\rm and} \qquad  \left.\frac{g^2_\UoD{D} M_\PlD{D}^{D-2}}{M^2_\Plinf}\right|_{p_\epsilon}<\epsilon\,, 
				\end{equation}
				where $M_\Plinf$ is the Planck scale of the asymptotic theory in the limit $\epsilon\rightarrow 0$.
				\item\label{case:claim1-ii} For every $p_1\in \mathcal{M}\backslash B$ and for every $\epsilon >0$, there exists a $p_\epsilon\in \mathcal{M}\backslash B$ with the property that 
				\begin{equation}\label{Claim1.2}
					\left.\left(g^2_\UoD{D} M_\PlD{D}^{D-4}\right)\right|_{p_\epsilon} >\epsilon^{-1}\qquad {\rm and} \qquad d(p_1,p_\epsilon) \,  < \delta_1\, M_\PlD{D}^{\frac{D-2}{2}} \,
				\end{equation}
				for some $\delta_1 < \infty$,
				where $d(p_1,p_\epsilon)$ denotes the geodesic distance between $p_1$ and $p_\epsilon$. 
			\end{enumerate}
		\end{claim}
	\end{mdframed}
	
	\vspace{2mm}
	
	This claim covers all possible cases, since we require that a tower of states can become particle-like in some region of the field space, and we just argued that this can happen only if there exists a weak coupling limit or a special strong coupling limit. In this sense, \eqref{Claim1.1} or~\eqref{Claim1.2} are necessary conditions for the tower WGC for particles in the sense of Definition \ref{Def-tower}. Moreover, \eqref{Claim1.1} is required so that the tower of light states is made up of particle-like states, also in the asymptotic description of the theory. However, it is still not a sufficient criterion because there exist gauge theories, such as the gauge theories on D-branes discussed in Section \ref{sec:ExistenceSuperExremalParticles}, which satisfy \eqref{Claim1.1}, but for which there is no tower of super-extremal states associated with them.
	
	On the other hand, \eqref{Claim1.2} ensures that the strong coupling limit is obtained by taking a finite distance limit. This does not exclude the possibility of having a tower of super-extremal particle states in an infinite distance limit as long as it is superimposed by a suitable finite distance limit leading to a strongly coupled gauge theory. Note that also Case \ref{case:claim1-ii} is not sufficient for the existence of a tower of states:  There  are gauge theories that become strongly coupled at finite distance in field space without having an infinite tower of super-extremal particle-like states charged under them. 
	
	To conclude this section, we illustrate the two possibilities in \cref{claim:existencetower}, again in the context of M-theory compactifications on Calabi--Yau threefolds $X_3$ with gauge groups $\U(1)_C$ as in \eqref{U1C}. It is convenient to choose the basis $\{C^a\}$ to be a basis of Mori cone generators, with a corresponding basis of $\U(1)$ gauge factors. Since at each boundary of the K\"ahler cone a Mori cone generator vanishes, this basis is particularly useful in light of \cref{claim:existencetower,claim:WGC}. Generically, we can differentiate between three different types of boundary of the K\"ahler cone \cite{Witten:1996qb} depending on whether 
	\begin{enumerate*}[before=\unskip{: }, itemjoin={{; }}, itemjoin*={{ or }},label={{\itshape\roman*}.},ref={{\itshape\roman*}}]
		\item\label{case:shrinkspoint}only a curve shrinks to a point 
		\item\label{case:shrinksdivisor}also a divisor shrinks to a point,\footnote{For our purposes a boundary at which a divisor shrinks to a curve belongs to the first type.} 
		\item\label{case:shrinksCY}even the entire Calabi--Yau shrinks to a curve or a surface.\footnote{Shrinking the Calabi-Yau to a point is expected to be obstructed by quantum corrections in the hypermultiplet moduli space.}
	\end{enumerate*}
	The last case corresponds to an asymptotic boundary, since we can only reach it at infinite distance in the vector multiplet moduli space. The WGC at these asymptotic boundaries has been investigated in detail in \cite{Cota:2022maf}. The results of \cite{Cota:2022maf} establish that, indeed, for any linear combination of $\U(1)$s that admits a weak coupling limit where 
	\begin{equation}\label{limitMtheory}
		\frac{\Lambda_\ttiny{WGC}}{M_\Plinf} \ll 1\,,
	\end{equation}
	there exists a tower of super-extremal states that become particle-like in the vicinity of the asymptotic boundary where \eqref{limitMtheory} is realized.\footnote{In \cite{Cota:2022maf}, the ratio $\Lambda_\ttiny{WGC}/\Lambda_\QG$ was considered. For decompactification limits, this agrees with \eqref{limitMtheory}, but for emergent string limits, $\Lambda_\ttiny{WGC}/\Lambda_\QG$ cannot be made parametrically large, unlike \eqref{limitMtheory}.} Therefore, in this case, Eq. \eqref{Claim1.1} is, in fact, sufficient to ensure the existence of a tower of states.

	According to our general discussion motivating \cref{claim:existencetower}, in order for any other $\U(1)$ to have a tower of super-extremal states that can become particle-like at some point in field space, we need strong coupling dynamics.  In general, a strong coupling limit in M-theory requires some divisor to shrink to zero size. This can be seen from the way in which divisor volumes appear in the gauge kinetic function. Shrinking divisors are encountered at the boundaries of Case \ref{case:shrinksdivisor} or at asymptotic boundaries. If there exists a divisor that only shrinks at an asymptotic boundary but not at finite distance, this means that shrinking the divisor requires making the dual curve very large. As a consequence, the states wrapping this curve become massive in this limit, and there are no light states charged under the strongly coupled gauge symmetry. This illustrates why particle-like towers for $\U(1)$s that become strongly coupled arise at finite distance in the moduli space, i.e. along the boundaries of Case \ref{case:shrinksdivisor}. These boundaries correspond to CFT boundaries at which infinitely many BPS states can become massless. The massless states at the boundary arise from M2-branes (multi-)wrapping curves that are contained in the divisor shrinking to a point. This potentially leads to a super-extremal BPS tower of particle states, thereby illustrating the second option in \cref{claim:existencetower}. However, notice that the second condition in \cref{claim:existencetower} is not sufficient, since along the CFT boundaries most $\U(1)$s become strongly coupled. However, if the associated curve is not contained in the shrinking divisors, all charged states have a mass at or above the QG cutoff. 
	
	\section{Weak Gravity Conjecture in Absence of Towers}
	\label{sec:absencetower}
	
	The results of the previous sections can be combined as follows: First, in Section \ref{sec:necessitytower} we have concluded that a general theory of gravity is expected to lead to a minimal radius upon circle compactification, with the notable exception of the heterotic string. Second, in \cref{claim:WGC} we have given sufficient conditions for when the WGC must be satisfied at the particle level, although our \cref{claim:existencetower} implies that these do not necessarily need to be part of a tower.

	Now, the main statement of this work is that if the WGC is satisfied at the particle level in a $D$-dimensional theory of gravity, these particle states are enough to satisfy the CHC also after dimensional reduction on a circle and that a tower of particles is present whenever it is needed for consistency. More precisely, we formulate the following conjecture:
	
	\begin{mdframed}[backgroundcolor=white,shadow=true,shadowsize=4pt,shadowcolor=seccolor,roundcorner=6pt]
		\begin{conjecture} \label{Conjecture1}
			Consider a $D$-dimensional $\U(1)_D$ gauge theory in a $D$-dimensional theory of quantum gravity such that the WGC is realized by a set of particle-like excitations below the black hole threshold. Then, in the $(D-1)$-dimensional theory obtained by dimensional reduction on an $S^1$ the CHC for $\U(1)_D\times \U(1)_\KK$ is satisfied by the KK replicas of the $D$-dimensional super-extremal particle states for any value of the circle radius which allows for an interpretation as a circle reduction of the $D$-dimensional gauge theory. This holds irrespective of whether the particles are part of a tower in the $D$-dimensional theory.
		\end{conjecture}
	\end{mdframed}
	\vspace{2mm}
	
	In the following, we will illustrate this conjecture in explicit string theory setups. In Section \ref{subsec_6t5FM} we consider the circle reduction of six-dimensional theories of gravity with $\U(1)$ gauge groups arising from F-theory on elliptically fibered Calabi--Yau threefolds. The effective five-dimensional theory is dual to M-theory on the same Calabi--Yau threefold, for which we also test Conjecture~\ref{Conjecture1} upon circle compactification in Section \ref{subsec:5t4MA}. We find that, indeed, the CHC for $\U(1)_D\times \U(1)_\KK$ is always satisfied even if no (known) tower of super-extremal particles exists in the higher-dimensional theory. In the presence of 2-forms in the higher-dimensional theory, there are further winding $\U(1)_w$s in the lower-dimensional theory. In \cref{sec:6dto5dheterotic,sec:windingU1}, we discuss the CHC involving these winding $\U(1)_w$s. 
	
	What remains as an interesting question is whether the converse is also true, i.e. whether super-extremal particle towers, whenever present, are needed for consistency of the CHC. We will come back to this question in Section \ref{sec:Conclusions}. 
	
	\subsection{From Six to Five Dimensions in Heterotic String Theory}
	\label{sec:6dto5dheterotic}
	
	The subtleties involved in the notion of a minimal radius for perturbative string theories were already mentioned at the beginning of Section \ref{sec:necessitytower}. We now elaborate on this further and verify the need for perturbative string towers for consistency of the WGC, while non-perturbative sectors are consistent even without a tower. For definiteness, we focus on compactifications of heterotic string theory to six dimensions and reduce this theory on a circle of radius $r_{S^1_\het}$.

	Due to T-duality, there are two possible interpretations of what the minimal radius for this theory should be, both leading to the same conclusion in regard to consistency of the WGC.
	First, one can take the standpoint that T-duality allows us to restrict to the regime $r_{S^1_\het} \geq  r_{S^1_\het}^\ttiny{min.} $ with $r_{S^1_\het}^\ttiny{min.} $ the self-dual radius 
	\begin{equation}\label{eq:minhetradius}
		r_{S^1_\het}^\ttiny{min.}  = \sqrt{\alpha'_\het} \,. 
	\end{equation}
	According to this logic, one takes $ r_{S^1_\het}^\ttiny{min.}$ as the minimal radius which can be probed by the theory.
	
	For a perturbative heterotic abelian gauge theory in a six-dimensional compactification on $K3$, the
	gauge coupling is related to the six-dimensional Planck scale as \cite[Eq. (2.50)]{Lee:2018urn} 
	\begin{equation}
		g_\ttiny{$\U(1)_{\rm pert.}$, 6}^2 M_\PlD{6}^4 = \frac{1}{4m}\frac{16\pi^2}{\ell^2_\het} = \frac{1}{m}\frac{1}{\alpha'_\het}\coma
	\end{equation}
	where
	\begin{equation}\label{eq:Plhet6d}
		M_\PlD{6}^4 = 4\pi \frac{(\text{Vol}(K3)M_\het^4)}{g_\het^2} M_\het^4\fstop
	\end{equation}
	Here the integer $m$ is the level of the perturbative gauge  algebra, and in our conventions
	$\ell_\het^2 = (2\pi)^2\alpha'_\het = \frac{2 \pi}{T}$, with $T$ the heterotic string tension. 
	
	The spectrum of the perturbative heterotic string in six dimensions includes special
	non-BPS states at the excitation level $n=m k^2$ with charge (see \cite{Lee:2018spm,Lee:2018urn})
	\begin{equation}\label{superhetero}
		q_n^2 = 4  m n 
	\end{equation}
	and mass
	\begin{equation}\label{eq:hetstates}
		m_{6,n}^2 = 8\pi T(n-1) = \frac{4(n-1)}{\alpha'_\het}\fstop
	\end{equation}
	These are super-extremal as follows from
	the charge-to-mass ratio $z_6$ in \eqref{eq:zDvector}, 
	\begin{equation}
		z_{6,n}^2 = g_\ttiny{$\U(1)_{\rm pert.}$, 6}^2 M_\PlD{6}^4  \frac{|q_n|^2}{\gamma m_{6,n}^2} = \frac{1}{\gamma}\frac{n}{n-1}>1\coma
		\label{eq:hetlim}
	\end{equation}
	where we used that $\gamma = 1$~\cite{Lee:2018spm}. 
	To evaluate the constraint imposed by the CHC after circle reduction, note first that the RHS of \eqref{eq:mDrvszD} becomes 
	\begin{equation}
		\frac{1}{4z_{6,n}^2(z_{6,n}^2-1)} + \frac{\fn_6(1-\fn_6)}{z_{6,n}^2} = \frac{n-1}{n} \left( \frac{n-1}{4} + \fn_6 (1- \fn_6)  \right) \,,
	\end{equation}
	where $\fn_6 = \frac{q_n \theta'}{2\pi}=q_n \theta$ and $0\leq \theta < 1$. 
	Therefore, the constraint \eqref{eq:mDrvszD} becomes
	\begin{equation}
		4 \frac{r_{S^1_\het}^2}{\alpha'_\het} \geq \frac{1}{n} \left( \frac{n-1}{4} + \fn_6 (1- \fn_6)  \right)\fstop
		\label{eq:rminforheteroticKK}
	\end{equation}
	For the minimal radius \eqref{eq:minhetradius} and the states with mass \eqref{eq:hetstates}, the LHS simplifies to
	\begin{equation}
		\left(m_{6,n} r_{S^1_\het}\right)^2 =  4 (n -1) \,,
	\end{equation}
	leading to the inequality
	\begin{equation}\label{CHCKKhet}
		4 \geq \frac{1}{n} \left( \frac{n-1}{4} + \fn_6 (1- \fn_6)  \right) \,.
	\end{equation}
	Clearly, this constraint can be easily satisfied. 
	We conclude that the CHC involving the KK $\U(1)_\KK$
	would be comfortably satisfied above the self-dual radius even in the absence of a tower of states --- a single super-extremal state in six dimensions is already sufficient.
	
\begin{figure}
		\centering
		\begin{tikzpicture}[baseline=0,scale=2]
			\draw[thick,|-|] (0,0) -- node[pos=0,below] {$0$} node[pos=1,below] {$r_{S^1_\het}^\ttiny{min.}=\sqrt{\alpha'_\het}$} (2,0);
			\draw[thick] (2,0) -- (3,0);
			\draw[thick,densely dashed,->] (3,0) -- node[pos=1,below] {$r$} (4,0);
			\node (U1w) at (1,0.5) {$\U(1)_w$};
			\node (U1KK) at (1,-0.5) {$\U(1)_\KK$};
			\node (U1KK2) at (3,0.5) {$\U(1)_\KK$};
			\node (U1w2) at (3,-0.5) {$\U(1)_w$};
			\draw[thick,Triangle-Triangle] (U1w) to[bend left] (U1KK2);
			\draw[thick,Triangle-Triangle] (U1w2) to[bend left] (U1KK);
		\end{tikzpicture}
		\caption{Interpretation of the KK $\U(1)_\KK$ and the winding $\U(1)_w$ for different values of $r$.}
		\label{fig:windingU1vsKKU1}
	\end{figure}
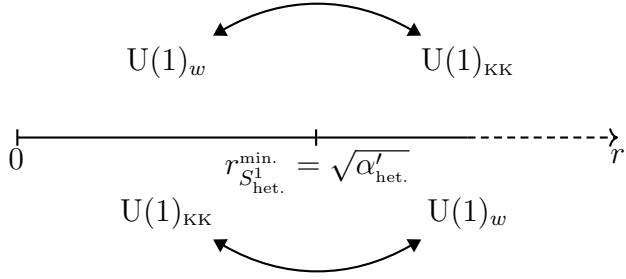

	However, so far we have neglected the perturbative winding $\U(1)_w$, which also exists as a consequence of T-duality. The KK reduced theory must satisfy the CHC for the $\U(1)_w \times \U(1)_\ttiny{pert.}$ pair of gauge groups above the self-dual radius.
	This is equivalent to demanding that the CHC for $\U(1)_\KK \times \U(1)_\ttiny{pert.}$ also holds in the regime $r_{S^1_\het} <\sqrt{\alpha'_\het}$ (see Figure \ref{fig:windingU1vsKKU1}). Given our discussion in Section~\ref{sec:necessitytower}, one might object that in the regime $r_{S^1_\het} <\sqrt{\alpha'_\het}$ the gauge theory $\U(1)_\KK$ cannot be viewed any longer as a KK $\U(1)_\KK$. To check whether this is true, we need to compare the mass of the KK $\U(1)_\KK$ with the mass of the minimal black hole. Equivalently, for large radii and $\U(1)_w$ we have to check whether the winding modes have mass below the black hole threshold. 
	
	Let us focus on the latter case. In the emergent string limit the black hole threshold is set by the smallest six-dimensional black holes with mass (cf. \eqref{Mminhet})
	\begin{equation}
		\frac{M_\ttiny{BH, min.}}{M_\PlD{6}}=\left(\frac{M_\PlD{6}}{M_\het}\right)^3\sim \frac{1}{g_\het^{3/2}}\fstop
	\end{equation}
	On the other hand, the mass of the winding states is\footnote{Analogously, one can compute
		\begin{equation}
			\frac{M_w}{M_\PlD{5}} \sim \left(\frac{r_{S}^1}{\sqrt{\alpha'_\het}}\right)^{2/3} g_\ttiny{het}^{2/3}\fstop
		\end{equation}
	}
	\begin{equation}
		M_w \sim \frac{r_{S^1_\het}}{\alpha'_\het} \Longrightarrow \frac{M_w}{M_\PlD{6}} \sim \frac{r_{S^1_\het}}{\sqrt{\alpha'_\het}}g_\het^{1/2}\fstop
	\end{equation}
	Comparing the two masses, we find that $\U(1)_w$ can be interpreted as a winding $\U(1)_w$ as long as
	\begin{equation}\label{maxradius}
		\frac{r_{S^1_\het}}{\sqrt{\alpha'_\het}} \leq \frac{{\cal O}(1)}{g^2_\het}\,.
	\end{equation}
	
	To check if the CHC is satisfied for the winding $\U(1)_w$ and the six-dimensional perturbative gauge theory $\U(1)_\ttiny{pert.}$, it is sufficient to evaluate \eqref{eq:mDrvszD} upon replacing 
	\begin{equation}\label{Tdualityhet}
		r_{S^1_\het} \longleftrightarrow \frac{\alpha'_\het}{r_{S^1_\het}}\fstop
	\end{equation}
	In this case \eqref{eq:rminforheteroticKK} becomes
	\begin{equation}
		4 \frac{\alpha'_\het}{r_{S^1_\het}^2} \geq \frac{1}{n} \left( \frac{n-1}{4} + \fn_6 (1- \fn_6)  \right)\,.
	\end{equation}
	We see a clear violation of the above equation for $r_{S^1_\het}^2$ sufficiently large, which can be achieved in the weak coupling limit $g^2_\het \to 0$. Avoiding a violation of the CHC between $\U(1)_w$ and $\U(1)_\ttiny{pert.}$ requires a tower of super-extremal charged states in six dimensions such that states with ever higher charge become particle-like as $g_\het\rightarrow 0$. This is exactly what happens for the tower of super-extremal excitations of the heterotic string, whose existence in six dimensions was shown in~\cite{Arkani-Hamed:2006emk,Heidenreich:2016aqi,Lee:2018spm,Lee:2018urn}. 
	
	We now come to the second, equivalent point of view:
	Via T-duality, \eqref{maxradius} translates into a minimal radius
	\begin{equation} \label{minrad-2}
		r_{S^1_\het}^\ttiny{min.} \sim g_\het^2\sqrt{\alpha'_\het}\,,
	\end{equation}
	below which $\U(1)_\KK$ cannot be viewed as a KK $\U(1)_\KK$ anymore because $M_\KK \geq M_\ttiny{BH, min.}$. In this interpretation, there is indeed a minimal radius, well below the self-dual radius for small string coupling. More importantly, it goes to zero in the perturbative limit. Also in this (T-dual) interpretation, a tower of perturbative super-extremal states is required by the CHC.
	
	So far we have focused on the perturbative gauge sector.
	What remains to investigate is whether also a tower of super-extremal states charged under a non-perturbative gauge group $\U(1)_\ttiny{n.p.}$ is needed for consistency of the CHC.
	As a shortcut, we can immediately work in the second picture, where the minimal radius is set by \eqref{minrad-2}, and focus on the CHC
	between $\U(1)_\KK$ and $\U(1)_\ttiny{n.p.}$ at small radii. 
	The main difference is that for $\U(1)_\ttiny{n.p.}$ the gauge coupling scales as
	\begin{equation}
		g_\ttiny{$\U(1)_{\rm n.p.}$, 6}^2 M_\het^2  \propto g_\het^{-2}\coma
	\end{equation}
	while the Planck mass is still \eqref{eq:Plhet6d}. This leads to a scaling
	\begin{equation}
		z_6^2 \propto g_\het^{-4}\fstop
	\end{equation}
	For simplicity focus on massless states charged under the non-perturbative gauge group.\footnote{At least for the non-Higgsable clusters, such states always exist in the form of the non-abelian gauge bosons.} Then \eqref{eq:mDrvszD} effectively reduces to
	\begin{equation}
		\frac{r_{S^1_\het}^2}{\alpha'_\het} \geq \frac{\fn_6(1-\fn_6)}{z_6^2}\propto g_\het^4 \fn_6(1-\fn_6)\fstop
	\end{equation}
	
	By comparison with \eqref{minrad-2}, there is no parametric clash in the perturbative regime and hence the CHC is obeyed only with a finite number of super-extremal states in the non-perturbative sector. Equivalently, one can focus on the radii above the self-dual radius and convince oneself that no parametric violation of the CHC occurs with the winding $\U(1)$.
	This conclusion is in agreement with the fact that in the six-dimensional parent theory, there is no tower of super-extremal particles charged under $\U(1)_\ttiny{n.p.}$ in the perturbative regime.
	
	\subsection{Perturbative Gauge Theories in Type II String Theory}
	\label{sec:pertDbranes}
	
	Instead of the heterotic string, let us now consider gauge theories realized on D-branes in perturbative Type II string theory. The coupling of a gauge theory on a space-time filling D-brane in a $D$-dimensional theory of gravity  scales as 
	\begin{equation}
		g^2_\UoD{D} M_s^{D-4} \sim g_s \,.
	\end{equation}
	In the $g_s\rightarrow 0$ limit, we find 
	\begin{equation}
		\frac{\Lambda_\WGC^2}{M_\PlD{D}^2} = g^2_\UoD{D} M_\PlD{D}^{D-4} \sim g_s \left(\frac{M_\PlD{D}}{M_s}\right)^{D-4} = g_s^{\frac{6-D}{D-2}}  \gg g_s^{\frac{4}{D-2}} = \frac{\Lambda_\QG^2}{M_\PlD{D}^2}\coma 
	\end{equation}
	for $g_s\ll 1$ and $D>2$. According to our discussion in Section~\ref{sec:ExistenceSuperExremalParticles}, these gauge theories are effectively decoupled from gravity, as one would expect for an open string gauge theory in the weak coupling limit. In this sense, not even the ordinary WGC in $D$ dimensions is strictly speaking required to hold.

	If we compactify on an additional $S^1$, then, according to our Conjecture~\ref{Conjecture1}, the CHC between a D-brane and the KK $\U(1)_\KK$ should nonetheless be satisfied for all values of the circle radius for which the theory can be interpreted as a compactification of a $D$-dimensional gauge theory. One might be tempted to follow the arguments in the previous section {\it verbatim} and conclude that in the $g_s\rightarrow 0$ limit there is no minimal radius for the additional circle; this would mean that, for consistency of the WGC, the $D$-dimensional theory must have an infinite tower of super-extremal particle states with arbitrary high charge. However, for D-brane gauge theories in perturbative Type II theory, such particle states do not exist \cite{Cota:2022yjw}, since the open string sector at best gives rise to a tower of states with constant charge. The difference from the heterotic models is explained by noticing that in the small radius limit the theory cannot be viewed as a circle compactification of a $D$-dimensional gauge theory: In the small radius limit, the relevant light states are the winding modes, which now are not charged under the D-brane gauge theory since open strings do not lead to winding modes. From the T-dual picture, it is clear that we are not dealing with a $D$-dimensional gauge theory, as under T-duality the spacetime filling D-brane gets mapped to a brane with $(D-1)$-dimensional worldvolume localized at a point on the circle. Thus, for $r_{S^1}<\sqrt{\alpha'}$, the theory cannot be viewed as a circle compactification of a $D$-dimensional gauge theory, and according to our Conjecture~\ref{Conjecture1} the CHC does not need to be satisfied.

	\subsection{From Six to Five Dimensions in F-/M-theory}
	\label{subsec_6t5FM}
	
	We now turn to non-perturbative compactifications of
	F-theory and their circle reductions, focusing for definiteness on six-dimensional models on Calabi--Yau threefolds. Their basic properties have already been introduced in Section~\ref{sec:minimalradiusFM}: Abelian gauge theories in F-theory are associated with additional sections and can be viewed as, roughly, arising from 7-branes that wrap the height-pairing divisor in the base $B_2$ of $\pi : X_3 \to B_2$ defined in \eqref{heightpairing}.

	In the following, we analyze two illustrative classes of such
	$\U(1)$ gauge theories from the point of view of the WGC.
	First we choose $B_2=\mathbb{P}^2$, with an abelian gauge theory realized on (a multiple of) the hyperplane class $h\subset \mathbb{P}^2$. In this theory, which admits neither a weak nor a strong coupling limit, the CHC after circle reduction is shown to be satisfied despite the absence of a charged super-extremal particle tower. 
	
	Next, we consider a non-Higgsable abelian gauge theory by choosing $B_2=dP_9$ with the abelian gauge theory realized on a fibral curve of the elliptic surface $dP_9$. 
	Here, even though no massless charged states exist, nor an asymptotic tower, we again confirm the CHC after circle compactification.
	
	Finally, recall that the case that the $\U(1)$ gauge theory is dual to a perturbative heterotic gauge group has been previously analyzed in great detail in \cite{Lee:2018spm,Lee:2018urn}, where the existence of an infinite tower of super-extremal states has been established in the weak coupling limits. 
	
	\subsubsection{Example: \texorpdfstring{$B_2= \mathbb{P}^2$}{P2}}\label{sec:P2}
	
	F-theory on $B_2=\mathbb{P}^2$ has no tensor branch, so the gauge coupling of a $\U(1)$ gauge theory realized on (a multiple of) the hyperplane class $h\subset \mathbb{P}^2$ is always $\mathcal{O}(1)$ in Planck units. Thus, the gauge theory does not admit a weak or strong coupling limit. According to \cref{claim:existencetower}, we, hence, do not expect a tower of particle-like states for this gauge theory. If we denote by $v_s$ the volume of the hyperplane class in units of $M_s$, then\footnote{Even though we work in F-theory, we display the IIB gauge coupling for illustration.} 
	\begin{equation}
		\frac{M_\PlD{6}^4}{M_s^4}= 4\pi \text{vol}(\mathbb{P}^2) \frac{M_s^4}{ g_\IIB^2} =\frac{4\pi}{ g_\IIB^2} \frac{1}{2} v^2_s \,. 
	\end{equation}
	In this setting charged states arise from $[p,q]$-string states or D3-branes wrapped on $n\,[H]$ with mass 
	\begin{equation}
		m \sim (2\pi)^2 v_s M_s^3 \,.
	\end{equation}
	For sufficiently large $n$, the latter gives rise to black strings that, when wrapped on a loop, yield charged black holes in 6d. Since $v_s$ has to be at least of $\mathcal{O}(1)$, the charged states arising from D3-branes or stretched $[p,q]$-strings always have a mass at or above the black hole threshold. Therefore, there is no infinite tower of light super-extremal states consistent with our \cref{claim:existencetower}. The only charged particle-like states are massless, and such states are guaranteed to exist because $\U(1)$ gauge theories in this geometry are Higgsable~\cite{Morrison:2016lix}. At the particle level, the WGC in the six-dimensional theory is trivially satisfied just by the massless states.
	
	For F-theory on $\pi : X_3 \to B_2$ with a gauge theory realized on some divisor $D\in H_2(B_2)$ the diameter $\ell_\ttiny{perp.}$ of the space transverse to $D$ can be estimated to be
	\begin{equation}\label{ellperpdef}
		\ell_\ttiny{perp.} = \max_{C \in \mathcal{I}_D} \left\{\left(\text{vol} (C)\right)^{1/2} \right\}\,,
	\end{equation}
	where the set $\mathcal{I}_D$ of curves is defined as
	\begin{equation}
		\mathcal{I}_D = \{C\;\text{generator of Mori cone of }B_2 \;|\; C\cdot D\geq 0\}\,. 
	\end{equation}
	For the gauge theory on $h$ the transverse space is hence given by 
	\begin{equation}
		\ell_\ttiny{perp.} M_\PlD{6}= \left(\text{vol}(h) M_s^2\right)^{1/2} \frac{M_\PlD{6}}{M_s} =v_s\,,
	\end{equation}
	which cannot be made smaller than the minimal length scale resolvable by the EFT 
	\begin{equation}\label{ellperpP2}
		\ell_\ttiny{min.} M_\PlD{6}= \frac{M_\PlD{6}}{\Lambda_\QG} = \frac{M_\PlD{6}}{M_s} \sim v^{1/2}_s\,,
	\end{equation}
	since $v_s\geq 1$. From the EFT point of view, the gauge theory should thus be viewed as a defect theory, as the transverse space to the 7-brane can be resolved within the EFT. The fact that at the particle level, the WGC is satisfied only by massless states is consistent with this picture. Nevertheless, according to our Conjecture~\ref{Conjecture1}, the WGC should continue to be satisfied upon circle compactification, and this is indeed a non-trivial prediction that must be checked. We now discuss the compactification of such a model on a circle. 
	
	For definiteness, consider a realization of $\pi: X_3\rightarrow \mathbb{P}^2$ with one extra section as a Morrison--Park model, in which the elliptic fiber is realized by the genus one curve locus on $\mathrm{Bl}_1(\mathbb{P}_{112})$~\cite{Morrison:2012ei}. 
	The Calabi-Yau has the following intersection data
	\begin{equation}
		\mathcal{I}(X_3) = 18 J_1 J_2^2+6 J_1 J_2 J_3+2 J_1 J_3^2+9 J_2^3+3 J_2^2 J_3 +J_2 J_3^2 
	\end{equation}
	and integrated second and third Chern class
	\begin{equation}\label{cherndataP2}
		\int_{X_3} c_2 (X_3)\wedge J_i = ( 24 , 126, 36) \coma \chi (X_3) = -216 \,.
	\end{equation}
	We can identify the zero section as given by $S_0 = J_2 -3J_3$, while the rational section is $S_Q = J_1 -J_2+3J_3$. The Shioda map \eqref{Shiodamap} is then
	\begin{equation}
		\sigma(S_Q) = J_1 -2J_2\,,
	\end{equation}
	and the shifted zero section \eqref{hatS0} is $\widehat{S}_0=J_2$. The elliptic fiber reads $\mathcal{E} = 2 \mathcal{C}^1 + \mathcal{C}^2$. The fiber degenerates to an $I_2$ fiber over points on the base, and in such a degenerate fiber the isolated fibral curve with charge $q=1$ is given by $C_z =  \mathcal{C}^1$. Finally, we identify $C_b=\mathcal{C}^3$ with the base curve in $\mathbb{P}^2$. 
	
	The Mori cone of $X_3$ is spanned by the base curve $C_b = S_0 \cdot \pi^*h$ --- with $h$ being the hyperplane curve in $\mathbb{P}^2$ --- as well as the rational curves $\{A_I, B_I\}$ and $\{A_{II}, B_{II}\}$ that appear in the two types of $I_2$ fibers over the codimension-two loci $C_I$ and $C_{II}$ on the base.
	The isolated fibral curves intersect the sections as displaced in Table \ref{tableP2}~\cite{Mayrhofer:2014haa}.
	\begin{table}[!t]
		\centering
		\begin{tabular}{ c|cccccc } 
			& $A_I$ & $B_I$ & $A_{II}$ & $B_{II}$ \\ 
			\hline
			$S_Q$ & $-1$ & 2 & 0 & 1 \\
			$S_0$ & 1 & 0 & 1 & 0 \\
			$S_Q+S_0 $& 0 & 2 & 1 & 1  \\
			$S_Q-S_0$ & $-2$ & 2 & $-1$ & 1   \\
		\end{tabular}
		\caption{$\U(1)$ charges of M2-branes wrapping fiber components over isolated fibral curves.}
		\label{tableP2}
	\end{table}
	Hence, we identify 
	\begin{equation}
		A_I = \mathcal{C}^2 \,, \quad B_{I} = 2\mathcal{C}^1  \,, \quad A_{II} = \mathcal{C}^1 +  \mathcal{C}^2 \,, \quad B_{II} = \mathcal{C}^1  \,. 
	\end{equation}
	Note that $A_I$ and $B_{II}$ are Mori cone generators of $X_3$. 
	The number of $\vert q \vert =1$ massless charged hypermultiplets in the $\mathcal{N}=(1,0)$ 
	six-dimensional theory can be obtained by calculating the Gopakumar--Vafa invariants $n_{A_{II}}^0 = n_{B_{II}}^0=144$~\cite{Oehlmann:2019ohh,Kashani-Poor:2019jyo}.
	Similarly, we obtain $n_{A_{I}}^0 = n_{B_{I}}^0=18$ massless charged hypermultiplets with $\vert q \vert =2$. 
	
	In the Shioda--Tate--Wazir basis $(\sigma(S_Q),\widehat{S}_0,J_3)$, the intersection data can be written as
	\begin{equation}
		\mathcal{I}(X_3) = 144 \sigma(S_Q)^3-36 \sigma(S_Q)^2\widehat{S}_0 -12 \sigma(S_Q)^2 J_3+9 \widehat{S}_0^3+3 \widehat{S}_0^2 J_3   +\widehat{S}_0 J_3^2 \,.
	\end{equation}
	After defining the K\"ahler form as (cf. \eqref{ShiodaTW})
	\begin{equation}
		J = \tau \widehat{S}_0 + z \sigma(S_Q) + v^3J_3\coma
	\end{equation}
	with $\tau = 2v^1+ v^2$ and $z = v^1$, the total volume is given by
	\begin{equation}
		\mathcal{V}_{X_3} = \left(24 \theta ^3-18 \theta ^2+\frac{3}{2}\right) \tau ^3+ \left(\frac{3 }{2}v^3-6 \theta ^2 v^3\right)\tau ^2+\frac{1}{2}(v^3)^2\tau\fstop 
		\label{eq:volP2intheta}
	\end{equation}
	Here we defined the Coulomb branch parameter $\theta$ as
	\begin{equation}
		\frac{z}{\tau}=\frac{v^1}{2v^1+v^2}=\frac{\theta'}{2\pi}=\theta\fstop
	\end{equation}
	\begin{figure}[!t]
		\centering
		\begin{subfigure}[t]{0.49\textwidth}
			\centering
			\begin{tikzpicture}[scale=3]
				\node[label={[xshift=0.9cm,yshift=-3cm]:{\small $B_2$}}] (B2) at (0,0) {
					\begin{tikzpicture}[scale=1.7]
						\coordinate (N) at (2.6,4.6);
						\coordinate (O) at (0,0.1);
						\coordinate (P) at (0.7,0);
						\coordinate (Q) at (1.7,0.6);
						\coordinate (R) at (2.5,0.9);
						\coordinate (S) at (2.2,1.6);
						\coordinate (T) at (1.5,1.5);
						\coordinate (U) at (0.5,1.2);
						\shade[ball color = prcolor, opacity = 0.3,draw = prcolor,draw opacity=1,thick] (O) to [pattern=north east lines, closed, curve through = {(O) (P)  (Q)  (R) (S)  (T) (U)}] (O);
						\node at (0,0.6) {
							\begin{tikzpicture}
								\pic[rotate=-30,draw=prcolor,thick] at (0,0) {hole={0.5cm}{0.75cm}};
							\end{tikzpicture}
						};
						\node at (0.9,0.5) {
							\begin{tikzpicture}
								\pic[rotate=0,draw=prcolor,thick] at (0,0) {hole={0.5cm}{0.5cm}};
							\end{tikzpicture}
						};
						\node at (2,1.1) {
							\begin{tikzpicture}
								\pic[rotate=30,draw=prcolor,thick] at (0,0) {hole={0.5cm}{1.25cm}};
							\end{tikzpicture}
						};
					\end{tikzpicture}
				};
				\node (tF) at (-0.8,1.2) {
					\begin{tikzpicture}[scale=1]
						\pic[rotate=-30,draw=seccolor,thick, fill=seccolor!40!white,fill opacity = 0.9] at (0,0) {torus={9mm}{3.6mm}{55}};
					\end{tikzpicture}
				};
				\draw[densely dashed,thick,-Triangle] (tF) -- (-0.5,0.3);
				\node[label={[xshift=0.4cm,yshift=-0.3cm]:{\small $2\mathcal{C}^1$}}] (Cr) at (0.615,1.3) {
					\begin{tikzpicture}[scale=1]
						\shade[ball color=seccolor,opacity=0.6,rotate=50,smooth,draw=seccolor,thick] (0,0) to [out=140,in=90] (-1,-1)
						to [out=-90,in=240] (0.8,-0.6)
						to [out=60,in=-60] (1.2,1.2)
						to [out=120,in=90] (0.3,0.7)
						to [out=-90,in=20] (0.3,0)
						to [out=200,in=-40] (0,0);    
					\end{tikzpicture}
				};
				\node[label={[xshift=-0.4cm,yshift=-0.3cm]:{\small $\mathcal{C}^2$}}] (Cl) at (0.25,1.3) {
					\begin{tikzpicture}[scale=0.7]
						\begin{scope}[xscale=-1]
							\shade[ball color=tercolor,opacity=0.6,rotate=35,smooth,draw=tercolor,thick] (0,0) to [out=140,in=90] (-1,-1)
							to [out=-90,in=240] (0.8,-0.6)
							to [out=60,in=-60] (1.2,1.2)
							to [out=120,in=90] (0.3,0.7)
							to [out=-90,in=20] (0.3,0)
							to [out=200,in=-40] (0,0);    
						\end{scope}
					\end{tikzpicture}
				};
				\foreach \i in {0,195}
				{
					\begin{scope}[rotate around ={\i:(Cl)}] 
						\draw[<-] (0.275,1.5) -- (0.275,1.65);
					\end{scope}
				}
				\foreach \i in {0,180}
				{
					\begin{scope}[rotate around ={\i:(Cr)}] 
						\draw[->] (0.6,1)-- (0.6,0.85);
					\end{scope}
				}
				\draw[<-] (0.73,0.3) -- (0.88,0.3);
				\draw[<-] (0,0.25) -- (0,0.4);
				\draw[<-] (0,-0.11) -- (0,-0.26);
				\draw[<-] (-0.6,-0.1) -- (-0.75,-0.1);
				\draw[thick,-Triangle,densely dashed] (0.35,0.8) -- (0.28,0.33);
			\end{tikzpicture}
			\caption{}
			\label{sfig:CYP2-splitting}
		\end{subfigure}\hfill
		\begin{subfigure}[t]{0.49\textwidth}
			\centering
			\begin{tikzpicture}[scale=3.7]
				\node (K3) at (0,1.5) {
					\begin{tikzpicture}[scale=5.6]
						\draw[fill=seccolor,fill opacity=0.1,draw=seccolor,thick] (0.05,0) -- node[left,pos=0.8,seccolor,fill opacity=1] {\small $K3$} (0.05,0.5) -- (-0.45,0.25) -- (-0.45,-0.25) -- (0.05,0);
					\end{tikzpicture}
				};
				\node[label={[xshift=0.7cm,yshift=-0.3cm]:{\small $\mathcal{C}^1$}}] (C1) at (0,0) {
					\begin{tikzpicture}[scale=3.7]
						\shade[ball color=seccolor,opacity=0.6,smooth,draw=seccolor,thick] (0,0) circle (0.41);
						\draw[dashed] (-0.41,0) arc (180:0:0.41 and 0.05);
						\draw[dashed] (-0.41,0) arc (180:0:0.41 and -0.05);
						\foreach \i in {45,135,...,315}
						{
							\begin{scope}[rotate around ={\i:(0,0)}] 
								\draw[->] (0.25,0)-- (0.4,0);
							\end{scope}
						}
					\end{tikzpicture}
				};
				\foreach \i in {60,150,...,330}
				{
					\begin{scope}[rotate around ={\i:(0,1.5)}] 
						\draw[<-] (0.25,1.5)-- (0.4,1.5);
					\end{scope}
				}
				\draw[thick,-Triangle,densely dashed] (K3) -- (C1);
			\end{tikzpicture}
			\caption{}
			\label{sfig:CYP2-K3}
		\end{subfigure}
		\caption{An illustration of the geometry corresponding to F-theory on an elliptic fibration over $B_2=\mathbb{P}^2$ with extra section. Figure \ref{sfig:CYP2-splitting} shows the elliptic fibration that is adiabatic in the limit of a large base and a small fiber. As discussed in the main text, the fiber can be made arbitrarily large while satisfying $\cV_{X_3}=1$ if $\mathcal{C}^1$ is made large while $\mathcal{C}^2$ and $B_2$ are shrunk. Arrows indicate this limit. In this limit the geometry is better described by a K3 fibration with large base and small fiber as illustrated in Figure~\ref{sfig:CYP2-K3}.}
		\label{fig:CYP2}
	\end{figure}
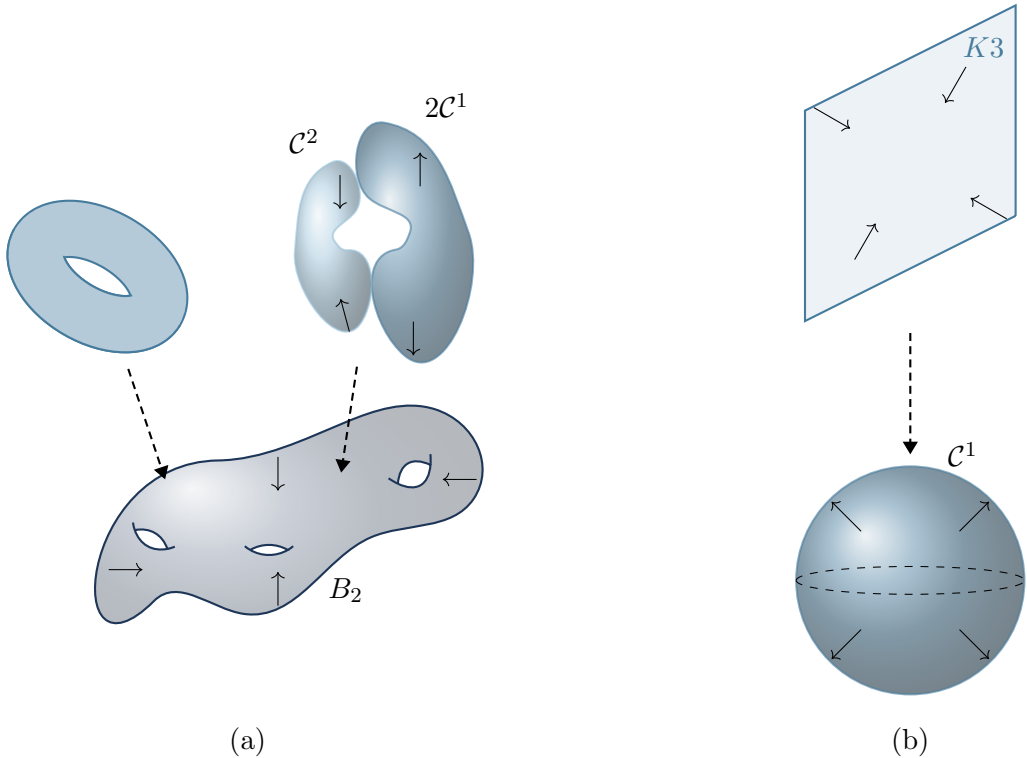
	We are now in a position to compute the minimal compactification radius $r_{\ttiny{min.}}$. According to the prescription of Section \eqref{sec:minimalradiusFM}, this requires finding the maximum volume of the elliptic fiber $\tau_\ttiny{max.}$, obtained by solving \eqref{maxtauprescr1} with respect to $\tau$. However, we first notice that the terms cubic and quadratic in $\tau$ vanish in \eqref{eq:volP2intheta} for $\theta =\frac{1}{2}$. This puts us in one of the situations described in Section~\ref{sec:minimalradiusFM} for which the limit $\tau\rightarrow \infty$ at finite threefold volume is allowed. In the present case, this can be achieved if the curve $A_I$ in Table~\ref{tableP2} shrinks to zero size. The resulting singular elliptic fibration~\cite{Mayrhofer:2014haa} can be deformed by a complex structure deformation into a smooth genus one fibered Calabi--Yau threefold with the same K\"ahler form.\footnote{The current geometry under consideration and its corresponding conifold transition have been studied in detail in~\cite{Cota:2019cjx}.} The limit $\tau\rightarrow \infty$ then corresponds to the limit in which the curves in class $[\mathcal{C}^1]$ become large. From the second Chern class data in~\eqref{cherndataP2}, we see that, in this limit, we indeed have a K3 fibration over a large base $\mathcal{C}^1$. This is illustrated in Figure \ref{fig:CYP2}. The large $\mathcal{C}^1$ limit then corresponds to a five-dimensional emergent string limit in which the quantum gravity cutoff and the black hole threshold are given by 
	\begin{equation}
		\frac{\Lambda_\QG^2}{M_\PlD{5}^2} \sim \frac{1}{v^1}\,,\qquad M_\ttiny{BH, min.} \sim v^1\,.  
	\end{equation}
	On the other hand, the mass scale of the M2-branes that wrap the curve $\mathcal{E}$ is given by 
	\begin{equation}
		\frac{m_\ttiny{M2$|\mathcal{E}$}}{M_\PlD{5}} \sim v^1 \,,
	\end{equation}
	which is at the black hole threshold. Indeed, in this limit, we cannot view these states as KK states of a circle reduction. Similarly, we notice that the diameter $\ell(\mathcal{C}^1)$ is of the order of the minimal scale resolvable in the EFT $\ell_\ttiny{min.}\sim \sqrt{v^1}$ such that from the perspective of the EFT the gauge theory obtained by reducing the M-theory 3-form on $\mathcal{C}^1$ can indeed be detected to be a higher-form symmetry. 
	
	In summary, for the present example the large $\tau$ limit, which is possible for $\theta=\frac12$, does not correspond to a small radius limit for a circle reduction of a six-dimensional theory since the would-be KK tower has mass above the black hole threshold and the six-dimensional gauge theory effectively becomes a higher-form symmetry consistent with our general discussion in Section~\ref{sec:necessitytower}. 
	
	With this in mind, we can now evaluate whether for $0\leq \theta < \frac{1}{2}$ the CHC in the five-dimensional theory is satisfied by the KK replicas of the massless states of the six-dimensional parent theory, i.e. we check the validity of \eqref{eq:mDrvszD}. To this end, it is more convenient to work in M-theory units and to rewrite
	\begin{equation}
		z_6^2 = g_\UoD{5}^2 M_\PlD{5}^3\frac{|q_n|^2}{\gamma m_6^2}\coma
	\end{equation}
	where we used 
	\begin{equation}
		\frac{1}{g_\UoD{5}^2} = \frac{2\pi r_{S^1}}{g_\UoD{6}^2} \coma M_\PlD{5}^3 = 2\pi r_{S^1} M_\PlD{6}^4\coma
	\end{equation}
	with
	\begin{equation}\label{gU15}
		g_\UoD{5}^2 = \frac{(2\pi)(4\pi)^{1/3}}{M_\PlD{5}} Q_\alpha f^{\alpha\beta}Q_\beta \coma 
		f^{\alpha\beta} = \mathcal{V}_{X_3}^{1/3}\left( \frac{1}{2}\frac{v^\alpha v^\beta}{\mathcal{V}_{X_3}}-\mathcal{V}^{\alpha\beta}\right) = \frac{1}{2}\hat{v}^\alpha \hat{v}^\beta-\hat{\mathcal{V}}^{\alpha\beta}\fstop
	\end{equation}
	In the following we focus on the massless states with charge $|q|=2$ and check whether \eqref{eq:mDrvszD} is satisfied for them for any allowed radius. From the M-theory perspective, these states correspond to M2-branes wrapped on $C_z$ in the Shioda--Tate--Wazir basis. For massless states, the CHC condition takes the form
	\begin{equation}
		r_\ttiny{min.}^2 \geq \frac{\gamma}{g_\UoD{5}^2M_\PlD{5}^3} \frac{q\theta(1-q \theta)}{|q_n|^2}\coma
		\label{eq:minrad5dP2}
	\end{equation}
	which, using \eqref{gU15}, can be further rewritten as
	\begin{equation}\label{conditionrewrittenP2}
		\frac{1}{\tau_\ttiny{max.}^2} \geq \left.\frac{\gamma}{2} \frac{1}{Q_\alpha f^{\alpha\beta}Q_\beta} \frac{1}{\mathcal{V}_{X_3}^{2/3}}\frac{q \theta(1-q \theta)}{|q|^2}\right|_{\tau \rightarrow \tau_\ttiny{max.}}\fstop
	\end{equation}
	Here, we used the fact that \eqref{circleradius} relates the minimal radius $r_\ttiny{min.}$ to $\tau_\ttiny{max.}$, which in turn is obtained by solving \eqref{maxtauprescr1}. Since we consider a model without tensors, the black hole extremality factor $\gamma$ is simply given by the value for extremal Reissner--Nordstr\"om black holes in six dimensions,
	\begin{equation}
		\gamma =\left.\frac{D-3}{D-2}\right|_{D=6}=\frac34\,. 
	\end{equation}
	The massless states are associated with the generator $\mathcal{C}^1$ and have charge $Q_\alpha =(q,0,0)$ for $q=2$. For these charges, the condition \eqref{conditionrewrittenP2} then becomes
	\begin{equation}
		\frac{1}{\tau_\ttiny{max.}^2} \geq \left.\frac{\gamma}{2} \frac{1}{f^{11}} \frac{1}{\mathcal{V}_{X_3}^{2/3}}\frac{q \theta(1-q \theta)}{|q|^4}\right|_{\tau \rightarrow \tau_\ttiny{max.}}\fstop
		\label{eq:masslessminrad}
	\end{equation}
	
	Solving \eqref{maxtauprescr1} for the volume of $X_3$ given in \eqref{eq:volP2intheta} we obtain 
	\begin{equation}
		\tau_\ttiny{max.} = \frac{2^{2/3} h(\theta,v^3)^2+2 h(\theta,v^3) \left(4 \theta ^2-1\right) v^3+8 2^{1/3} \theta ^2 (1-2 \theta )^2 (v^3)^2}{6 (1-2 \theta )^2 (4 \theta +1)h(\theta,v^3)}\coma
	\end{equation}
	where
	\begin{equation}
		\text{\small$\displaystyle
			\begin{aligned}
				h(\theta,v^3)  = & \, \left(18 (4 \theta +1)^2 (1-2 \theta )^4+(2 \theta -1)^3 \left(16 \theta ^3-6 \theta -1\right) (v^3)^3+\right.\\
				&+\left.\left((1-2 \theta )^7 (4 \theta +1)^2 \left((v^3)^3+18\right) \left(18 \left(6 \theta +1-32 \theta ^3\right)+(6 \theta +1) (v^3)^3\right)\right)^{1/2}\right)^{1/3}\fstop
			\end{aligned}$}
	\end{equation}
	Using this value of $\tau_\ttiny{max.}$, we find that \eqref{eq:masslessminrad} is always satisfied for any value of $0\leq \theta<\frac{1}{2}$, $v^3$, and $\gamma\simeq \frac{3}{4}$. In particular, this means that the condition \eqref{eq:mDrvszD} is satisfied in the regime where the five-dimensional theory allows for an interpretation as circle reduction of a six-dimensional theory. 
	
	In summary, we find that the CHC for $\U(1)_\ttiny{KK}\times \U(1)_6$ is always obeyed in the five-dimensional theory merely by the KK replicas of the massless states present in the six-dimensional theory. This is precisely what is claimed by Conjecture~\ref{Conjecture1}. Notice that the six-dimensional gauge theory can be viewed as a defect theory for which the WGC is trivially satisfied at the massless level. The WGC becomes a non-trivial condition only after circle compactification, where the $\U(1)_\KK$ factor is a genuine gauge theory coupled to gravity.
	
	\subsubsection{Example: Reduced Schoen Calabi--Yau with Extra Section}
	\label{sec:dP9}
	
	The second example that we study is F-theory over base space $B_2=dP_9$. 
	What makes this setup particularly interesting in the context of the WGC is that the $\U(1)$ gauge theories are non-Higgsable \cite{Morrison:2016lix} so that there are no massless charged states at a generic point in the F-theory field space. 
	Geometrically, this follows from the fact that the height pairing of the $\U(1)$ gauge theories is in the class of the elliptic fiber of the $dP_9$ surface, which has vanishing self-intersection.
	Furthermore, there is no strong coupling limit for the $\U(1)$ gauge theory at finite distance in field space, and hence no limit in which the extremality factor $\gamma\rightarrow \infty$. Therefore, for sufficiently large charges, extremal black holes cannot transition into particles without parametrically violating the black hole extremality bound already above the black hole threshold. As a consequence, there does not exist a tower of super-extremal particle states at any point in field space. 
	
	Since the curve hosting the 7-brane is a fibral curve of $dP_9$, finitely many charged states can arise either from $[p,q]$-string junctions in the base $\mathbb{P}^1_b$ of $dP_9$ stretched between different such 7-brane curves along $\mathbb{P}^1_b$
	or from a D3-brane wrapping $\mathbb{P}^1_b$. The mass of the $[p,q]$-string states can be estimated to be given by\footnote{The role of such non-critical strings for the WGC in F-theory compactifications to six dimensions has recently been also investigated in \cite{Hayashi:2023hqa}.} 
	\begin{equation}\label{eq:pqstringmass}
		M_{[p,q]}^2 \sim (2\pi)^2\text{vol}(\mathbb{P}^1_b) M_s^4\,. 
	\end{equation}
	The tension of the E-string obtained by wrapping a D3-brane on $\mathbb{P}^1_b$ is given similarly by 
	\begin{equation}
		T_{E} \sim 2\pi\text{vol}(\mathbb{P}^1_b) M_s^4\,. 
	\end{equation}
	For small $\mathbb{P}^1_b$, we expect to have charged light states of either origin. Let us focus on the D3-brane string. Since this string is strongly coupled, it does not have a perturbative excitation spectrum resembling that of the heterotic string discussed above. Instead, we can obtain charged particle states by wrapping the E-string on a small loop of length $2\pi \ell_s$ with mass 
	\begin{equation}\label{MassE}
		m_{E} = 2\pi  \ell_s T \sim (2\pi)^2 \text{vol}(\mathbb{P}^1_b) M_s^3 \,. 
	\end{equation}
	When $\mathbb{P}^1_b$ shrinks, the tension of the E string goes to zero, and naively all the particle-like states obtained from it become massless. However, as follows from our discussion above, we do not expect an infinite tower of super-extremal particle states.\footnote{This is consistent with the interpretation of the tensionless limit for an E-string as a small-instanton transition at which only finitely many states become massless.} 
	To avoid a contradiction, multi-loops of the E-string hence cannot produce light particles in the limit $\text{vol}(\mathbb{P}^1_b)\rightarrow 0$. Rather to account for multi-loops \eqref{MassE} should contain an offset term, and a natural guess would be
	\begin{equation}\label{mEn}
		m_{E,n} =  2\pi T n \ell_s + \frac{\alpha(n-1)}{\ell_s}\,,
	\end{equation}
	where $n$ is the wrapping number and $\alpha>0$ is some constant. This is in fact consistent with the idea that the E-string does not form bound states with itself, since the above equation implies that the bound state is disfavored over the two-particle state. This means that only for single loops we obtain states below the black hole threshold. A similar conclusion should hold if instead of the E-string, we consider the states obtained from the $[p,q]$-string junctions. Let us stress that one should not take \eqref{mEn} too literally, since the string is non-perturbative and, then, its would-be excitations are not stable. Therefore, these states are not states in the Hilbert space, but simply correspond to peaks in the S-matrix. Since the theory is not weakly coupled, these peaks are spread over a large energy range, and we cannot associate a well-defined mass to these states.
	
	An exception to this are the lightest states with the smallest $\U(1)$ charge. Since there is no massless charged matter for $B_2=dP_9$ these states cannot decay and are stable. For these we can hence define a mass that, following our discussion above, is just set by the volume of $\mathbb{P}^1_b$. In the limit of small $\mathbb{P}_b^1$ we hence have a finite number of light, stable charged states below the black hole threshold, since the quantum gravity cutoff remains constant in this limit. Furthermore, since this is not a strong coupling limit, the extremality factor $\gamma$ remains of the order of the unity, as well as when shrinking $\mathbb{P}^1_b$. We conclude that the light states arising from the E-string are indeed super-extremal for small $\mathbb{P}^1_b$. 
	
	On the other hand, in the limit of large $\mathbb{P}^1_b$ the would-be excitations have masses above the Planck scale and are hence not expected to be particle-like. Indeed, taking the base $\mathbb{P}^1_b$ large has two effects: First, we are moving away from the finite distance boundary at which the E-string becomes tensionless. Second, the gauge coupling of the $\U(1)$ becomes large in Planck units. In this limit, we do not have a particle-like state that satisfies the WGC. We can explain this observation for the $\U(1)$ in question by noticing that in the limit of large $\mathbb{P}^1_b$, the gauge theory effectively becomes a defect theory. To see this, we notice that the diameter $\ell_\ttiny{perp.}$ of the space transverse to the 7-brane on the generic fiber of $dP_9$ is by \eqref{ellperpdef} given by
	\begin{equation}\label{ellperpF}
		\ell_\ttiny{perp.} M_s = \left(\text{vol}(\mathbb{P}^1_b )M_s^2\right)^{1/2}\,. 
	\end{equation}
	In the large $\mathbb{P}^1_b$ limit, we can compare $\ell_\ttiny{perp.}$ to the length scale set by the quantum gravity cutoff
	\begin{equation}
		\ell_\ttiny{min.} M_\PlD{6} = \frac{M_\PlD{6}}{\Lambda_\QG} \sim \left(\text{vol}(\mathbb{P}^1_b) M_s^2\right)^{1/4}\,,
	\end{equation}
	where we used that for large base $M_\PlD{6}^4/M_s^4 = \text{vol}(\mathbb{P}^1_b) M_s^2$ and we identify $M_s$ as the quantum gravity cutoff. For the large base limit, we, then, find
	\begin{equation}
		\ell_\ttiny{perp.} \gg \ell_\ttiny{min.}\,.
	\end{equation}
	Hence, in the EFT we can resolve the space transverse to the brane, and the gauge theory effectively becomes a defect theory, for which the WGC does not necessarily have to hold at the particle level.

	In addition to making the base curve $\mathbb P^1_b$ large, we can also shrink the generic elliptic fiber of $dP_9$. 
	As a consequence, the string wrapping this elliptic fiber, which is dual to a fundamental Type II string, becomes light, i.e. 
	\begin{equation}
		\frac{T_\ttiny{II}}{M_s^2} \sim \text{vol}(T^2) M_s^2 \sim \left(\text{vol}(\mathbb{P}^1_b) M_s^2\right)^{-1} 
		\fstop
	\end{equation}
	In this case, the tension of the fundamental string sets the species scale such that the tension of the smallest black string is given by 
	\begin{equation}
		\frac{T_\ttiny{BS, min.}}{M_\PlD{6}^2} \sim \left(\frac{M_\PlD{6}}{\Lambda_\QG}\right)^2 \sim \frac{T_\ttiny{E}}{M_\PlD{6}^2}\,. 
	\end{equation}
	Accordingly, the tension of the field theory string is at the black string threshold such that it cannot have particle-like excitations. Hence, the WGC is no longer satisfied by particle-like excitations. In this limit, the scaling of the gauge coupling on the 7-branes leads to
	\begin{equation}
		\Lambda_\ttiny{WGC}^2 = g_{\UoD{6}}^2 M_\PlD{6}^2 \sim \text{vol}(\mathbb{P}^1_b) M_s^2 \gg \Lambda_\QG\,,
	\end{equation}
	signaling a decoupling of the gauge sector from gravity. For this reason, according to our~\cref{claim:WGC}, we do not expect the WGC to be satisfied at the particle level. 
	
	With this preparation, we can now test whether in the regime where we expect the WGC to be satisfied at the particle level, i.e. in the limit of small $\mathbb{P}^1_b$, the CHC for $\U(1)_6\times \U(1)_\KK$ is satisfied after dimensional reduction on a circle. We first construct the corresponding Calabi--Yau threefold for the M-theory compactification. A general class of Calabi--Yau threefolds has been described in~\cite{Morrison:2016lix},  
	yielding $\mathcal{N}=(1,0)$ six-dimensional effective theories with no massless charged matter. 
	For concreteness, we choose a subclass of these models that have an explicit Batyrev--Borisov toric construction~\cite{Batyrev:1994pg} which in essence is a smooth Schoen Calabi--Yau manifold.

	As a starting point, let us consider a rational elliptic surface $p: S \to \mathbb{P}^1$, 
	which is isomorphic to the blowup of $\mathbb{P}^2$ in the nine basepoints of a cubic pencil~\cite{MR2732092} 
	\begin{equation}
		\{ u_0 P_0   +u_1 P_1 =0\} \subset \mathbb{P}^2 \times \mathbb{P}^1 \coma
	\end{equation}
	where $P_a \in \mathcal{O}_{\mathbb{P}^2}(\bar{K})$ and $(u_0,u_1) \in \mathbb{P}^1$. 
	For each del Pezzo surface $dP_n = \text{Bl}_{p_1 \cdots p_n} (\mathbb{P}^2)$ one can find
	an associated cubic pencil and thus obtain that $dP_9$ is isomorphic to $S$. 
	The bundle $p: S \to \mathbb{P}^1$ is a torus fibration over $\mathbb{P}^1$ in which 
	its anticanonical class $\bar{K}$ is non-trivial but satisfies $\bar{K}^2 = 0$.  
	This relation is evident by inspecting the anticanonical class of a del Pezzo surface $dP_n$ 
	that reads
	\begin{equation}
		\bar{K}= 3h - \sum_{i=1}^n e_i \coma
	\end{equation}
	where $h$ is the divisor associated with hyperplane class in $\mathbb{P}^2$ and $e_i$ are the blowup divisors of $dP_n$. 
	Here, the intersection relations among the latter divisors are
	\begin{equation}
		h^2 = 1\coma   h \cdot e_i = 0\coma  e_i \cdot e_j = -\delta_{ij}\fstop
	\end{equation}
	Moreover, the torus fiber class $f$ is identified with the anticanonical class of $dP_9$, 
	while the blowup divisors give rise to sections of the torus fibrations that, together, span the Mordell--Weil group $\mathrm{MW}(S)$.  
	
	Let us take two rational elliptic surfaces $S_a = dP_9$ with $a = 1, 2$. The Schoen Calabi--Yau threefold is the fibered product
	\begin{equation}
		X_3 = S_1 \times_{\mathbb{P}^1_b} S_2  = \{ (x,y) \in S_1 \times S_2  \mid \pi_1 (x) = \pi_2 (y) \}\fstop
	\end{equation}
	It satisfies the following commutative diagram of fibrations:
	\begin{equation}
		\begin{tikzcd}
			&\ar[ ld,"\pi_1" ' ]  X_3 \ar[dd, "\rho"] \arrow[rd,"\pi_2" ] \\ 
			S_1 \ar[rd,"p_1" '] & &S_2  \ar[ ld, "p_2" ]  \\
			& \mathbb{P}^1_b
		\end{tikzcd}
	\end{equation}
	Here $\pi_a: X_3 \to S_a $ are the torus fibrations induced by $p_a : S_a \to \mathbb{P}^1_b$. 
	Note that when a given manifold admits multiple torus fibrations, 
	the physics of an F-theory model depends on the choice of fibration $\pi_a$.  
	In the following, we consider a reduced Schoen Calabi--Yau manifold $X_3$ with an extra visible section, in which only a few K\"ahler moduli are explicitly realized in the Batyrev--Borisov construction.

	Inspired by~\cite{Hosono:1997hp}, we now make an extension of their restricted Schoen Calabi--Yau toric construction 
	in which not the full lattice $\mathrm{Pic}(X_3)$ is realized torically, but a only a sublattice thereof, 
	which descends from the embedding of $X_3$ into its  toric ambient space. 
	For our purposes it will be sufficient to restrict our analysis of the K\"ahler moduli space to the moduli associated with this toric sublattice.
	With this in mind, we consider the following complete intersection Calabi--Yau that is defined by the codimension-two loci 
	\begin{equation}
		X_3 = \{ f_1  = 0 \} \cap \{ f_2 = 0 \}  \subset dP_n \times dP_m \times \mathbb{P}^1_b \coma
		\label{eq:SchoenLoci}
	\end{equation}
	where $f_1 \in \mathcal{O}_{dP_n} (\bar{K}) \oplus \mathcal{O}_{\mathbb{P}^1_b}(-1)$ and $f_2 \in \mathcal{O}_{dP_m} (\bar{K}) \oplus \mathcal{O}_{ \mathbb{P}^1_b}(-1)$ with $0\leq n,m \leq 9$. 
	Here, we identify each surface $\{f_a = 0\}$ with the pencil that descends from a rational elliptic surface $S_a$. 
	
	For concreteness, we now consider an explicit example of this construction, in which the toric ambient space is $dP_1 \times \mathbb{P}^2 \times \mathbb{P}^1$, i.e. $n=1$ and $m=0$ in \eqref{eq:SchoenLoci}. 
	The corresponding complete intersection Calabi--Yau 
	is realized by the following polytope generated by the convex hull of points:
	\begin{align}
		\begin{blockarray}{crrrrrl}
			\begin{block}{c(rrrrr)l}
				{x_0}& -1 &  0 &  0 & 0&0 & \Delta_1 \\ 
				{x_1} & -1 & 1&  0 & 0 &0 & \Delta_1\\
				{x_2} &  0& -1 &  0 & 0 & 0  & \Delta_1 \\
				{x_3} & 0 &  0 & -1& -1 & 0  & \Delta_2 \\
				{x_4} &  0 &  0& 0 & 0  &-1 &  \Delta_1 \\
				{x_5}&   0&   0 & 0 & 0 &  1 &  \Delta_2 \\
				{x_6} & 0&  0 & 0 & 1 &  0 &  \Delta_2 \\
				{x_7} & 0&  0 & 1 & 0 &  0 & \Delta_2 \\
				{x_8} & 1&  0 & 0 & 0 &  0 &  \Delta_1 \\
			\end{block}
		\end{blockarray}\,.
		\label{eqn:toricdP9}
	\end{align}
	Here ${x_k}$ denotes the toric coordinates associated with the ray generator that is determined by the polytope point on the right side. 
	On the further right side,  $\Delta_i$ denotes the chosen toric divisors for the nef-partition components. 
	Moreover, the Stanley--Reisner ideal is in our case
	\begin{equation}
		\mathrm{SR} = \{x_0 x_8, x_1 x_2, x_4 x_5, x_3 x_6 x_7 \}  \fstop
	\end{equation}
	
	Due to the chosen ambient space $dP_1 \times \mathbb{P}^2 \times \mathbb{P}^1$ the subgroup of $\mathrm{Pic}(X_3)$ realized torically has dimension four. We only give the intersection numbers of the toric subspace.\footnote{To reconstruct the entire Picard group of rank 19 of the Schoen Calabi--Yau torically, one has to perform a succession of blowups on the ambient space.} 
	The triple intersection data read
	\begin{equation}
		\mathcal{I}(X_3) = 9 J_1 J_2 J_3+6 J_1 J_2 J_4+3 J_2^2 J_3 +2 J_2^2 J_4 +3  J_2 J_3^2+3 J_2 J_3 J_4 \coma 
	\end{equation}
	its second Chern class intersection pairings with the K\"ahler cone generators $J_i$ are 
	\begin{equation}
		\int_{X_3}c_2(X_3)\wedge J_i = (0,36,36,24)\coma 
	\end{equation}
	and its Euler characteristic is $\chi(X_3) = 0$, as expected for a Schoen Calabi-Yau manifold. 
	Here, the K\"ahler cone generator $J_1$ is the abelian surface fiber $F$ of the fibration $\rho : X_3 \to \mathbb{P}^1_b$. 
	Moreover, $X_3$ admits two different torus fibrations $\pi_a : X_3 \to S_a$, where $S_a \simeq dP_9$ with $a=1,2$.  
	However, we are interested in the elliptic fibration $\pi: X_3 \to dP_9$ 
	in which $J_2$ is the class $\pi^*h$  in $H_2(X_3, \mathbb{Z})$ with $h$ being the hyperplane class in $dP_9$.  
	We can further identify the zero section $S_0 = J_3 - J_4 -J_1$ and an additional rational two-section $S_Q = J_4 +J_1$ from this fibration. 
	The Shioda map 
	is $\sigma(S_Q) =3 J_4-2J_3$ and the shifted zero section  $\widehat{S}_0=J_3 - J_4$. We then expand the K\"ahler form in the Shioda--Tate--Wazir basis $(J_1,J_2,\widehat{S}_0,\sigma(S_Q))$ as 
	\begin{equation}\label{STWbasisSchoen}
		J =   v^1 J_1  +v^2 J_2+ \tau \widehat{S}_0 +  z \sigma(S_Q)  \fstop 
	\end{equation}
	Here $z$ is the volume of the fibral curve $ C_z = \mathcal{C}^3+ \mathcal{C}^4$ and $\tau$ is the volume of the elliptic curve $\mathcal{E} = 3\mathcal{C}^3 + 2 \mathcal{C}^4$. Furthermore, $v^1$ is the volume of the base curve $\mathbb{P}^1_b$ and $v^2$ is the volume of the curve $\pi^{-1}h$. The intersection ring in this basis is
	\begin{equation}
		\mathcal{I}(X_3) = 3  J_1 J_2 \widehat{S}_0+J_2^2 \widehat{S}_0-3 J_2 \widehat{S}_0^2 +9 J_2 \widehat{S}_0   \sigma(S_Q) -24 J_2 \sigma(S_Q)^2 \fstop    
	\end{equation}
	Let us define the Coulomb branch parameter
	\begin{equation}
		\frac{z}{\tau} = \frac{v^3+v^4}{3v^3+2v^4}=\frac{\theta'}{2\pi}=\theta\coma
		\label{eq:thetadP9}
	\end{equation}
	and let us write the volume of $X_3$ in the Shioda--Tate--Wazir basis using $\theta$:
	\begin{equation}
		\mathcal{V}_{X_3} = \tau ^2 \left(9 \theta  v^2-12 \theta ^2 v^2-\frac{3 v^2}{2}\right)+\tau  \left(\frac{(v^2)^2}{2}+3 v^1 v^2\right)\fstop
		\label{eq:VX3dP9}
	\end{equation}
	Note that, from \eqref{eq:thetadP9}, $\frac{1}{3}\leq \theta \leq \frac{1}{2}$. From \eqref{eq:VX3dP9}, we can compute the value of $\tau_\ttiny{max.}$ through \eqref{maxtauprescr1} to obtain
	\begin{equation}
		\tau_\ttiny{max.}(\theta,v^1,v^2) = \frac{\sqrt{v^2 \left(v^2 \left(6 v^1+v^2\right){}^2-24 \left(8 \theta ^2-6 \theta +1\right)\right)}+(v^2)^2+6 v^1 v^2}{6 \left(8 \theta ^2-6 \theta +1\right) v^2}\fstop
	\end{equation}
	The states we can consider in the six-dimensional theory are $[p,q]$-strings or the excitations of the E-string. We can argue for the existence of these charged states in this geometry by considering \textit{geometric string junctions}~\cite{Grassi:2014ffa}, which can indeed be interpreted as $[p,q]$-strings. In our Schoen Calabi--Yau construction, in which there are no codimension-two singularities, there is a special kind of two-cycles that realize string junctions. Specifically, those given by paths connecting singular points $I_1$ and pinched fibers in one cycle. 
	A detailed geometrical construction of such two-cycles can be found in \cite[Appendix A]{Braun:2018fdp}.  The upshot is that a set of string junctions with asymptotic charge zero is in one-to-one correspondence with topological two-spheres in the Mordell--Weil group $\mathrm{MW}(S)$ of $S \simeq dP_9$~\cite{Grassi:2014ffa}.
	
	Consider a curve $C$ in $\mathrm{MW}(S)$ that embeds into $H_2(X_3,\mathbb{Z})$.  
	If its  genus-zero Gopakumar--Vafa invariant $n^0_{C}$ is non-zero, $C$ has a non-trivial curve class in $H_2(X_3,\mathbb{Z})$ that follows the relation \cite{Hori:2003ic}
	\begin{equation}
		n_C^0 = (-1)^{\text{dim}\left(\widehat{\mathcal{M}}_C\right)} \chi(\widehat{\mathcal{M}}_C)\,,
	\end{equation} 
	where $\widehat{\mathcal{M}}_C$ is the moduli space of curves in the class $[C]\in H_2(X_3,\mathbb{Z})$, 
	together with a choice of $\U(1)$ flat connection.  
	Thus, due to isomorphism, $n_C^0 \neq 0$ --- and in consequence non-trivial moduli space of curves in a given class in $[C] \in H_2(X_3,\mathbb{Z})$ --- implies an equivalent description of $C$ in terms of non-trivial string junctions.\footnote{Here we are not interested in interpreting such Gopakumar--Vafa invariants as numbers that count five-dimensional BPS invariants.}
	
	In our current discussion example, the curve $C_z$ descends from a blow-up divisor curve of a rational elliptic surface $S$ with elliptic fiber $\mathcal{E}$, such that $C_z \cdot_{S} C_z = -2$. 
	We can then associate string junctions to the class of curves $C_{[p,q]} = \mathbb{P}^1_b + n \mathcal{E} + m C_z$, such that $ n = m^2$~\cite{Grassi:2014ffa}.
	Let us now consider the genus zero curves generating function of the form
	\begin{equation}
		\mathcal{F}_{\mathbb{P}^1_b}^{(0)} =   \sum_{n, m \in \mathbb{Z} }  n_{\mathbb{P}^1_b + n \mathcal{E} + m C_z}^0 q^n \zeta^m \,.
	\end{equation}
	Here $q= \exp(2\pi i \tau)$ and $\zeta = \exp(2\pi i z)$.   A direct computation for our example reveals that \cite{Lee:2018urn}\footnote{For an independent discussion relating the counting of E-strings with $[p,q]$-strings see~\cite{Huang:2013yta}.} 
	\begin{equation}\label{F0example}
		\mathcal{F}_{\mathbb{P}^1_b}^{(0)}  = 9 q^{\frac{1}{2}}\frac{\Theta_{E_8}(\tau,z) }{\eta^{12} (\tau) } = 
		9 + \left(612+ 504 \zeta^\pm +252 \zeta^{\pm2}+72 \zeta^{\pm 3}
		\right) q+ \mathcal{O}(q^2) \,. 
	\end{equation}
	Here $\zeta^\pm = \zeta + \zeta^{-1}$ and 
	\begin{equation}
		\Theta_{E_8} (\tau, z) = \frac{1}{2} \sum_{(a,b) \in (\mathbb{Z}/2\mathbb{Z})^2 }\prod_{i=1}^8 \vartheta_{a,b}(\tau, z_i)\,, 
	\end{equation} 
	where $\vartheta_{a,b}$ are the classical Jacobi theta functions, $z_i = z$ for $i =1, \ldots 7$, and $z_8 = -z$. Therefore, there are indeed states charged under the $\U(1)$ gauge theory, for which those with minimal charge are expected to be stable. The mass of these states is given in \eqref{eq:pqstringmass} in F-theory units by
	\begin{equation}
		m_6^2 \sim (2\pi)^2\text{vol}(\PP^1_b) M_s^4\fstop
		\label{eq:massm6dP9}
	\end{equation}
	This can be expressed in M-theory units via \eqref{eq:M11dinMsgIIB} as
	\begin{equation}
		m_6^2 \sim (2\pi)^2 g_\IIB \tau v^1 M_\ttiny{11d} \fstop
		\label{eq:massm6dP9-Mth}
	\end{equation}
	We are now ready to compute \eqref{eq:mDrvszD} for the minimal radius $r_\ttiny{min.}$,
	\begin{equation}
		(r_\ttiny{min.}m_6)^2 \geq  \frac{\gamma m_6^2}{g_\UoD{5}^2 M_\PlD{5}^3}\left(\frac{\gamma  m_6^2}{|q|^2 \left(|q|^2 g_\UoD{5}^2 M_\PlD{5}^3-\gamma  m_6^2\right)}+\frac{  q \theta  (1-\theta  q)}{|q|^2 }\right) \,, 
		\label{eq:genexprmindP9}
	\end{equation}
	where we expressed $z_6$ in M-theory units as we did in Section \ref{sec:P2}. Via the relation 
	\begin{equation}
		r_\ttiny{min.} = \frac{1}{2\pi \tau_\ttiny{max.}M_\ttiny{11d}}\coma
	\end{equation} 
	we can now evaluate \eqref{eq:genexprmindP9}  for states with charge $q$ and mass \eqref{eq:massm6dP9}, obtaining
	\begin{equation}
		\begin{split}
			\frac{1}{\tau_\ttiny{max.}^2}  \geq &\, \frac{\gamma }{2Q_\alpha f^{\alpha\beta}Q_\beta\mathcal{V}_{X_3}^{2/3}}\left(\frac{\gamma  g_\IIB v^1\tau}{ |q|^2  \left(2 |q|^2 Q_\alpha f^{\alpha\beta} Q_\beta    \mathcal{V}_{X_3}^{2/3}-\gamma  g_\IIB v^1\tau\right)}+\left. \frac{    q \theta(1-\theta  q)}{ |q|^2    }\right)\right|_{\tau\rightarrow \tau_\ttiny{max.}}\fstop
		\end{split}
		\label{eq:taumaxreldP9}
	\end{equation}
	In five dimensions, such states arise from curves intersecting only the divisor $\sigma(S_Q)$. 
	In the following, we denote the corresponding entrance in the gauge kinetic matrix $f^{\alpha\beta}$ in the Shioda--Tate--Wazir as
	$f^{44}$, and the states have charges $Q_\alpha = (0,0,0,q)$ in this $\U(1)$ basis. 
	Notice that for $\theta =\frac{1}{2}$, the term in \eqref{eq:VX3dP9} quadratic in $\tau$ vanishes and $\tau$ can be arbitrarily large on the  $\mathcal{V}_{X_3} = 1$ locus. To achieve $\theta=\frac12$, we need to tune $v^2=0$ so that $\tau\rightarrow \infty$ corresponds to $v^1\rightarrow \infty$. This limit again corresponds to a heterotic emergent string limit, i.e. a limit in which we have an adiabatically-fibered K3 over $\mathbb{P}^1$ with the latter in class $\mathcal{C}^1$. The situation is akin to the $\tau\rightarrow \infty$ limit discussed for the $\mathbb{P}^2$ example in the previous section, and again in this limit we lose the interpretation of the five-dimensional theory as a KK reduction of a higher-dimensional theory. 
	On the other hand, when setting $\theta = \frac{1}{3}$ we recover the model of~\cite{Hosono:1997hp}, which effectively behaves as a double genus-one fibered Calabi--Yau threefold with no sections but three-sections. Still, we can check the validity of the bound \eqref{eq:taumaxreldP9}, which we rewrite as 
	\begin{equation}
		\frac{1}{\tau_\ttiny{max.}^2}  \geq  \frac{\gamma }{2 f^{44}}\left(\frac{\gamma    g_\IIB v^1\tau}{ |q|^4  \left(2 |q|^4 f^{44}     -\gamma   g_\IIB v^1\tau\right)}+\left. \frac{    q \theta(1-\theta  q)}{ |q|^4    }\right)\right|_{\tau\rightarrow \tau_\ttiny{max.}}\fstop
		\label{eqn:CHCSchoen}
	\end{equation}
	Since the gauge theory of interest is never weakly coupled, we expect only the states with the smallest charge to be stable with mass as in \eqref{eq:pqstringmass}. We should then consider a state with $q=1$ that indeed exists, as is clear from \eqref{F0example}. For this charge, we find that \eqref{eqn:CHCSchoen} is indeed satisfied for all values of $\frac{1}{3}\leq \theta\leq \frac{1}{2}$, $g_\IIB$ and $\gamma \simeq \frac{3}{4}$. 
	
	We see that the CHC in the compactified theory is satisfied by the KK replicas of the finite number of charged particle-like states present in the parent theory for a small base $\mathbb{P}^1_b$, consistent with Conjecture~\ref{Conjecture1}.

	\subsection{From Five to Four Dimensions in M-theory/Type IIA} 
	\label{subsec:5t4MA}

	We now compactify the five-dimensional theory discussed in the previous section on an additional circle. This gives M-theory on Calabi--Yau threefold times a circle, which is dual to Type IIA string theory on the same Calabi--Yau threefold. As discussed in Section \ref{sec:minradius5to4}, due to the $\alpha'$-corrections, again there effectively exists a minimal radius below which we cannot view the four-dimensional theory as a circle reduction of a five-dimensional theory. This removes the immediate need for a tower of super-extremal particles in five dimensions to satisfy the CHC in four dimensions, at least for those gauge theories that cannot become asymptotically weakly coupled.
	
	To illustrate this, consider the vector of charge-to-mass ratios ${\bf{z}}$ in the KK reduced theory \cite{Heidenreich:2015nta}
	\begin{equation}
		{\bf{z}} = \frac{1}{\left(m_{5}^2r_{S^1}^2+\left(q_\KK-\fn_5\right)^2\right)^{1/2}} \left(m_{5}r_{S^1}z_5, q_\KK - \fn_5 \right)\,,
	\end{equation}
	where $\fn_5 = \frac{q \theta'}{2\pi}=q\theta$, and 
	\begin{equation}
		z_5 = \left(g_{\UoD{5}} M_{\PlD{5}}^{1/2}\right)M_{\PlD{5}} \gamma^{-1/2} \frac{|q|}{m_5}\fstop
	\end{equation}
	For M-theory on Calabi--Yau threefolds, $m_5$ is the mass of an M2-brane wrapping a two-cycle and the denominator in ${\bf{z}}$ can be identified with the mass of a (BPS) D2-brane in Type IIA with $q_\KK$ units of D0-brane charge,
	\begin{equation}
		r_{S^1}^2 M_{D2+q_\KK D0}^2 = \left(\text{vol}(C)M_{\ttiny{11d}}^2\right)^2r_{S^1}^2M_{\ttiny{11d}}^2 + \left(q_\KK-\fn_5\right)^2 
		\fstop
	\end{equation}
	The CHC in four dimensions, as analyzed in the regime of validity of the KK ansatz, requires
	\begin{equation}
		(m_5r_{S^1})^2 \geq \frac{1}{4z_5^2(z_5^2-1)} + \frac{\fn_5(1-\fn_5)}{z_5^2} \,.
		\label{eq:m5rvsz5}
	\end{equation}
	According to our arguments in Section \ref{sec:minradius5to4}, the minimal radius up to which the CHC must hold in this form is given by \eqref{eq:rmin5d}. For $r_{S^1} = r_{S^1}^\ttiny{min.}$, as in \eqref{eq:rmin5d}, the constraint becomes
	\begin{equation}
		\frac{m_5^2\alpha^{\frac{2}{3}}}{(2 \pi)^2 \left(\mathcal{V}_{X_3}\right)^{\frac{2}{3}}M_\ttiny{11d}^2}  \geq  \frac{\gamma m_5^2}{g_\UoD{5}^2 M_\PlD{5}^3}\left(\frac{\gamma  m_5^2}{|q|^2 \left(|q|^2 g_\UoD{5}^2 M_\PlD{5}^3-\gamma  m_5^2\right)}+\frac{  q \theta  (1-\theta  q)}{|q|^2 }\right)\coma
	\end{equation}
	leading to
	\begin{equation}
		\alpha^{\frac{2}{3}}  \geq  \frac{\gamma }{2Q_\alpha f^{\alpha\beta}Q_\beta}\left(\frac{\gamma  m_5^2}{ |q|^2  \left(2(2 \pi)^2 |q|^2 Q_\alpha f^{\alpha\beta} Q_\beta M_\ttiny{11d}^2   \mathcal{V}_{X_3}^{2/3}-\gamma  m_5^2\right)}+ \frac{    q \theta(1-\theta  q)}{ |q|^2    }\right)\fstop
	\end{equation}
	The mass $m_5$ of the states in the five-dimensional theory is 
	\begin{equation}
		m_5^2=(2\pi)^2\mathcal{V}_C M_\ttiny{11d}^2\coma
	\end{equation}
	where $C$ is some curve in $X_3$. With this input, the inequality becomes
	\begin{equation}
		\begin{split}
			\alpha^{\frac{2}{3}}  \geq &\, \frac{\gamma }{2Q_\alpha f^{\alpha\beta}Q_\beta}\left(\frac{\gamma  \mathcal{V}_C}{ |q|^2  \left(2 |q|^2 Q_\alpha f^{\alpha\beta} Q_\beta\mathcal{V}_{X_3}^{2/3} -\gamma  \mathcal{V}_C\right)}+ \frac{    q \theta(1-\theta  q)}{ |q|^2    }\right)\fstop
		\end{split}
		\label{eq:alphaconditions}
	\end{equation}
	For asymptotically weakly coupled gauge groups, a non-trivial constraint arises, and consistently, for such groups a tower of super-extremal particle states has been in established in \cite{Cota:2022maf}.
	For gauge groups whose couplings remain of unit order, by contrast, the constraint can in principle be satisfied without the need of a tower of super-extremal particle states in five dimensions because the parametric violation of the constraint in the limit of vanishing radius has been resolved.

	Whether the CHC in four dimensions is satisfied now still depends on the value of $z_5$, i.e. on the charge-to-mass ratio in the five-dimensional theory. As discussed in \cite{Heidenreich:2015nta}, if $z_5=1$, the CHC in the lower-dimensional theory is violated for any radius $r_{S^1}$ unless there exists an infinite tower of states with $z_5=1$. This is, for example, the case for the gauge groups analyzed in \cite{Alim:2021vhs}. Other rays in the charge lattice support only a finite number of BPS states with $z_5>1$. In this situation, our above argument suggests that a tower of super-extremal states is not necessary as long as the constraint \eqref{eq:m5rvsz5} is obeyed for the minimal radius
	by
	one of the finitely many super-extremal states. \\
	
	To test this idea, we focus on simple Calabi--Yau threefolds $X_3$ that arise as complete intersections of hypersurfaces on products of (weighted) projective spaces. To determine the parameter $\alpha$ in \eqref{eq:rmin5d}, we switch to the mirror dual Type IIB compactifications and analyze the complex structure moduli space of the mirror manifolds $Y_3$ of $X_3$. We parametrize the complex structure moduli space by coordinates $\phi^i$ such that the large complex structure point is located at $\phi^i=\infty$, where the mirror map identifies the K\"ahler moduli of $X_3$ as 
	\begin{equation}
		t^i = \frac{i}{2\pi} \log \phi^i + \dots \,. 
	\end{equation}
	The models that we consider have in common that for small $|\phi^i|$ there is a Landau--Ginzburg phase. Since this phase corresponds to a formally infinitely negative volume on the K\"ahler side, $e^{-K_\ttiny{c.s.}(Y_3)}$, defined in \eqref{eq:eKY3}, is expected to be minimized at $\phi^i=0$, $\forall i$.\footnote{As discussed in \cite{vandeHeisteeg:2022btw,vandeHeisteeg:2023dlw} in general the Landau--Ginzburg/Orbifold points corresponds to the desert point~\cite{Long:2021jlv} in moduli space where the species scale is maximized. Therefore, it is expected that $e^{-K_\ttiny{c.s.}(Y_3)}$ is minimized at that point, since it sets the ratio between the Planck scale and the string scale.} If we fix a point in the five-dimensional M-theory moduli space, then changing the radius of the circle compactification, or equivalently $g_\IIA$, corresponds to homogeneously shrinking the curve volumes of $X_3$ in string units (see the discussion in Section~\ref{sec:minradius5to4}). In the mirror picture, this means that we are effectively moving from $\phi^i\rightarrow \infty$ to $\phi^i\rightarrow 0$ for all $i$. Hence, in shrinking the M-theory circle, we are automatically led to the Landau--Ginzburg phase. To compute $\alpha$, we now evaluate $e^{-K_\ttiny{c.s.}(Y_3)}$ in the Landau--Ginzburg phase and check if \eqref{eq:alphaconditions} holds in simple examples. 
	
	\subsubsection{One-modulus Example}
	We first choose $X_3$ to be the quintic whose mirror $Y_3$ is the one-parameter Calabi--Yau threefold defined by the zero locus of the following polynomial in $\PP^4$,
	\begin{equation}\label{defeqn}
		p(x_i,\phi) = \sum_{k=1}^5x_k^5-\phi x_1x_2x_3x_4x_5\coma
	\end{equation}
	where $\phi$ parametrizes the only complex structure deformation. From the perspective of \eqref{eq:alphaconditions} this example is not particularly interesting since the only gauge group in M-theory on $X_3$ is the graviphoton $\U(1)$, and hence all wrapped M2-branes giving BPS states have $z_5=1$. In fact, for the single curve class, the BPS invariants are non-vanishing for any multi-wrapping. Still, we can compute $\alpha$ to confirm the existence of a minimal radius for the circle compactification of M-theory on $X_3$. 
	
	In the present case, as pioneered in \cite{Candelas:1990rm}, one can use \eqref{defeqn} to define the periods of the Calabi--Yau. Around the Landau--Ginzburg point at $|\phi|\rightarrow 0$, the periods take the following form (see also \cite{Blumenhagen:2018nts}),
	\begin{equation}
		\varpi_k(\phi)=(5\phi)^k\sum_{n=0}^\infty \left(\frac{\Gamma\left(\frac{k}{5}+n\right)}{\Gamma\left(\frac{k}{5}\right)}\right)^5\frac{\Gamma(k)}{\Gamma(k+5n)}(5\phi)^{5n}\coma
	\end{equation}
	with $k =1,\ldots,4$. The K\"ahler potential is now computed by evaluating
	\begin{equation}
		e^{-K_\ttiny{c.s.}} = i\int_{Y_3}\Omega \wedge \bar{\Omega} = -i \Pi^\dagger \Sigma \Pi\coma
	\end{equation}
	where
	\begin{equation}
		\Pi = \left(\begin{array}{c}
			F_\Lambda \\
			X^\Lambda
		\end{array}\right) \text{ and } \Sigma =\left(\begin{array}{cc}
			0 & \ID \\
			-\ID  & 0
		\end{array}\right)\fstop
	\end{equation}
	Here $\Pi$ are the periods of an integral symplectic homology basis of $H^{2,1}(Y_3)$ with fixed asymptotic behavior at the large complex structure point. The periods are related to $\varpi$ through a transition matrix, $\Pi_i = m_i\,^j\varpi_j$. For the quintic the inverse of the transition matrix $m$ between the integer basis and the $\varpi$ basis adapted to the Landau--Ginzburg point has been computed explicitly in \cite[(A.3)]{CaboBizet:2016uzv},\footnote{In \cite{CaboBizet:2016uzv}, the definition of $\Pi$ is $(X^\Lambda,F_\Lambda)^T$, and their transition matrix is with respect to that convention.}
	and the K\"ahler potential at the Landau--Ginzburg point has been evaluated in \cite[(4.4)]{Blumenhagen:2018nts}. To find the correct normalization, we must divide by $|X^0|^2$, which leads to
	\begin{equation}
		\alpha_{\PP^4_{11111}[5]} = \lim_{|\phi| \rightarrow 0}  \frac{1}{|X^0|^2} e^{-K_\ttiny{c.s.}} \simeq 3.08\fstop 
	\end{equation}
	By contrast, a similar computation at the conifold point reveals that $\left.\alpha_{\PP^4_{11111}[5]} \right|_{\mathrm{c}} \simeq 13.98$\,. Indeed, there is a minimal radius for the circle compactification of M-theory on the quintic, which is determined by its minimal quantum volume at the Landau--Ginzburg point. 
	
	\subsubsection{Two-moduli Example}
	
	We now turn to two-moduli examples. We are particularly interested in two-parameter models that feature conifold curves, i.e. curves that do not allow for multi-wrapping. In these cases, there is only a finite number of super-extremal charged states, and \eqref{eq:alphaconditions} becomes a non-trivial condition. This situation is obtained for two-parameter models that are K3 fibrations over $\mathbb{P}^1$. The class of the base $\mathbb{P}^1$ does not allow for multi-wrappings such that in the limit of small $\mathbb{P}^1$ there only exists a finite number of super-extremal particle states in the parent five-dimensional M-theory.\footnote{At least at the level of known states which are technically under control.}
	
	Explicit examples are
	the complete intersection Calabi--Yau threefolds $\PP^4_{11222}[8]$ and $\PP^4_{11226}[12]$ in a weighted projective space. The periods of the mirror manifolds of these geometries were first studied in \cite{Candelas:1994hw}. Here, we use the convention of \cite{Blumenhagen:2018nts}. We can proceed again as for the quintic example and relate the local periods at the Landau--Ginzburg point to the integral basis through the transition matrices found in \cite{Candelas:1994hw}. 
	Computing $e^{-K_\ttiny{c.s.}(Y_3)}$ at the Landau--Ginzburg point yields
	\begin{equation}\label{alphatwomoduli}
		\alpha_{\PP^4_{11222}[8]} \simeq 2.83 \coma \alpha_{\PP^4_{11226}[12]} \simeq 6.00\fstop 
	\end{equation}
	With this value for $\alpha$ we can now test \eqref{eq:alphaconditions}. To this end, we need the corresponding intersection rings of the respective geometries, which are given by
	\begin{equation}
		\mathcal{I}(\PP^4_{11222}[8]) = 8J_1^3+4J_1^2 J_2 \coma \mathcal{I}(\PP^4_{11226}[12]) = 4J_1^3+2J_1^2 J_2\fstop 
	\end{equation}
	Here $J_1$ denotes the hyperplane class divisor of the ambient weighted projective space and $J_2$ is the exceptional divisor class associated to the blow-up of the orbifold singularities. 
	As mentioned before, in M-theory, the $\PP^1$ base curve in the Calabi--Yau with volume $v^2$ yields a BPS state only for single wrapped M2-branes. Hence, along this ray in the charge lattice there only exists a state with charge $Q_\alpha = (0,1)$, which for small $v^2$ has $|z_5|\gg 1$. For this state, we can now evaluate \eqref{eq:alphaconditions}. For $\theta =\frac{1}{2}$, for which the function $\theta(1-\theta)$ is maximized, and for $\gamma \simeq \frac{2}{3}$, one finds numerically that the maximum possible value of the RHS of \eqref{eq:alphaconditions}  is
	\begin{equation}
		\text{RHS}_{\PP^4_{11222}[8]} \simeq 0.17 \coma \text{RHS}_{\PP^4_{11226}[12]} \simeq 0.10\coma 
	\end{equation}
	Hence the constraint \eqref{eq:alphaconditions} is obeyed for the corresponding values of $\alpha$.

	It is interesting to note that while $\alpha$ increases in the two Calabi--Yau models compared to the one-parameter quintic example, the maximal value of the RHS of \eqref{eq:alphaconditions} has decreased.
	
	Two comments are in order: First, we notice that $\alpha$ obtained in \eqref{alphatwomoduli} is a lower bound. The interpretation of the four-dimensional theory as a circle reduction of a five-dimensional theory potentially breaks down even before we reach the Landau--Ginzburg point once we leave the geometric Calabi--Yau phase. Second, the Landau--Ginzburg point can only be reached when setting the axionic partner of $v^2$ to $\frac12$. Hence, the absolute minimum for the radius is obtained for $\theta=\frac12$. Similarly, in the F-theory examples discussed in the previous section, we observed that $\tau$ can be made the largest if we tune $\theta=\frac12$. This seems to be a general pattern that the parameter that we asymptotically interpret as the circle can be made the smallest if we tune $\theta=\frac12$, even though it might not have this geometric interpretation in these limits. This is not surprising since, e.g. from orbifolds in string theory, we know that shrinking the orbifold curves requires tuning the axion to $\frac12$. It is interesting to see that a similar pattern holds for the radius of circle compactifications from $6$d $\to5$d and $5$d $\to4$d.

	\subsection{Winding \texorpdfstring{$\U(1)$}{U(1)}s}
	\label{sec:windingU1}
	
	In Section \ref{sec:6dto5dheterotic}, we noticed that, when a $D$-dimensional theory is reduced on a circle, additional $\U(1)$s can arise apart from the $\U(1)_{\ttiny{KK}}$ if the $D$-dimensional parent theory has 2-form potentials. Such winding $\U(1)_w$s may give rise to additional CHC constraints in the $(D-1)$-dimensional theory. This may require the existence of a tower of super-extremal particle states, even if the CHC with respect to the $\U(1)_\KK$ is satisfied already by a finite number of states. We will now address this issue further, focusing for definiteness on the circle compactifications of Calabi--Yau threefold compactifications of F-theory. 
	
	To this end, we differentiate between self-dual and anti-self-dual 2-forms in the six-dimensional parent theory. Since the states charged under the $\U(1)$s arising from these higher-form symmetries are wrapped strings, one can equivalently classify the possibilities based on the properties of the strings in six dimensions. For a string charged under a 1-form symmetry in six dimensions with charge $Q$ one distinguishes three cases
	\begin{enumerate*}[before=\unskip{: }, itemjoin={{, }}, itemjoin*={{, and }},label={{\Roman*}.},ref={{\Roman*}}]
		\item\label{case:Q2=0}$Q\cdot Q=0$ 
		\item\label{case:Q2>0}$Q\cdot Q>0$ 
		\item\label{case:Q2<0}$Q\cdot Q<0$.
	\end{enumerate*}
	
	For $Q\cdot Q = 0$, the string charged under the 2-form in six dimensions is a critical string, and after we switch to the duality frame of this string, the analysis in Section \ref{sec:6dto5dheterotic} applies:
	As concluded there, the CHC for a perturbative heterotic $\U(1)_6$ and the winding $\U(1)_w$ requires a tower of super-extremal states, identified as the excitations of the heterotic string. On the other hand, the CHC is not violated if one simply restricts to the KK $\U(1)_\KK$ for radii larger than the minimal radius. By contrast, the CHC between the non-perturbative $\U(1)_\ttiny{n.p.}$ and the KK or winding $\U(1)$s is violated only when the tower of winding states have already crossed the black hole threshold.
	
	In this section, we turn to the remaining two types of winding $\U(1)_w$ and to whether the induced CHC requires an infinite tower of super-extremal particle states in six dimensions. 
	
	\subsubsection{Comments on \texorpdfstring{$Q\cdot Q>0$}{Q.Q>0}: Supergravity String} 
	\label{sec:windingQ>0}
	
	We begin with a $\U(1)_w$ inherited from a self-dual 2-form in six dimensions. 
	States in the five-dimensional theory charged under both $\U(1)_6$ and $\U(1)_w$ arise by dimensionally reducing strings carrying charge $Q$ under the self-dual 2-form in six dimensions. Strings of this type have been analyzed in \cite{Heidenreich:2021yda,Cota:2022yjw} in the context of the (tower) WGC.  Here, we point out that their tension is bounded from below by the Planck scale, so that they are always in the black string region in six dimensions. In fact, they become black strings with a horizon geometry AdS$_3\times S^3$ for a large charge \cite{Haghighat:2015ega}. Therefore, these strings do not give rise to particle-like excitations in six dimensions. Since there exists a minimal radius for the circle compactification, the mass of the wrapped strings can never drop below the black hole threshold. In fact, the WGC for the associated winding $\U(1)_w$ is not expected to be satisfied at the particle level. This is because, in six dimensions, the origin of the 2-form as a reduced 4-form is visible in the EFT. Indeed, the curve giving rise to the self-dual 2-form has a diameter of the order of the minimal radius (see, e.g. \eqref{ellperpP2}) and the would-be 2-form can be resolved within the EFT as arising from a 4-form. In the circle reduced theory, the WGC is not expected to be satisfied at the particle level either, and the same is true for the CHC with the $\U(1)_6$. 
	
	The fact that the absence of a tower of super-extremal particle states in six dimensions does not lead to a violation of the CHC in five dimensions is due to the lack of a T-duality that exchanges the winding $\U(1)_w$ and the KK $\U(1)_\KK$. Without such a duality, there is no way to effectively describe the radii of $S^1$ that are parametrically smaller than $\tau_\ttiny{max}^{-1}$. 
	
	\subsubsection{Comments on \texorpdfstring{$Q\cdot Q<0$}{Q.Q<0}: Field Theory String}
	\label{sec:windingQ<0}
	
	The last class of 2-forms in six dimensions is the anti-self-dual 2-form, whose charges satisfy $Q\cdot Q<0$. The charged strings are field theory strings that become tensionless at finite distance in the moduli space of the six-dimensional theory. The anti-self-dual 2-forms do not admit weak coupling limits, in agreement with the non-perturbative nature of the charged strings throughout the moduli space. 
	Suppose that the six-dimensional theory has an abelian gauge theory $\U(1)_6$ that satisfies the WGC at the particle level. Upon compactification on a circle, the anti-self-dual 2-form gives rise to an additional winding $\U(1)_w$. 
	We now argue that the CHC for $\U(1)_6\times \U(1)_w$ does not require a tower of super-extremal states. 
	
	To this end, consider the circle-reduced theory described by M-theory on an elliptically fibered Calabi--Yau threefold $X_3$. The states charged under $\U(1)_w$ correspond to M2-branes on a curve, $C_B$, with negative self-intersection in the base $B_2$ of $X_3$. The states charged under $\U(1)_6$ are associated with a curve, $C_z$, in the fiber of $X_3$. Both kinds of M2-branes are BPS.\footnote{Notice that even in the case $B_2=dP_9$ discussed in Section~\ref{sec:dP9} in M-theory, the curve $C_z$ can be wrapped by M2-branes leading to BPS states in five dimensions even though the GV invariants for $C_z$ vanish due to enhanced supersymmetry.} We claim that the states carrying charge just under either of the $\U(1)$ gauge factors are already sufficient to satisfy the CHC for $\U(1)_6\times \U(1)_w$. To see this, we now show that the BPS bound for this combination of gauge factors is a straight line in the $(Q_6/M, Q_w/M)$-plane. Since the black hole region must lie within the BPS bound, if the BPS bound is a straight line, the CHC is satisfied by the state with charge $(Q_6,0)$ already present in six dimensions and a BPS state with charge $(0,Q_w)$. 
	
	To verify that the BPS bound is a line in the $(Q_6/M, Q_w/M)$-plane, compactify further to four dimensions. The resulting F-theory on $X_3\times S^1_1 \times S^1_2$ is dual to Type IIA on $X_3$. In the Type IIA duality frame, the states charged under $\U(1)_6$ and $\U(1)_w$ correspond, respectively, to D2-branes wrapped on $C_z$ and $C_B$. For these charges, the BPS bound is a straight line if, in the relevant limit of moduli space, the D2-branes are mutually BPS, i.e. their central charges have the same phase; see \cite{Gendler:2020dfp,Bastian:2020egp} for a detailed analysis of the extremality bound in 4d $\mathcal{N}=2$ theories. 
	
	The limit to consider is the F-theory limit of Type IIA, realized as the small fiber limit with the volume of the base co-scaled to large values. To determine whether the D2-branes in question are mutually BPS, we should analyze the monodromy around the infinite distance divisor in moduli space. If the log-monodromy does not exchange the two states, they are mutually local, and hence the BPS bound is a straight line in the plane of charge-to-mass ratios. For the case at hand, it is, in fact, easier to map the F-theory limit to the large volume limit by means of two T-dualities along the elliptic fiber. The D2-brane on $C_z$ is mapped to itself, while the D2-brane on $C_B$ gets mapped to a D4-brane on $\pi^*(C_B)$ where $\pi:X_3 \rightarrow B_2$ is the projection associated to the elliptic fibration. On the other hand, the KK states for the compactification from six to five dimensions, i.e. D2-branes on the generic elliptic fiber, are mapped to D0-branes.

	In the chosen basis, when acting upon the D4-brane on $\pi^*(C_B)$ the large volume monodromies do not induce any D2-brane charge along $C_z$ but only induce D2-brane charge on the generic elliptic fiber $\mathcal{E}$. To see this, we notice that the central charge for a D4-brane on $\pi^*(C_B)$ generically has the structure 
	\begin{equation}
		Z\left(D4|_{\pi^*(C_B)}\right) = t_\mathcal{E} t_B + q(t_\mathcal{E},t_z)\,,
	\end{equation}
	where $t_\mathcal{E}, t_B, t_z$ are, respectively, the complexified volumes of the generic elliptic fiber $\mathcal{E}$, the curve $C_B$ and additional fibral curves and $q$ is some quadratic function in its arguments whose exact form depends on the precise geometry. The large volume monodromies now act by sending $t_I\rightarrow t_I+1$. When acting on the D4-brane central charge, these monodromies induce D2-brane charge. The monodromy with respect to any fibral curve necessarily induces a D2-brane charge along $C_B$, while the monodromy with respect to $t_B$ only induces a D2-brane charge on the generic elliptic fiber $\mathcal{E}$. As a consequence, none of the log-monodromies maps the D4-brane on $\pi^*(C_B)$ to a D2-brane on $C_z$. Therefore, the D4-brane on $\pi^*(C_B)$ and the D2-brane on $C_z$ are mutually BPS and hence their BPS bound is a straight line in the charge-to-mass ratio plane. Notice that, similarly, log-monodromies also do not map the D4-brane on $\pi^*(C_B)$ to a D0-brane, implying that these two states are mutually BPS. In contrast, the D2-brane on ${\cal E}$ is mapped to a D0-brane by log-monodromy matrices. Since the D0-branes are KK states, this means that the BPS region between $\U(1)_6$ and $\U(1)_\KK$ is curved, whereas the BPS region between $\U(1)_w$ and $\U(1)_\KK$ is a straight line. This is consistent with the analysis of the black hole region for these kinds of $\U(1)$s in \cite{Heidenreich:2015nta}. 
	
	We can translate these results back to the F-theory limit of Type IIA on $X_3$, where we conclude that the BPS bound for $\U(1)_6$ and $\U(1)_w$ is a straight line. This implies that the black hole region is also bounded by this straight line. 
	Thus, we have shown that, as long as there is a single super-extremal charged state under $\U(1)_6$, the CHC for $\U(1)_6$ and $\U(1)_w$ is always satisfied for any radii. This is similar to the discussion in \cite{Heidenreich:2015nta} that showed that the CHC between a KK $\U(1)_\KK$ and a winding $\U(1)_w$ is always satisfied due to the diamond-shaped black hole region. While we argued for this using a chain of dualities, it would be interesting to confirm our result directly in F-theory.

	\section{Discussion and Conclusions}
	\label{sec:Conclusions}

	The goal of this work has been to understand
	the minimal requirements imposed by the WGC on the particle spectrum of a gauge theory coupled to quantum gravity.
	Two main questions have guided our investigation: First, under which conditions must the WGC already be satisfied at the level of particle states, i.e. by states with mass below the black hole threshold?
	Second, when does consistency of the WGC under dimensional reduction necessarily demand the existence of a \textit{tower} \cite{Heidenreich:2015nta,Heidenreich:2016aqi,Montero:2016tif,Andriolo:2018lvp} of super-extremal particle states? 
	
	We have proposed the following answers to these questions.
	First, in \cref{claim:WGC} we conjecture a sufficient condition for when the WGC must hold at the level of particle-like states, whether as part of a tower or as finitely many particles: The gauge theory coupled to gravity should be
	a genuine 0-form theory, as opposed to a defect or a higher-form symmetry in disguise.
	Furthermore, in \cref{claim:existencetower}  we formulate general necessary criteria for when a super-extremal tower of particle
	states can exist below the black hole threshold: This is the case for all weak- and for certain finite distance strong coupling limits.
	Our principal result is encapsulated in Conjecture \ref{Conjecture1}: Whenever the WGC must be satisfied at the level of particle states in $D$ dimensions, 
	these states are inherently sufficient for the WGC to maintain consistency under dimensional reduction to $D-1$ dimensions, even without being part of a tower. 
	This implies that if the theory does not fit the classes outlined in \cref{claim:existencetower}, the WGC remains consistent under circle reduction despite the absence of a tower of super-extremal particle states.
	
	To circumvent a violation of the CHC \cite{Cheung:2014vva} after compactifying such theories on a circle \cite{Heidenreich:2015nta}, we argue for the necessity of a minimal value for the circle radius. Below this threshold, the theory cannot be accurately described as a KK theory, offering a workaround to the need for a tower of super-extremal states, as previously envisioned in~\cite{Heidenreich:2015nta,Andriolo:2018lvp}. 
	In our work, we have argued that such a minimal radius arises whenever the KK tower reaches the black hole threshold, as the circle becomes too small.
	Our arguments, presented in Section \ref{sec:necessitytower}, establish the existence of a finite minimal radius for the circle compactification unless the
	black hole threshold can be increased indefinitely in Planck units. This, in turn, requires a parametrically lowered quantum gravity cutoff, either as a result of the
	limit $r_{S^1} \to 0$ itself or because the original theory is situated near a suitable asymptotic boundary of its moduli space.
	The first scenario occurs when the small radius limit is an emergent string limit, and the second is the case for circle compactifications of perturbative string theories. 
	Following the logic of the Emergent String Conjecture \cite{Lee:2019wij}, these are the only theories where there is no minimal radius.
	For circle compactifications of perturbative heterotic string theories, the minimal radius vanishes in the weak coupling limit. This aligns with the existence of a tower of super-extremal particle states below the black hole threshold, ensuring that the WGC is consistent. More subtle are the emergent string limits corresponding to the regime $r_{S^1} \to 0$.
	In Section \ref{subsec:5t4MA}, we provide independent evidence that the WGC remains consistent under circle reduction even without a particle tower prior to compactification.
	
	We have substantiated these general claims in explicit string theory setups arising from F- and M-theory compactifications on Calabi--Yau threefolds. In such settings, a subset of the charged massive spectrum can be described in a controlled way thanks to current techniques of BPS state counting.
	However, these theories also permit gauge sectors for which establishing a tower of super-extremal states is not straightforward~\cite{Alim:2021vhs,Cota:2022maf}. 
	Moreover, we have argued that certain gauge theories inherently prohibit the emergence of a tower of super-extremal states below the black hole threshold.
	This includes scenarios like
	strong coupling limits in six-dimensional F-theory, where non-critical strings become tensionless.
	In line with our general Conjecture \ref{Conjecture1}, we have identified a minimal radius for circle compactifications in such theories, and we have confirmed in examples that the minimal radius is attained comfortably before any
	clash arises with the CHC in the dimensionally reduced theory. 
	
	Our results suggest that quantum gravity provides a tower of super-extremal particle states whenever it is needed for consistency of the WGC under dimensional reduction. However, one natural question is if all super-extremal towers, if present, are required for the WGC to be consistent under dimensional reduction.
	To answer this, we return to \cref{claim:existencetower}, specifying when super-extremal particle towers
	can occur.
	First, all weakly coupled super-extremal towers, according to Case \ref{case:claim1-i}, are
	required by consistency of the WGC under dimensional reduction:
	Such towers are expected to be either Kaluza--Klein towers themselves or heterotic emergent string towers --- at least according to the Emergent String Conjecture --- and, as explained, in such theories super-extremal towers are needed. 
	This leaves us with the interesting question whether also strongly coupled super-extremal towers as in Case \ref{case:claim1-ii} are actually required for the CHC upon dimensional reduction. 
	Our minimal radius argument should generally hold in such theories, suggesting that any necessity would stem from reasons other than a parametric violation as $r_{S^1} \to 0$.
	There is one instance, however, where even a minimal radius falls short in preventing an inconsistency for the WGC in the absence of an infinite number of super-extremal states.
	This occurs if all the (super-)extremal states are strictly extremal. Examples of this scenario are certain BPS states for which the extremality bound coincides with the BPS bound. In such theories, the CHC is violated after circle reduction for \textit{all} values of the radius, unless an infinite tower of extremal states with increasing charge exists~\cite{Heidenreich:2015nta}. 
	
	Consequently, a strongly coupled tower is needed for the CHC if it exclusively comprises extremal states, with no other super-extremal states in the mix. 
	In M-theory on a Calabi-Yau threefold, strongly coupled super-extremal towers are BPS, although it remains unclear whether for them the BPS bound and extremality bound align~\cite{Alim:2021vhs}. If they do, then the tower of strongly coupled states is indeed necessary to satisfy the CHC upon circle reduction. 
	Resolving this question requires a closer examination of the relationship between extremality and the BPS bound for strongly coupled M-theory states.
	Even if the BPS and extremality differ, it is essential to note that upon circle compactification, the central charge of the strongly coupled states receives non-trivial quantum corrections. This contrasts with the behavior of the central charge of conifold curves upon circle reduction, which does not receive quantum corrections.
	Understanding whether this qualitative difference is linked to the necessity of the tower of super-extremal states in cases where BPS and extremality bounds do not align is a compelling avenue for further exploration.
	
	Subject to this significant uncertainty, \cref{claim:existencetower}, coupled with the Emergent String Conjecture and the existence of a minimal radius as discussed in Section \ref{sec:necessitytower},  suggests a stronger version of
	Conjecture~\ref{Conjecture1}: 
	
	\begin{mdframed}[backgroundcolor=white,shadow=true,shadowsize=4pt,shadowcolor=seccolor,roundcorner=8pt]
		\begin{conjecture}[Minimal Weak Gravity Conjecture]\label{Conjecture2}
			Towers of (super-)extremal particle states below the black hole threshold exist
			\textit{if and only if} they are required by consistency of the WGC under dimensional reduction. This is the case for either emergent string limits, Kaluza--Klein reductions with KK gauge bosons, or strongly coupled limits with exactly extremal states.
		\end{conjecture}
	\end{mdframed}
	
	\vspace{2mm}
	
	We hope to further put such a \textit{Minimal Weak Gravity Conjecture} to the test in the future.

	\acknowledgments
	
	The authors thank Rafael \'Alvarez-Garc\'ia, Ivano Basile, Florent Baume, Ralph Blumenhagen, Niccolò Cribiori, Mateo Galdeano, Naomi Gendler, Alvaro Herráez, Dieter L\"ust, Fernando Marchesano, Paul-Konstantin Oehlmann, Marco Scalisi, Damian van de Heisteeg,  Cumrun Vafa and Xin Wang for helpful discussions. A. M. thanks LMU and MMP for the hospitality during the finalization of this work. 
	C. F. C., A. M. and T. W. are supported in part by Deutsche Forschungsgemeinschaft under Germany's Excellence Strategy EXC 2121  Quantum Universe 390833306 and by Deutsche Forschungsgemeinschaft through a German-Israeli Project Cooperation (DIP) grant ``Holography and the Swampland". M. W. is supported in part by a grant from the Simons Foundation (602883, CV) and also by the NSF grant PHY-2013858. 
	
	\appendix

	\section{Conventions}
	\label{app:Conventions}
	Let us write down explicitly our conventions for the map between M-theory and Type IIA. 
	Let us first follow the convention by \cite{Polchinski:1998rr}, and write the bosonic part of the $11$d M-theory effective action, with $2\kappa_{\ttiny{11d}}^2 = M_{\ttiny{11d}}^{-9}(2\pi)^8$, as
	\begin{equation}
		S_\ttiny{11d} = \frac{M_\ttiny{11d}^9}{(2\pi)^8} \int_{\RR^{1,10}}\left(R\star \ID -\frac{1}{2}d C_3\wedge \star d C_3\right)+\ldots \,.
		\label{eq:S11}
	\end{equation}
	In this way, the Type IIA effective action is obtained by compactifying M-theory on a circle $S^1_M$ with radius $r_{S^1_M}$, and identifying
	\begin{equation}\label{eq:mapMtoIIA-1}
		M_s^2 = \frac{1}{\alpha'} = M_{\ttiny{11d}}^3r_{S^1_M} \coma g_\IIA^2  = M_{\ttiny{11d}}^3 r_{S^1_M}^3  \fstop
	\end{equation}
	In particular, we have $2\kappa_{s,\,10}^2 = M_s^{-8}(2\pi)^7$ in string frame and $2\kappa_{10}^2 = M_s^{-8}(2\pi)^7g_\IIA^2$ in Einstein frame, so that
	\begin{equation}
		\frac{M_{\PlD{10}}^8}{2} = \frac{M_s^8}{(2\pi)^7g_\IIA^2}\fstop 
		\label{eq:MPl10-IIA}
	\end{equation}
	Another convention for the effective actions (which is the one we have used in the main text) is obtained by rescaling $M_{\ttiny{11d}}$ and $M_s$ by a $2\pi$ factors, so that \eqref{eq:S11} is 
	\begin{equation}
		S_\ttiny{11d} = 2\pi M_\ttiny{11d}^9 \int_{\RR^{1,10}}\left(R\star \ID -\frac{1}{2}d C_3\wedge \star d C_3\right)+\ldots \,,
		\label{eq:S11-2}
	\end{equation}
	and analogously $2\kappa_{s,\,10}^2 = M_s^{-8}(2\pi)^{-1}$. By this redefinition, \eqref{eq:mapMtoIIA-1} becomes
	\begin{equation}
		M_s^2 = \frac{1}{(2\pi)^2\alpha'} = 2\pi M_{\ttiny{11d}}^3r_{S^1_M} \coma g_\IIA^2  = (2\pi)^3M_{\ttiny{11d}}^3 r_{S^1_M}^3  \fstop
		\label{eq:mapMtoIIA-2}
	\end{equation}
	In this way, \eqref{eq:MPl10-IIA}, which now is 
	\begin{equation}
		\frac{M_{\PlD{10}}^8}{2} = 2\pi \frac{M_s^8}{g_\IIA^2}\coma
		\label{eq:MPl10-IIA-2}
	\end{equation}
	is obtained by the compactification of \eqref{eq:S11-2} on a circle. 
	Both \cref{eq:mapMtoIIA-1,eq:mapMtoIIA-2} give
	\begin{equation}
		M_\ttiny{11d} = \frac{M_s}{g_\IIA^{1/3}}\fstop
		\label{eq:M11dinMsgIIA}
	\end{equation}
	With this convention, since $\ell_s = 2\pi \sqrt{\alpha'}$, we can identify
	\begin{equation}
		M_s = \frac{1}{\ell_s} \coma M_\ttiny{11d} = \frac{1}{\ell_\ttiny{11d}}\coma
	\end{equation}
	and express dimensional volumes in the corresponding units using $\ell_{\ldots}$ and $M_{\ldots}$ interchangeably.
	
	We can now consider the duality between M-theory and Type IIB string theory, via T-duality between Type IIA and Type IIB. Consider M-theory compactified on a circle $S^1_M$ of radius $r_{S^1_M}$. This can be mapped to type IIA via \eqref{eq:mapMtoIIA-2}. Then, compactify on another circle $S^1_\IIA$ of radius $r_{S^1_\IIA}$. This realizes M-theory compactified on a torus $T^2$ defined by the radii $r_{S^1_M}$ and $r_{S^1_\IIA}$. Type IIA compactified on a circle $S^1_\IIA$ is T-dual to Type IIB compactified on a circle $S^1_\IIB$ defined by \cite{Polchinski:1998rr}
	\begin{equation}
		r_{S^1_\IIB} = \frac{\alpha'}{r_{S^1_\IIA}} \coma \frac{r_{S^1_\IIB}}{g_\IIB^2} = \frac{r_{S^1_\IIA}}{g_\IIA^2}\fstop
		\label{eq:IIAtoIIB}
	\end{equation}
	These relations allow relating the radius $r_{S^1_\IIB}$ in Type IIB with the torus $T^2$ simply replacing $M_s$ and $g_\IIA$ in M-theory units using \eqref{eq:mapMtoIIA-2}. Eventually, we find
	\begin{equation}
		g_\IIB =\frac{r_{S^1_M}}{r_{S^1_\IIA}} \coma   r_{S^1_\IIB}M_{\ttiny{11d}} = \frac{1}{2\pi\text{vol}(T^2)M_\ttiny{11d}^2}\coma 
		\label{eq:M-thvsTypeIIB}
	\end{equation}
	where we defined $\text{vol}(T^2)=\left(2\pi r_{S^1_M}\right)\left(2\pi r_{S^1_\IIA}\right)$. Finally, we can express $g_\IIA^2$ in terms of Type IIB data using \eqref{eq:IIAtoIIB}, that is
	\begin{equation}
		g_\IIA = \frac{g_\IIB}{2\pi M_sr_{S^1_\IIB}}\coma
		\label{eq:gIIAintypeIIB}
	\end{equation} 
	and, using respectively \eqref{eq:M11dinMsgIIA} and \eqref{eq:gIIAintypeIIB} in \eqref{eq:M-thvsTypeIIB}, one obtains the relation between $r_{S^1_\IIB}$ and the volume of $T^2$ completely in Type IIB units, i.e.
	\begin{equation}
		r_{S^1_\IIB} = \frac{1}{2\pi\text{vol}(T^2)M_\ttiny{11d}^3} \stackrel{\text{\eqref{eq:M11dinMsgIIA}}}{=}\frac{g_\IIA}{2\pi\text{vol}(T^2)M_s^3} \stackrel{\text{\eqref{eq:gIIAintypeIIB}}}{\Longrightarrow} r_{S^1_\IIB}^2M_s^2 = \frac{g_\IIB}{(2\pi)^2\text{vol}(T^2)M_s^2}\fstop
		\label{eq:rIIBvsT2inIIBunits}
	\end{equation}
	
	\cref{eq:M11dinMsgIIA,eq:gIIAintypeIIB} are also useful to write $M_\ttiny{11d}$ only in Type IIB units, namely,
	\begin{equation}
		M_\ttiny{11d} =\frac{(2\pi)^{1/3}M_s^{4/3} r_\IIB^{1/3}}{  g_\IIB^{1/3}}\fstop
	\end{equation}
	Moreover, using \eqref{eq:rIIBvsT2inIIBunits}, we obtain
	\begin{equation}
		M_\ttiny{11d} = \frac{M_s^{2/3}}{ g_\IIB^{1/6} \text{vol}(T^2)^{1/6}}\fstop
	\end{equation}
	If we define $\tau =\text{vol}(T^2)M_\ttiny{11d}^2$ and $\tau_s=\text{vol}(T^2)M_s^2$, we finally get
	\begin{equation}
		M_\ttiny{11d} = \frac{M_s}{ g_\IIB^{1/4} \tau^{1/4}}\quad \text{ or }\quad M_\ttiny{11d} = \frac{M_s}{ g_\IIB^{1/6} \tau_s^{1/6}}\fstop
		\label{eq:M11dinMsgIIB}
	\end{equation}

	
\newpage

\bibliographystyle{JHEP}
\bibliography{mybib}
	
\end{document}